\documentclass[12pt]{book}

\usepackage{susspss} 
\usepackage{graphicx}     
\usepackage{epsfig}
\usepackage{makeidx}
\makeindex

\begin{document}

\headings{{\boldmath $B$} Phenomenology} 
{$B$ Phenomenology}
{Sheldon Stone}
{Department of Physics\\
Syracuse University, USA, 13244-1130\\
Email: Stone@physics.syr.edu}

\centerline{ABSTRACT}
{Many topics are discussed in $b$ physics including lifetimes, decay
mechanisms, determinations of the CKM matrix elements $|V_{cb}|$ and
$|V_{ub}|$, facilities for $b$ quark studies, neutral $B$ meson mixing, rare
$b$ decays, hadronic decays and CP violation. I review techniques for finding
physics beyond the Standard Model and describe the complimentarity of
$b$ decay measurements in
elucidating new physics that could be found at higher energy machines.}
\vspace{35mm} 
\begin{flushleft}
.\dotfill .
\end{flushleft}
Lectures presented at 55th Scottish Universities Summer School in Physics on
{\it Heavy Flavour Physics}, A NATO Advanced Study Insititute, St. Andrews,
Scotland, August, 2001.
\tableofcontents
\section{Introduction: The Standard Model and $B$ Decays}

Studies of $b$ and $c$ physics are focused on two main goals. The first is to
look for new phenomena beyond the Standard Model. The second is to measure
Standard Model parameters including CKM elements and decay constants. These
lectures concern ``$B$ Phenomenology," a topic so broad that it can include
almost anything concerning $b$ quark decay or production. I will cover
an eclectic ensemble of topics that I find interesting and hope will be
educational.

\subsection{Theoretical Basis}
The physical states of the ``Standard Model"  are comprised of  left-handed 
doublets containing leptons and quarks and right handed singlets (Rosner 2001) 
\begin{eqnarray}
&\left(\begin{array}{c}u\\d\end{array}\right)_L
\left(\begin{array}{c}c\\s\end{array}\right)_L
\left(\begin{array}{c}t\\b\end{array}\right)_L,&~u_R,~d_R,~c_R,~s_R,~t_R,~b_R\\
&\left(\begin{array}{c}e^-\\\nu_e\end{array}\right)_L
\left(\begin{array}{c}\mu^-\\\nu_{\mu}\end{array}\right)_L
\left(\begin{array}{c}\tau^-\\\nu_{\tau}\end{array}\right)_L,&~e^-_R,
~\mu^-_R,~\tau^-_R,~{\nu_e}_R,~{\nu_{\mu}}_R,~{\nu_{\tau}}_R.
\end{eqnarray}
 
The gauge bosons, $W^{\pm}$, $\gamma$ and $Z^o$ couple to  
mixtures of the physical $d,~ s$ and $b$ states. This mixing is described
by the Cabibbo-Kobayashi-Maskawa (CKM) matrix (see below) (Kobayashi 1973).

The Lagrangian for charged current weak decays is
\begin{equation}
L_{cc}=-{g\over\sqrt{2}}J^{\mu}_{cc}W^{\dagger}_{\mu}+h.c., \label{eq:lagrange}
\end{equation}
where
\begin{equation}
J^{\mu}_{cc} =\left(\bar{\nu}_e,~\bar{\nu}_{\mu},~\bar{\nu}_{\tau}\right)
\gamma^\mu V_{MNS}\left(\begin{array}{c}e_L\\ \mu_L\\ \tau_L\\\end{array}\right) +
\left(\bar{u}_L,~\bar{c}_{L},~\bar{t}_{L}\right)\gamma^\mu V_{CKM} 
\left(\begin{array}{c}d_L\\  s_L\\ b_L\\ \end{array}\right) 
\end{equation}
and
\begin{equation}
V_{CKM} =\left(\begin{array}{ccc} 
V_{ud} &  V_{us} & V_{ub} \\
V_{cd} &  V_{cs} & V_{cb} \\
V_{td} &  V_{ts} & V_{tb}  \end{array}\right).\end{equation}

Multiplying the mass eigenstates $(d_L, s_L, b_L)$ by the CKM matrix leads to the
weak eigenstates. $V_{MNS}$ is the analogous matrix required for massive
neutrinos (we will not discuss this matrix any further). There are nine complex CKM elements. These 18 
numbers can be reduced to four independent quantities by applying unitarity 
constraints and the fact that the phases of the quark wave functions are arbitrary. 
These four remaining numbers are {\bf fundamental constants} of nature that 
need to be determined from experiment, like any other
fundamental constant such as $\alpha$ or $G$. In the Wolfenstein 
approximation the matrix is written in order $\lambda^3$ for the real part
and $\lambda^4$ for the imaginary part as (Wolfenstein 1983)
\begin{equation}
V_{CKM} = \left(\begin{array}{ccc} 
1-\lambda^2/2 &  \lambda & A\lambda^3(\rho-i\eta)(1-\lambda^2/2) \\
-\lambda &  1-\lambda^2/2-i\eta A^2\lambda^4 & A\lambda^2(1+i\eta\lambda^2) \\
A\lambda^3(1-\rho-i\eta) &  -A\lambda^2& 1  
\end{array}\right).
\end{equation}
The constants $\lambda$ and $A$ are determined from charged-current weak 
decays. The measured values are $\lambda = 0.2205\pm 0.0018$ and
A=0.784$\pm$0.043. There are constraints on $\rho$ and $\eta$ from other
measurements that we will discuss. Usually the matrix is viewed only up to
order $\lambda^3$. To explain CP violation in the $K^o$ system the term of
order $\lambda^4$ in $V_{cs}$ is necessary. 

\subsubsection{Determination of $G_F$}
\begin{figure}[hbtp]
\vspace{-.2cm}
\centerline{\epsfig{figure=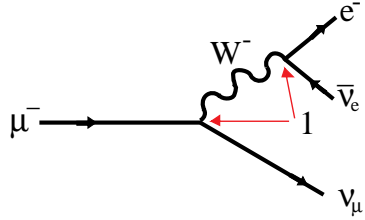, width=2.2in}}
\vspace{-1.2cm}
\caption{\label{mu_decay}Diagram for muon decay.}
\end{figure}

Muons, being lighter than the lightest hadrons, must decay purely into leptons.
The process is $\mu^-\to e^-\overline{\nu}_e \nu_{\mu}$ as shown in 
Figure~\ref{mu_decay}.

 The total width for
this decay process is given by 
\begin{equation}
\Gamma_{\mu}={G_F^2\over 192\pi^3} m^5_{\mu}\times {\rm (phase~space)}
\times{\rm (radiative~corrections)}~~.
\label{eq:mudecay}
\end{equation}
Since $\Gamma_{\mu}\cdot \tau_{\mu}=\hbar$, measuring the muon lifetime gives a
direct measure of $G_F$.

\subsubsection{Determination of {\boldmath $|V_{us}|$}}

\begin{figure}[hbtp]
\vspace{-.4cm}
\centerline{\epsfig{figure=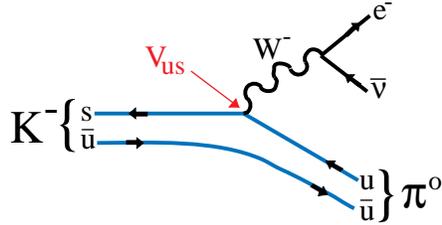, width=2.7in}}
\vspace{-.2cm}
\caption{\label{k_decay}Semileptonic $K^-$ decay diagram.}
\end{figure}

A charged current decay diagram for strange quark decay is shown in
Figure~\ref{k_decay}. Here 
the CKM element $V_{us}$ is present. The decay rate is given by a formula similar 
to equation~(\ref{eq:mudecay}), with the muon mass replaced by the $s$-quark mass and an 
additional factor of $|V_{us}|^2$. Two complications arise since we are now 
measuring a decay process involving hadrons, $K^-\to \pi^o e^-\bar{\nu}$ rather 
than elementary constituents.  One is that the $s$-quark mass is not well  defined 
and the other  is that we must make corrections for the probability that the 
$\bar{u}$-spectator-quark indeed forms a $\pi^o$ with the ${u}$-quark  from the 
$s$-quark decay. These considerations will be discussed in greater detail in the 
semileptonic $B$ decays section. Fortunately there are theoretical calculations
that allow for a relatively precise measurement because they deal with hadron rather
than quark masses and have good constraints on the form-factors; using the
models we have 
$\lambda = V_{us}=0.2205\pm 0.0018$. $A$ is found by measuring $V_{cb}$ in
semileptonic $b$ decays and 
constraints on $\rho$ and $\eta$ are found from other measurements. These will
also be discussed later.

\subsubsection{Plan for Using Semileptonic \boldmath{$B$} Decays to Determine 
\boldmath{$V_{cb}$} and \boldmath{$V_{ub}$}}

Semileptonic $b$ decays  arise from a similar diagram to Figure~\ref{k_decay}, 
where the $b$ quark replaces the $s$ quark. In this case the $b$ quark can
decay into either the $c$ quark or the the $u$ quark, so  
we can use these decays to determine
$V_{cb}$ and $V_{ub}$ providing we have three ingredients
\begin{enumerate}
\item $B$ lifetimes
\item Relevant $B$  branching fractions
\item Theory or model to take care of hadronic physics.
\end{enumerate}

\subsection{Lifetime Measurements}
The ``$b$-lifetime" was first measured at the 30 GeV $e^+e^-$ colliders
PEP and PETRA where $b$-quarks were produced via the diagrams shown in
Figure~\ref{epemtobbX}. The measured the average lifetime of all $b$-hadron
species. The distribution they found most useful was the ``impact parameter,"
which is the minimum distance of approach of a track from the primary
production vertex.  This distance is related to the lifetime (Atwood 1994). 

A more direct measurement would be to measure the actual 
decay distance $L$ and the momentum of the $b$ hadron. Then, since
$L=\gamma\beta c t$, where $t$ is the decay time of the 
individual particle (also called the proper time)
$\beta=p/E$ and $\gamma=E/m_b$, the distribution of decay
times $t$ can be derived
Events will be distributed exponentially in $t$ as $e^{-t/\tau}$, with $\tau$
being the lifetime.
Uncertainty results from errors on $L$, momenta and contributions of 
backgrounds.

\begin{figure}[hbtp]
\centerline{\epsfig{figure=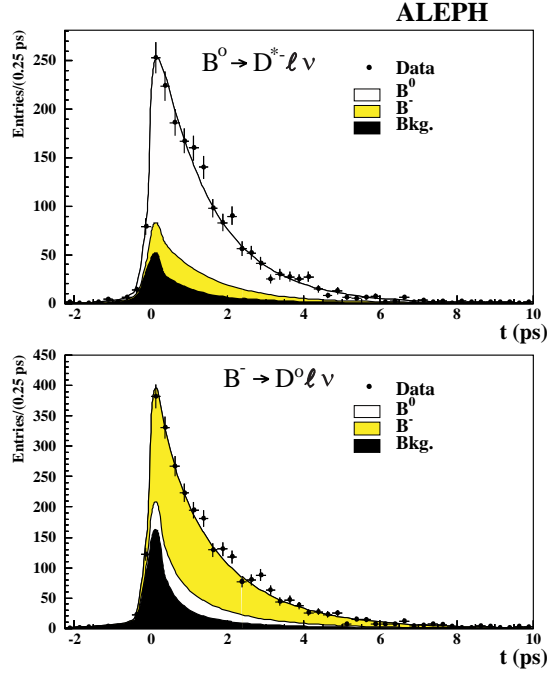,width=3in}}
\caption{\label{aleph_life} Proper time distributions of exclusive semi-leptonic
decays from ALEPH.}
\end{figure}

Precision lifetimes of individual $b$-flavored hadrons have been measured at LEP where
the production process is
$e^+e^-\to Z^o\to b\overline{b}$ and at CDF in 1.8 TeV $p\overline{p}$
collisions. Large samples of semileptonic decays have been used to determined
the $B^o$ and $B^-$ lifetimes. (Note that the CPT theorem guarantees that
the lifetime of the anti-particles is the same as the particles.)
The decay distributions for two semileptonic $B$ decay channels are shown in 
Figure~\ref{aleph_life}. The $B^o\to D^{*-}\ell^+\nu$ channel has mostly signal
with some background from $B^-$ decays and other backgrounds as indicated in
the figure. It takes a great deal of careful work to accurately estimate these
background contributions. The clear exponential lifetime shapes can be seen in
these plots. Some data has also been obtained using purely hadronic final
states (Sharma 1994).

\begin{figure}[hbtp]
\centerline{\epsfig{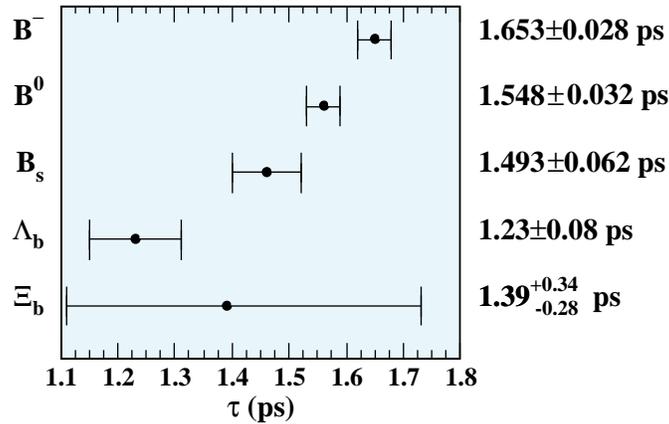}}
\caption{\label{lepblife_all}Measured lifetimes of different $B$ species.}
\end{figure}

A summary of the lifetimes of specific $b$-flavored hadrons is given in 
Figure~\ref{lepblife_all} (Groom 2001). Note that the ratio of $B^+$ to $B^o$ lifetimes 
is 1.074$\pm$0.028, a 2.6$\sigma$ difference from unity, bordering on
significance. 
Also, the $\Lambda_b$ lifetime is much shorter than
the $B^o$ lifetime. According to proponents of the 
Heavy Quark Expansion model, there should be at most a 10\% difference between
them (Bigi 1997). To understand lifetime differences we must first analyze
hadronic decays.

\subsection{{\boldmath $B$} Decay Mechanisms}

\begin{figure}[hbtp]
\centerline{\epsfig{figure=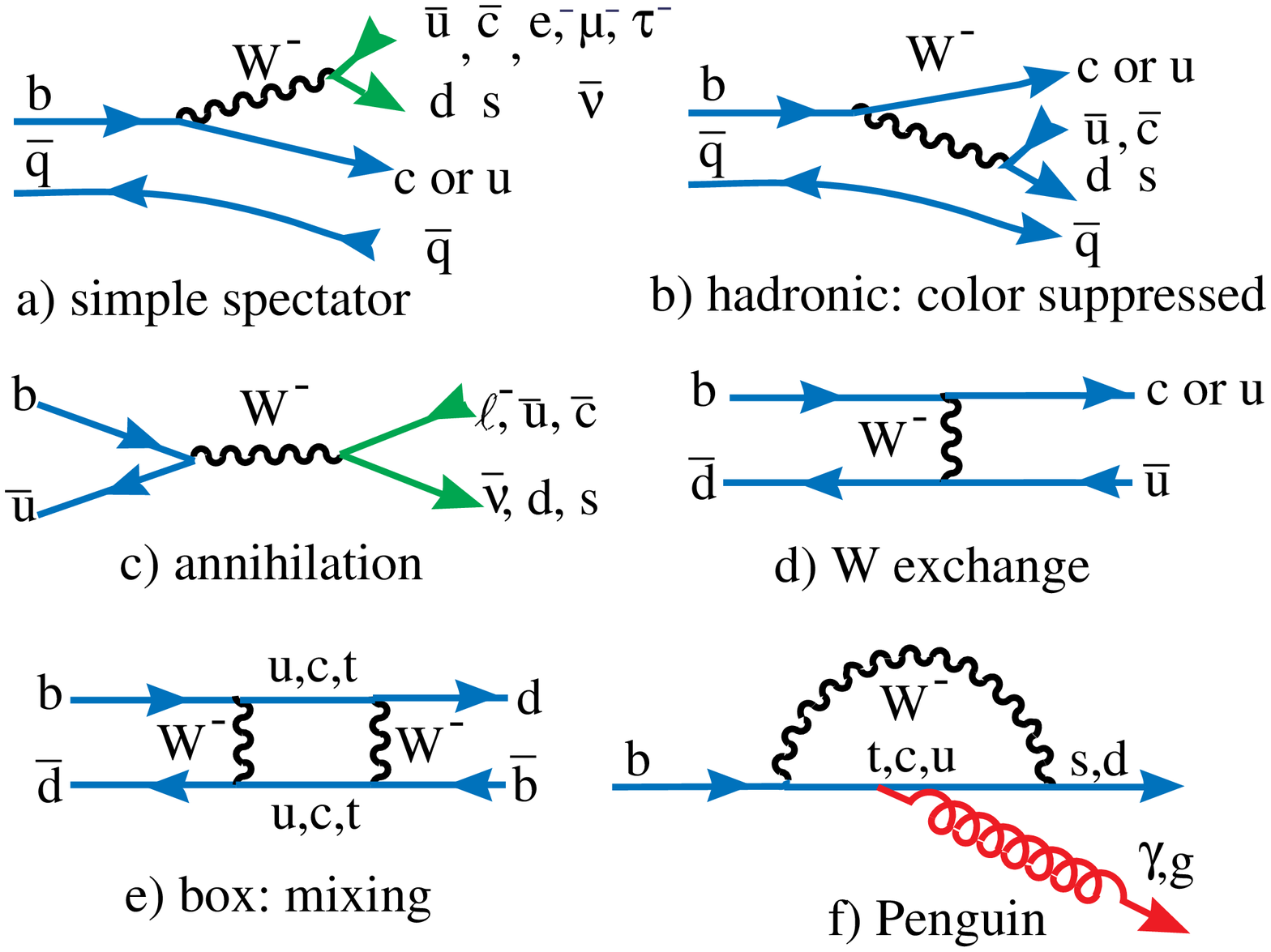,width=5in}}
\vspace{-.1cm}
\caption{\label{Bdecaymech}Various mechanisms for $B$ meson decay.}
\end{figure}

Figure~\ref{Bdecaymech} shows sample diagrams for $B$ decays. Semileptonic
decays  are depicted in Figure~\ref{Bdecaymech}(a), when the virtual
$W^-$ materializes as a lepton-antineutrino pair. The name ``semileptonic" is
given, since  there are both hadrons and leptons in the final state.  The
leptons arise from the  virtual $W^-$, while the hadrons come from the coupling
of the spectator anti-quark with either the $c$ or $u$ quark from the $b$ quark
decay.  Note that the $B$ is  massive enough that all three lepton species can
be produced. The simple spectator diagram for hadronic decays
(Figure~\ref{Bdecaymech}(a)) occurs when the virtual  
$W^-$ materializes as a quark-antiquark pair, 
rather than a lepton pair.  The terminology  {\it simple spectator} comes from
viewing the decay of the $b$ quark, while  ignoring the presence of the {\it
spectator} antiquark. If the colors of the quarks  from the virtual $W^-$ are
the same as the initial $b$ quark, then the color  suppressed diagram, 
Figure~\ref{Bdecaymech}(c), can occur. While the amount of  color suppression is
not well understood, a good first order guess is that these modes  are
suppressed in amplitude by the color factor 1/3 and thus in rate by 1/9, with 
respect to the non-color suppressed spectator diagram.

The annihilation diagram shown in Figure~\ref{Bdecaymech}(c) occurs when the $b$ 
quark and spectator anti-quark find themselves in the same space-time region and 
annihilate by coupling to a virtual $W^-$. The probability of such a wave function 
overlap between the $b$ and $\bar{u}$-quarks is proportional to a numerical 
factor called $f_B$.  The decay amplitude is also proportional to the coupling 
$V_{ub}$. The mixing and penguin diagrams will be discussed later.

Each diagram contributes differently to the decay width of the individual
species. Diagram (a) is expected to be dominant. There are even more diagrams
expected for baryons. 
Currently there is no direct evidence for diagrams (c) and (d), although (c) is
expected to occur, indeed it would be responsible for the purely leptonic decay
$B^-\to\tau^-\overline{\nu}$. 

The semileptonic decay width, $\Gamma_{sl}$, is defined as the decay rate in
units of inverse seconds into a hadron (or hadrons) plus a lepton-antineutrino
pair. (Decay rates can also be expressed in units of MeV by multiplying by
$\hbar$.) $\Gamma_{sl}$ is related to the semileptonic branching ratio
${\cal B}_{sl}$ and the lifetime $\tau$ as
\begin{equation}
\Gamma_{sl}= {\cal B}_{sl}\cdot \Gamma_{total}={\cal B}_{sl}/\tau~~.
\end{equation}
The semileptonic width should be equal for all $b$ species. This is true for
$D^o$ and $D^+$ mesons, even though their lifetimes differ by more than a
factor of two. Thus, it is differences in the hadronic widths among the
different $b$ species that drive the lifetime differences.

\begin{figure}[hbtp]
\centerline{\epsfig{figure=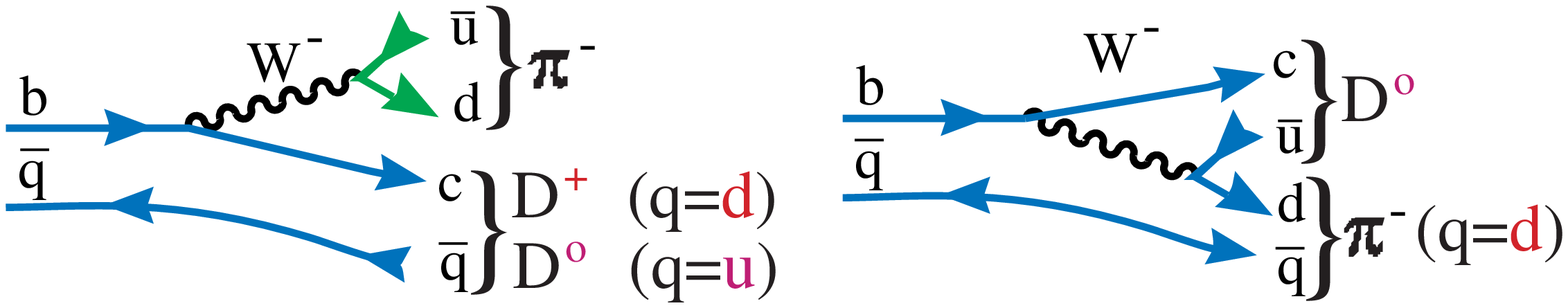,width=4.5in}}
\vspace{-.1cm}
\caption{\label{DSPI} (left) Spectator diagram for $\overline{B}^o\to D^+\pi^-$
and $B^-\to D^o\pi^-$. (right) Color suppressed spectator diagram for
$B^-\to D^o\pi^-$ only.}
\end{figure}

Let us now consider the case of $\overline{B}^o$ and $B^-$ lifetime
differences. There is some indication that the $\overline{B}^o$ has a shorter
lifetime, that would imply that there are more decay channels available.
Figure~\ref{DSPI} shows the color allowed and color suppressed decay diagrams for
{\it two-body} decays into a ground-state charmed meson and a $\pi^-$. The
color suppressed diagram only exists for the $B^-$. The relative rate
\begin{equation}
{\Gamma(B^-\to D^o\pi^-)\over \Gamma(\overline{B}^o\to D^+\pi^-)}=1.8\pm 0.3~~,
\end{equation}
and the same is true for all other similar two-body channels, such as
$D^*\rho^-$.
Thus we would expect, if
most $B$ decays are given by these diagrams that the $B^-$ would have a shorter
lifetime than the $\overline{B}^o$, opposite of what the data suggests.
An explanation is that this ratio reverses for higher multiplicity decays, but
this is an interesting discrepancy that needs to be kept in mind.

\section{Semileptonic $B$ Decays}
\subsection{Formalism of Exclusive Semileptonic $B$ Decays}

The same type of semileptonic charged current decays used to find $V_{us}$ are
used to find $V_{cb}$ and $V_{ub}$. The basic diagram is shown in 
Figure~\ref{Bdecaymech}(a). We can use either inclusive decays, where we look
only at the lepton and ignore the hadronic system at the lower vertex, or
exclusive decays where we focus on a particular single hadron. Theory currently
can  predict either the inclusive decay rate, or the exclusive decay rate when
there is only a single  hadron in the final state.  

Now let us briefly go through the mathematical formalism of semileptonic
decays. We start with pseudoscalar $B$ to pseudoscalar $m$ transitions. The decay
amplitude is given by (Grinstein 1986)(Gilman 1990) 
\begin{equation}
A(\bar{B}\to me^-\bar{\nu})={G_F\over\sqrt{2}}V_{ij}L^{\mu}H_{\mu}{\rm ,~where} 
\end{equation}
\begin{equation}
L^{\mu}=\bar{u}_e\gamma^{\mu}\left(1-\gamma_5\right)v_{\nu} {\rm ,~and}
\end{equation}
\begin{equation}
H_{\mu}=\langle m|J^{\mu}_{had}(0)|\overline{B}\rangle=
f_+(q^2)(P+p)_{\mu}+f_-(q^2)(P-p)_{\mu},
\end{equation}
where $q^2$ is the four-momentum transfer squared between the $B$ and the $m$,
and $P(p)$ are four-vectors of the $B(m)$.
$H_{\mu}$ is the most general form the hadronic matrix element can have. It
is written in terms of the unknown $f$ functions that are called ``form-factors." 
It turns out that the term multiplying the $f_-(q^2)$ form-factor is the mass
of lepton squared. Thus for electrons and muons (but not $\tau$'s), the decay 
width is given by
\begin{equation}
{d\Gamma_{sl}\over dq^2} = {G_F^2|V_{ij}|^2K^3\over 
24\pi^2}|f_+(q^2)|^2, {\rm ~~where}
\end{equation}
\begin{equation}
K={1\over 2M_B}\left[\left(M_B^2+m^2-q^2\right)^2-4m^2M_B^2\right]^{1/2}
\end{equation}
is the momentum of the particle $m$ (with mass $m$) in the $B$ rest frame.
In principle, ${d\Gamma_{sl}/ dq^2}$ can be measured over all $q^2$. Thus
 the shape of $f_+(q^2)$ can be determined experimentally. However, the 
normalization, $f_+(0)$ must be obtained from theory, for $V_{ij}$ to
be measured. In other words,
\begin{equation}
\Gamma_{SL} \propto |V_{ij}|^2|f_+(0)|^2{1\over \tau_B}\int K^3 g(q^2)dq^2,
\end{equation} 
where $g(q^2) =f_+(q^2)/f_+(0)$. Measurements of semileptonic $B$ decays 
give the integral term, while the lifetimes are measured separately, allowing
the product $|V_{ij}|^2|f_+(0)|^2$ to be experimentally determined. 

For pseudoscalar to vector transitions there are three independent
form-factors whose shapes and normalizations must be determined
(Richman 1995).

\subsection{Measurement of {\boldmath $|V_{cb}|$}}

\subsubsection{Heavy Quark Effective Theory and  
$\bar{B}\to D^{*}\ell^-\bar{\nu}$}

We can use exclusive $B$ decays to find $V_{cb}$ coupled with
``Heavy Quark Effective Theory" (HQET) (Isgur 1994).
We start with
a quick introduction to this theory. It is difficult to solve QCD at long distances, but it is
possible at short distances. Asymptotic freedom, the fact that the strong coupling 
constant 
$\alpha_s$ becomes weak in processes with large $q^2$, allows perturbative 
calculations. Large distances are of the order $\sim 1/\Lambda_{QCD}\sim$1 fm, 
since
$\Lambda_{QCD}$ is about 0.2 GeV. Short distances, on the other hand, are of the 
order
of the quark Compton wavelength; $\lambda_Q\sim 1/m_Q$ equals 0.04 fm for the 
$b$
quark and 0.13 fm for the $c$ quark. 

For hadrons, on the order of 1 fm, the light quarks are sensitive only to the heavy 
quark's
color electric field, not the flavor or spin direction. Thus, as $m_Q\to\infty$, 
hadronic
systems which differ only in flavor or heavy quark spin have the same configuration of 
their light degrees of freedom. The following two predictions follow 
immediately
(the actual experimental values are shown below):
\begin{eqnarray}
m_{B_s}-m_{B_d} &= &m_{D_s}-m_{D^+}\\
(90.2\pm 2.5) {\rm ~MeV}  &    & (99.2\pm 0.5){\rm ~MeV  ~, ~and} \nonumber \\
m^2_{B^*}-m^2_{B} &= &m^2_{D^*}-m^2_{D}  .\\
0.49 {\rm ~GeV^2}  &    & 0.55 {\rm ~GeV^2}. \nonumber
\end{eqnarray}

The agreement is quite good but not exceptional.
Since the charmed quark is not that heavy, there is some heavy quark symmetry 
breaking. This must be accounted for in quantitative predictions, and can
probably explain the discrepancies above. The basic idea is that
if you replace a $b$ quark with a $c$ quark moving at the same {\bf velocity},
there should only be  small and calculable changes.

In lowest order HQET there is only one form-factor function $F$ which is a 
function of
the Lorentz invariant four-velocity transfer $\omega$, where
\begin{equation}
\omega = {{M^2_B+M^2_{D^*}-q^2}\over {2M_BM_{D^*}}}.
\end{equation}
The point $\omega$ equals one corresponds to the situation where the $B$ decays to a
$D^*$ which is at rest in the $B$ frame. Here the ``universal" form-factor function
$F(\omega)$ has the value, $F(1)=1$, in lowest order. This is the point in phase space 
where the $b$ quark changes to a $c$ quark with zero velocity transfer.  The idea
is to measure the decay rate at this point, since we know the value of the form-factor,
namely unity, and then apply the hopefully small and hopefully well understood 
corrections.  Although this analysis can be applied to 
$\bar{B}\to D\ell^-\bar{\nu}$, overall decay rate is only
40\% of $D^*\ell^-\bar{\nu}$ and the decay rate vanishes at $\omega$ equals 1 
much faster, making the measurement worse.  

\subsubsection{Detection of {\boldmath $B\to D^*\ell\nu$}}
Since this is a semileptonic final state containing a missing neutrino, the
decay cannot be identified or reconstructed by merely measuring the 4-vectors
of the final state particles. One technique used in the past relies on
evaluating the missing mass ($MM$) where
\begin{eqnarray}
MM^2&=&(E_{B}-E_{D^*}-E_{\ell})^2-(\overrightarrow{p}_{\!B}-
\overrightarrow{p}_{\!D^*}-\overrightarrow{p}_{\!\ell})^2\\
&=& M^2_B+M^2_{D^*}+M^2_{\ell}-2E_B\cdot(E_{D^*}+E_{\ell})+2E_{D^*}E_{\ell}
\nonumber \\
& &-2\overrightarrow{p}_{\!D^*}\cdot \overrightarrow{p}_{\!\ell}
+2\overrightarrow{p}_{\!B}\cdot
(\overrightarrow{p}_{\!D^*}+\overrightarrow{p}_{\!\ell})~~. \nonumber    
\end{eqnarray}
For experiments using $e^+e^-\to\Upsilon(4S)\to B\overline{B}$,  
the $B$ energy, $E_B$, is set equal to the beam energy, $E_{beam}$,  and
all quantities are known except the angle between the $B$ direction and the
sum of the $D^*$ and lepton 3-vectors (the last term). A reasonable estimate
of $MM^2$ is obtained by setting this term to zero. The signal for the
$D^*\ell\nu$ final state should appear at the neutrino mass, namely at
$MM^2=0$.
In an alternative technique $MM^2$ is set
to zero and the angle between the $B$ momentum and the sum of the $D^*$ and
lepton 3-vectors is evaluated as
\begin{equation}
\cos\left(\Theta_{B\cdot D^*\ell}\right)={{2E_B(E_{D^*}+E_{\ell})-M^2_B-
M^2_{D^*\ell}}\over {2\left|\overrightarrow{p}_B\right|
\left|\left(\overrightarrow{p}_{\!D^*}+\overrightarrow{p}_{\!\ell}\right)\right|}}~~,
\end{equation}
where $M_{D^*\ell}$ indicates the invariant mass of the $D^*$-lepton
combination.

\begin{figure}[hbtp]
\centerline{\epsfig{figure=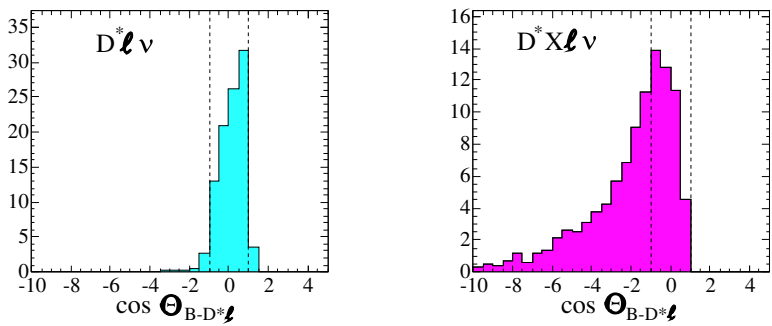,width=4.in}}
\caption{\label{Vcb_4} The cosine of the angle between the $B$ momentum
vector and the sum of $D^*$ and lepton momentum vectors for (left) the final
state $D^*\ell\nu$ and (right) $D^* X\ell\nu$, where $X$ refers to an
additional pion.}
\end{figure}

A Monte-Carlo simulation of $\cos\left(\Theta_{B\cdot D^*\ell}\right)$ is given
for the final state of interest and for the main background reaction in
Figure~\ref{Vcb_4}. For the correct final state only a few events are outside the
``legal" region of $\pm 1$, while when there are extra pions in the final state
the shape changes and many events are below $-1$.

\begin{figure}[hbtp]
\centerline{\epsfig{figure=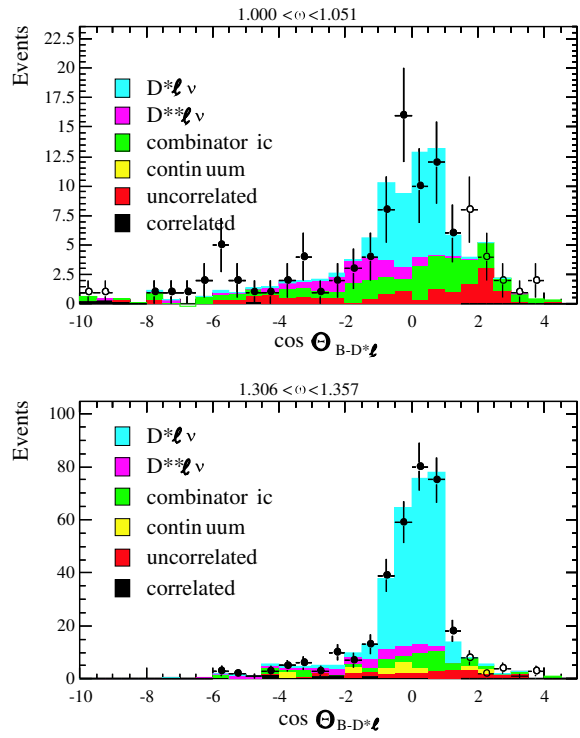,width=3.in}}
\caption{\label{Vcb_3} The signal and background contributions in
two different $\omega$ bins for the final state $D^*\ell\nu$.}
\end{figure}

\begin{figure}[hbtp]
\centerline{\epsfig{figure=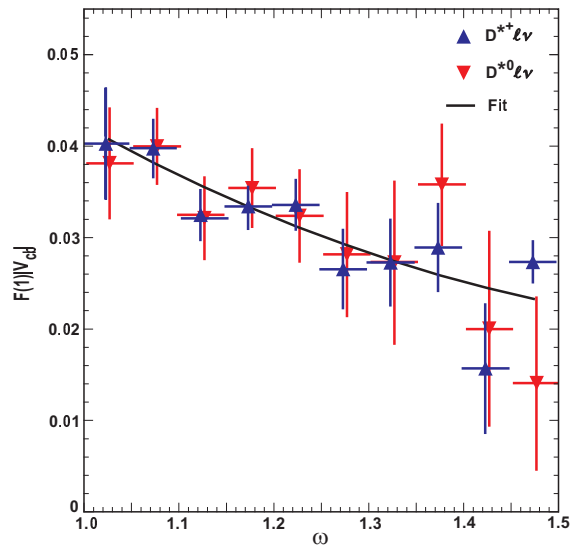,width=3.in}}
\vspace{-.5cm}
\caption{\label{CLEO_Vcb} The CLEO results for $D^*\ell\nu$ for
both $\overline{B}^o$ and $B^-$. The curve is a fit to a shape suggested in
(Caprini 1998).}
\end{figure}

Recent CLEO data has been analyzed with such a technique. The data in two
specific $\omega$ bins is shown in Figure~\ref{Vcb_3}. The final result for
all $\omega$ bins is shown in Figure~\ref{CLEO_Vcb}. The result is characterized
by both a value for $F(1)|V_{cb}|$ and a shape parameter $\rho^2$.

\subsubsection{Evaluation of {\boldmath $V_{cb}$} Using 
{\boldmath $B\to D^*\ell \nu$}}

\begin{figure}[hbtp]
\centerline{\epsfig{figure=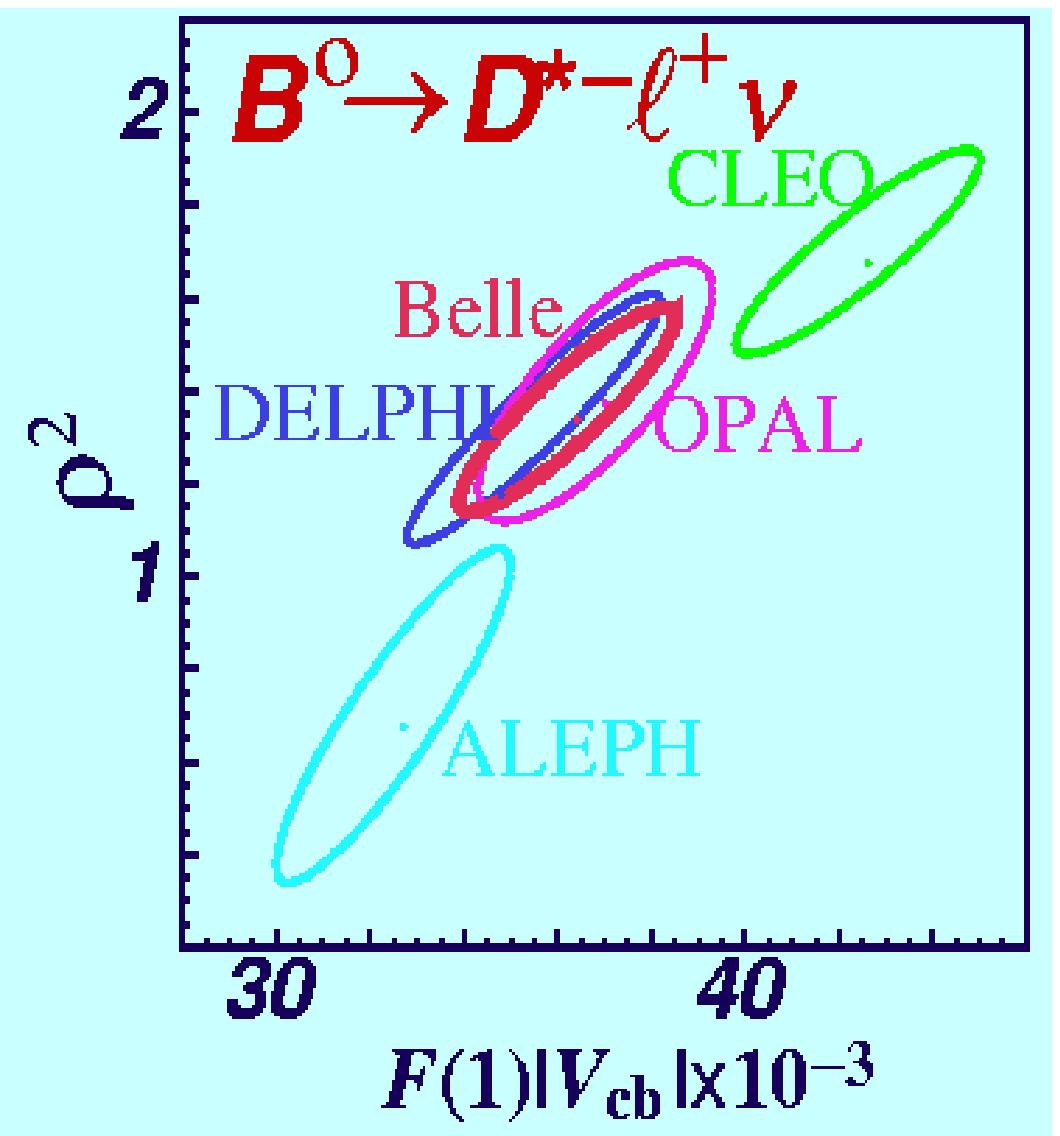,width=3.in}}
\vspace{-.5cm}
\caption{\label{Vcb_sum} The correlation between the slope parameter
$\rho^2$ and $F(1)|V_{cb}|$. The contours are for a change in the
fit $\chi^2$ of one unit. }
\end{figure}

Figure~\ref{Vcb_sum} and
Table~\ref{tab:Vcb} give recent experimental results on exclusive
$B\to D^*\ell\nu$ decays. The CLEO results are not in particularly good
agreement with the rest of the world including the BELLE results.
\begin{table}[th]
\vspace{-2mm}
\begin{center}
\caption{Modern Determinations of $F(1)|V_{cb}|$ using 
$B\to D^*\ell^-\overline{\nu}$ decays
\label{tab:Vcb}}
\vspace*{2mm}
\begin{tabular}{lcc}\hline\hline
Experiment & $F(1)|V_{cb}|$ $(\times 10^{-3})$ & $\rho^2$\\\hline
ALEPH (Buskulic 1997) & $33.0\pm 2.1 \pm 1.6$ & $0.74\pm 0.3\pm 0.4$ \\
BELLE (Tajima 2001) & $35.4\pm 1.9 \pm 1.9$ & $1.35\pm 0.17\pm 0.18$ \\
CLEO (Heltsley 2001) & $42.2\pm 1.3 \pm 1.8 $ & $1.61\pm 0.09\pm 0.21$ \\
DELPHI (Abreu 2001)& $34.5\pm 1.4 \pm 2.5$ & $1.2\pm 0.1 \pm 0.4$ \\
OPAL($\pi\ell$) (Abbiendi 2000) & $37.9\pm 1.3 \pm 2.4$ &$1.2\pm 0.2\pm 0.4$  \\
OPAL (Abbiendi 2000)& $37.5\pm 1.7 \pm 1.8$ & $1.4\pm 0.2 \pm 0.2$\\\hline
Average & $37.8\pm 1.4$ & 1.37$\pm$ 0.13\\
 \hline\hline
\end{tabular}
\end{center}
\end{table}

To extract the value of $|V_{cb}|$ we have to determine the corrections
to $F(1)$ that lower its value from unity.
The corrections are of two types: quark mass, characterized as some
coefficient  times $\Lambda_{QCD}/m_Q$, and hard gluon, characterized as
$\eta_A$. The value of the form-factor can then be expressed as (Neubert 1996)
\begin{equation}
F(1)=\eta_A\left(1+0\cdot \Lambda_{QCD}/m_Q + 
c_2\cdot \left(\Lambda_{QCD}/m_Q\right)^2+....\right)= \eta_A(1+\delta).
\end{equation}
The zero coefficient in front of the $1/m_Q$ term reflects the fact that the
first order correction in quark mass vanishes at $\omega$ equals one. This is called
Luke's Theorem (Luke 1990). Recent estimates are 0.967$\pm$0.007 and $-0.55\pm$0.025
for $\eta_A$ and $\delta$, respectively. The value predicted for $F(1)$ then
is 0.91$\pm$0.05. This is the conclusion of the PDG review done by
Artuso and Barberio (Artuso 2001). 
There has been much
controversy surrounding the theoretical prediction of this number.
In the future Lattice-Gauge Theory calculations will presumably become
accurate when unquenched. (Quenched calculations are those performed without
light quark loops.) Current lattice calculations give
$F(1)=0.913^{+0.024}_{-0.017}\pm
0.016^{+0.003+0.000+0.006}_{-0.014-0.016-0.014}$, where the uncertainties
come respectively, from statistics and fitting, matching lattice gauge theory
to QCD, lattice spacing dependence, light quark mass effects and the
quenching approximation. (Hashimoto 2001).

Using the Artuso-Barberio value for $F(1)$ we have
\begin{equation}
\left|V_{cb}\right|=(41.5\pm 1.5\pm 2.3)\times 10^{-3}~~,
\end{equation}
where the first error is experimental and the second the error on the 
calculation of $F(1)$.

\subsubsection{\boldmath{$|V_{cb}|$} from Inclusive Semileptonic Decays}
\label{section:Vcbinc}
The inclusive semileptonic branching ratio $b\to X\ell\nu$ has been measured by
both CLEO and LEP to reasonable accuracy. CLEO finds 
$(10.49\pm 0.17\pm 0.43)$\%, while LEP has $(10.56\pm 0.11\pm 0.18)$\%. These
are not quite the same quantities as the CLEO number is an average over $B^-$
and $B^o$ only, while the LEP number is a weighted average over all $b$ hadron species
produced in $Z^o$ decays. Thus using the LEP number one should use the average
``$b$ quark" lifetime of 1.560$\pm$0.014 ps. 

Using the  Heavy Quark Expansion, HQE, model (Bigi 1997)
we can relate the total semileptonic decay rate at the quark level
to $|V_{cb}|$ as
\begin{eqnarray}
\left|V_{cb}\right|^2 &=& h(\lambda_1,\lambda_2,\overline{\Lambda})\times \Gamma(b\to c\ell\nu) \\
& =&  h(\lambda_1,\lambda_2,\overline{\Lambda})\times 
{\cal{B}}(b\to c\ell\nu)/\tau_b \nonumber~~, 
\end{eqnarray}
where ${\cal{B}}(b\to c\ell\nu)$ is the inclusive semileptonic branching ratio
minus a small $b\to u\ell \nu$ component. (It is precisely
the decay of a $B$ meson to a lepton-antineutrino pair plus any charmed
hadron.) $\tau_b$ is the lifetime of that particular meson or average 
lifetime of the combination of the $b$-flavored hadrons used in the analysis,
suitably weighted.
In HQE the semileptonic rate
is described to order $\left(\Lambda_{QCD}/m_b\right)^2$ by the parameters:
\begin{itemize}
\item $\lambda_1={M_B\over 2}\langle B(v)\left|\overline{h_v}
(iD)^2h_v\right|B(v)\rangle$, is the kinetic energy of the residual motion of
the $b$ quark in the hadron
\item  $\lambda_2=-{M_B\over 2}\langle B(v)\left|\overline{h_v}
{g\over 2}\cdot \sigma^{\mu\nu}G_{\mu\nu}  h_v\right|B(v)\rangle$,
is the chromo-magnetic coupling of the $b$ quark spin to the gluon field.
\item $\overline{\Lambda}=\overline{M_B}-m_b+{\lambda_1 \over 2M_B}$, is 
the strong interaction coupling where
$\overline{M_B}$ is the spin averaged $B$ meson mass, $(M_B+3M_{B^*})/4$.
\end{itemize}

These parameters are further related as
\begin{eqnarray}
M_B=m_b+\overline{\Lambda}-{{(\lambda_1+3\lambda_2)}\over {2m_b}}  \\
M_{B^*}=m_b+\overline{\Lambda}-{{(\lambda_1-\lambda_2)}\over {2m_b}}~~. \nonumber 
\end{eqnarray}
These relations allow us to determine $\lambda_2$ from the $M_{B^*}-M_B$ mass
splitting as 0.12 GeV$^2$.
The function $h(\lambda_1,\lambda_2,\overline{\Lambda})$ 
can be calculated from the Heavy Quark Expansion (HQE) model.  This involves both
perturbative and non-perturbative pieces.

Although we will go through this
example there is a  disturbing aspect of assuming quark-hadron duality; the idea of duality is that if you integrate over
enough exclusive charm bound states and enough phase space, the inclusive
hadronic result will match the quark level calculation. However, we do not know
what size is the uncertainty associated with the duality {\it assumption}.
In fact, Isgur said ``I identify a source of $\Lambda_{QCD}/m_Q$ corrections to
the assumption of quark-hadron duality in the application of heavy quark
methods to inclusive heavy quark decays. These corrections could substantially
affect the accuracy of such methods in practical applications and in particular
compromise their utility for the extraction of the CKM matrix element $V_{cb}$"
(Isgur 1999). 

Let us move to the details of the calculation. In one implementation
$\lambda_1$ and $\overline{\Lambda}$ are derived; 
the relationship between the
inclusive $b\to c \ell\nu$ branching fraction (about 99\% of $b\to
X\ell\nu$) is given as
\begin{eqnarray}
\left|V_{cb}\right| =& & 0.0411\sqrt{{{\cal{B}}(b\to X_c \ell\nu)}\over{0.105}}
\sqrt{1.55~{\rm ps}\over\tau_b}\left(1-0.012{{\mu_{\pi}^2-0.5~{\rm GeV}^2}
\over{0.1~\rm{GeV}^2}}\right) \\
& &\times \left(1\pm 0.015_{pert.}\pm 0.010_{m_b}\pm 0.012_{1/m_Q^3}\right)~~,
\nonumber \\
\end{eqnarray}
where $\mu_{\pi}^2$ is the negative of $\lambda_1$ modulo QCD corrections and
is taken as $(0.5\pm 0.1)$ GeV$^2$ (Bigi 1997).

This leads to a value of
\begin{equation}
\left|V_{cb}\right|=(40.7\pm 0.5 \pm 2.4)\times 10^{-3}
\end{equation}
from the LEP data alone, with a similar value from CLEO. The first error
contains the statistical and systematic error from the experiments while the
second error contains an estimate of the theory error from sources other
than duality.
 
In another implementation the parameters $\lambda_1$ and $\overline{\Lambda}$
are obtained from data. Here we use the HQE formula
\begin{eqnarray}
\lefteqn{\Gamma_{sl}={{G_F^2\left|V_{cb}\right|^2M_B^5}\over
{192\pi^3}}0.369\times}\\
& & \left[1-1.54{\alpha_s\over\pi}-1.65{{\overline{\Lambda}}\over M_B}
\left(1-1.087{\alpha_s\over\pi}\right) 
-0.95{\overline{\Lambda}^2\over M_B^2}-3.18{\lambda_1\over M_B^2}
+0.02{\lambda_2\over M_B^2}
\right]~. \nonumber
\end{eqnarray}

Determining $\lambda_1$ and $\overline{\Lambda}$
can be accomplished by measuring, for example, the ``moments" of
the hadronic mass produced in $b\to c\ell\nu$ decays. The first moment is
defined as the deviation from the $D$ mass ($M_D$) as 
$\langle M_X^2-\overline{M}_D^2\rangle$ and the second moment as
$\langle \left( M_X^2-\overline{M}_D^2\right)^2\rangle$. It is also possible
to use the first and second moments of the lepton energy distribution in these
decays, or moments of the photon energy in the process $b\to s\gamma$. In fact any
two distributions can be used; in practice it will be critical to use all of
them to try and ascertain if any violations of quark-hadron duality appear and
to check that terms of order $\left(\Lambda_{QCD}/m_b\right)^3$ are not
important.

\begin{figure}[hbtp]
\centerline{\epsfig{figure=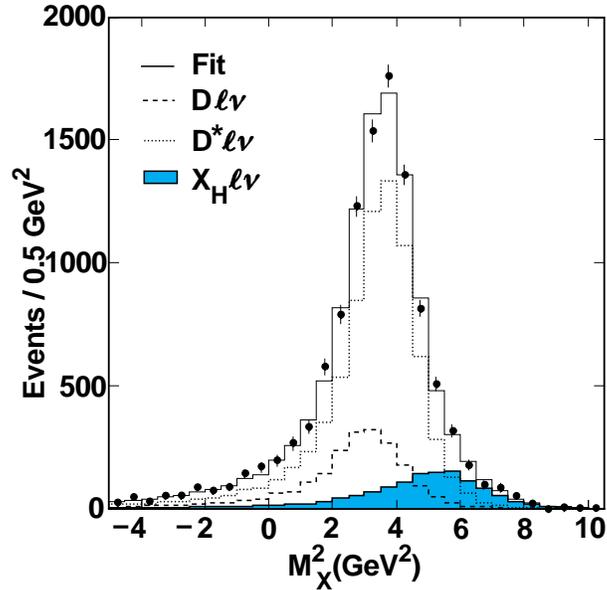,width=3.2in}}
\vspace{-.3cm}
\caption{\label{massmoment} The mass distribution for $b\to c\ell\nu$
events from CLEO. X$_{\rm H}$ indicates an additional pion plus a $D$
or $D^*$.}
\end{figure}

CLEO has used the first and second moments of the hadron mass in $b\to c\ell\nu$
decays. They find the $M_X$ distributions by using missing energy and
momentum in the event to define the $\nu$ four-vector. Then detecting only the
lepton and requiring it to have a momentum above 1.5 GeV/c, they calculate:
\begin{eqnarray}\label{eq:mx}
M_X^2&=&\left(E_B-E_{\ell}-E_{\nu}\right)^2
-\left(\overrightarrow{p}_{\!B}-\overrightarrow{p}_{\!\ell}-
\overrightarrow{p}_{\!\nu}\right)^2  \\
&=& M_B^2+M^2_{\ell\nu}-2E_B\left(E_{\ell}+E_{\nu}\right)+2\overrightarrow{p}_{\!B}
\cdot\left(\overrightarrow{p}_{\!\ell}-\overrightarrow{p}_{\!\nu}\right)
\\
&\approx& M_B^2+M^2_{\ell\nu}-2E_BE_{\ell\nu}~~, \nonumber
\end{eqnarray}
where $M_{\ell\nu}$ is the invariant mass of the lepton-neutrino pair.
The measured $M_X^2$ distribution is shown in Figure~\ref{massmoment}. We do not
see distinct peaks at the mass of the $D$ and $D^*$ mesons because ignoring
the last term in equation~(\ref{eq:mx}) causes poor resolution. This
term must be ignored, however, using this technique because we do not know
the direction of the $B$ meson. 

\begin{figure}[hbtp]
\vspace{-.1cm}
\centerline{\epsfig{figure=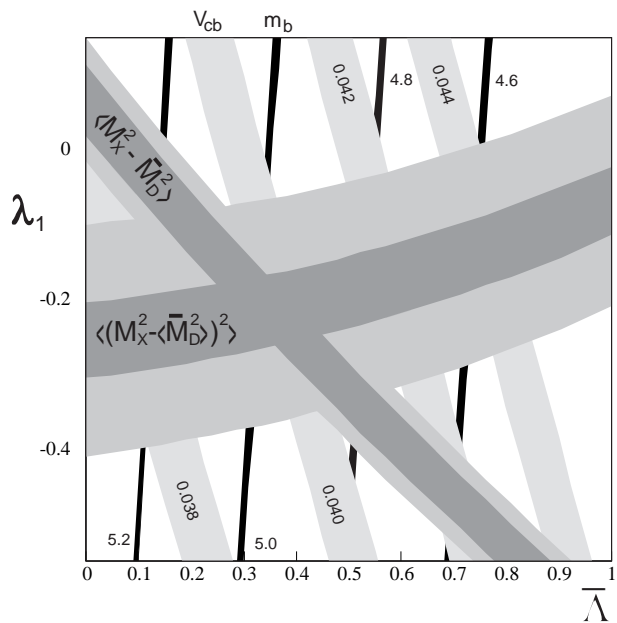,width=3.5in}}
\vspace{-.3cm}
\caption{\label{moments_had} Constraints in $\lambda_1$ versus
$\overline{\Lambda}$ from CLEO measurements of first and second hadronic
moments in semileptonic decay. The darker band gives experimental uncertainties
alone, while the lighter outer band also includes uncertainties from unknown
3rd order theoretical parameters. Bands of constant $V_{cb}$ and $b$ quark
mass, $m_b$,
are also shown, where the band widths represents theoretical uncertainties due
to unknown 3rd order parameters.}
\end{figure}

CLEO finds values of the first and second moments of
(0.287$\pm$0.023$\pm$0.061) GeV$^2$ and (0.712$\pm$0.056$\pm$0.176) GeV$^4$,
respectively. These lead to the determination of $\lambda_1$,
$\overline{\Lambda}$ and $V_{cb}$ shown in Figure~\ref{moments_had}. Later we will
see a different determination using $b\to s\gamma$ (section~\ref{section:Vcbsg}).

In summary the exclusive measurements of $V_{cb}$ are to be trusted while the
inclusive determination, though consistent, has an unknown source of systematic
error and should not be used now.

\subsection{Measurement of {\boldmath $|V_{ub}|$}}
This is a heavy to light quark transition where HQET cannot be used directly as
in finding $V_{cb}$.
Unfortunately the theoretical models that can be used to 
extract a value from the data do not currently give precise predictions. 

\begin{figure}[hbtp]
\vspace{-.1cm}
\centerline{\epsfig{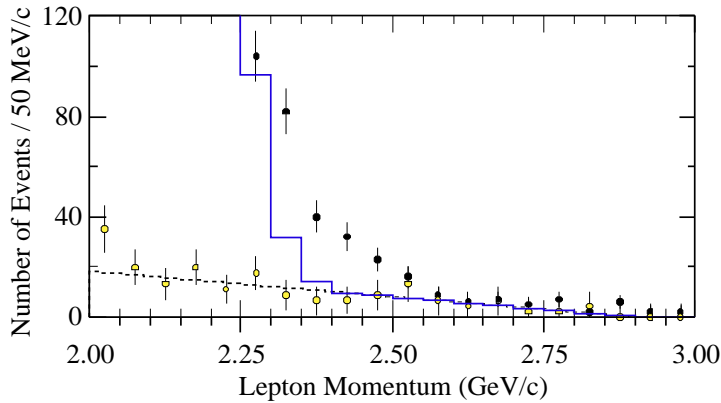}}
\caption{\label{cleo_end} Sum of inclusive electron and muon distributions
from CLEO. The solid points are data taken on the peak of the $\Upsilon(4S)$
while the open circles are data taken on the continuum 30 MeV below the
resonance(suitably normalized). The dashed line is a fit to the continuum data
and the solid line is the predicted curve from $b\to c\ell\nu$ dominated by
$B\to D\ell\nu$ near the end of the allowed lepton spectrum.}
\end{figure}

Three techniques have been used. The first measurement of $V_{ub}$ done by CLEO
(Fulton 1990) and subsequently
confirmed by ARGUS (Albrecht 1990), used only leptons which were more energetic than those that
could come from $b\to c\ell^- \bar{\nu}$ decays. These  
``endpoint leptons'' can occur $b\to c$ background free at the
$\Upsilon (4S)$, because the $B$'s are almost at rest. The CLEO data are
shown in Figure~\ref{cleo_end}. Since the lepton momentum for $B\to D\ell\nu$
decays is cut off by phase space, this data provides incontrovertible evidence
for $b\to u\ell\nu$ decays.

 Unfortunately, there is
only  a small fraction of the $b\to u \ell^-\bar{\nu}$ lepton spectrum that
can be seen this way, leading to model dependent errors. The models used are
either inclusive predictions, sums of exclusive channels, or both
(Isgur 1995) (Bauer 1989) (K\"orner 1989) (Melikhov 1996) (Altarelli 1982)
(Ramirez 1990). The average among the models is $|V_{ub}/V_{cb}|=0.079\pm 0.006$,
without a model dependent error.
These models differ by at most 11\%, making it tempting to assign a $\pm$6\%
error. However, there is no quantitative way of estimating the error.

\begin{figure}[bt]
\vspace{-9mm}
\centerline{\epsfig{figure=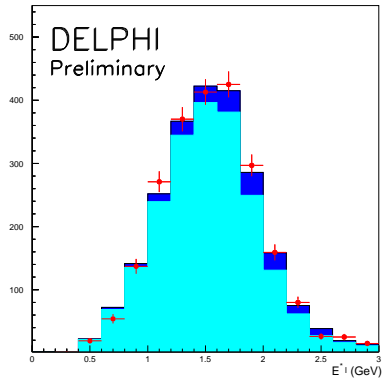,width=2.in}}
\vspace{-4mm}
\caption{\label{delphi_vub}The lepton energy distribution in the $B$ rest
frame from DELPHI. The data have been enriched in $b \to u$ events, and the 
mass of the recoiling hadronic system is required to be below 1.6 GeV. The 
points indicate data, the light shaded region, the fitted background and the 
dark shaded region, the fitted $b \to u \ell \nu$ signal. }
\end{figure}

ALEPH (Barate 1999), L3 (Acciarri 1998)
and DELPHI (Abreu 2000) isolate a class of events where the hadron 
system associated
with the lepton is enriched in $b\to u$ and thus depleted in $b\to c$.     
They define a likelihood that hadron tracks come from $b$ decay by using a large 
number of variables including, vertex information, transverse momentum, not 
being a kaon etc.. Then they require the hadronic mass to be less than 1.6 GeV, which 
greatly reduces $b\to c$, since a completely reconstructed $b\to c$ decay has a 
mass greater than that of the $D$ (1.83 GeV). They then examine the lepton 
energy distribution, shown in Figure~\ref{delphi_vub} for
DELPHI.

The average of all three results is $|V_{ub}|=(4.13^{+0.42+0.43+0.24}_{-0.47-0.48-0.25}\pm 0.20)\times
10^{-3}$, resulting in a value for $|V_{ub}/V_{cb}|=0.102 \pm 0.018$, using 
$|V_{cb}|=0.0405\pm 0.0025$. I have several misgivings about this result.
First of all the experiments have to understand the systematic errors
very well. To understand semileptonic $b$ and $c$ decays and thus find
their $b\to u\ell\nu$ efficiency, they employ
different models and Monte Carlo manifestations of these models. To find the 
error they take half the spread that different models give. This alone may be a
serious underestimate. Secondly they use one model,
the HQE model, to translate their measured rate to a value for $|V_{ub}|$. This model assumes duality,
and there are no successful experimental checks: The model
fails on the $\Lambda_b$ lifetime prediction. Furthermore, the quoted
theoretical error, even in the context of the model, has been estimated by
Neubert to be much larger at 10\% (Neubert 2000). Others have questioned the
effect of the hadron mass cut and estimate 10-20\% errors due to this alone
(Bauer 2001).

It may be possible to use the spectrum of photons in $b\to s\gamma$ to reduce
the theoretical error in the endpoint lepton method or to make judicious cuts
in $q^2$ instead of hadronic mass to help reduce the theoretical errors. See (Wise 2001) for an erudite
discussion of these points.

\begin{figure}[hbtp]
\vspace{-.1cm}
\centerline{\epsfig{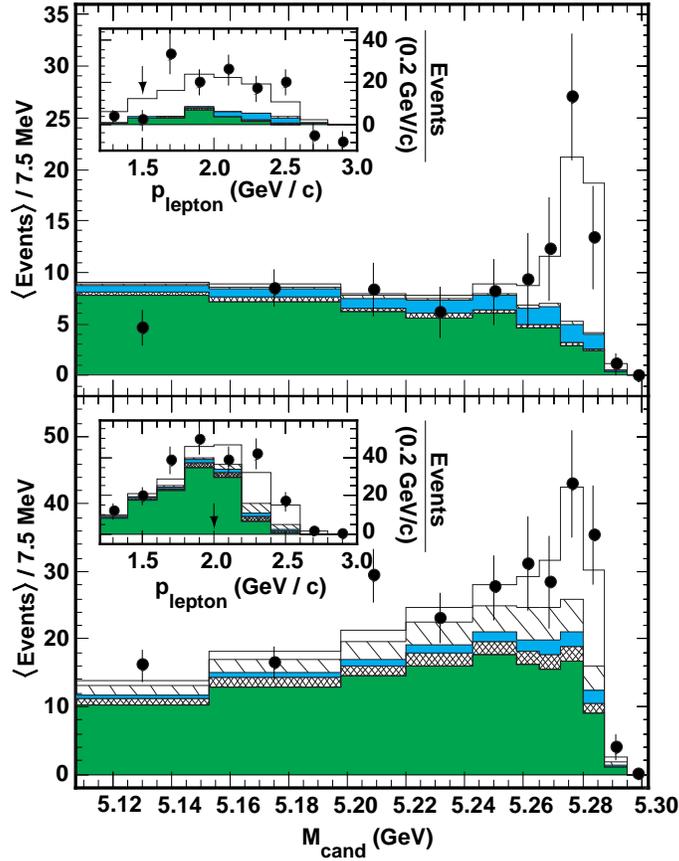}}
\caption{\label{pilnxx} The $B$ candidate mass distributions and the
signal bin lepton momentum spectra (insets) for the pion modes (top) and the
sum of $\rho$ and $\omega$ (vector) modes (bottom). The points are the data after
continuum and fake background subtractions; the dark-shaded, cross-hatched and
unshaded histograms are $b\to cX$, $b\to u\ell\nu$ feed-down and signal,
respectively. For the $\pi$ (vector) modes, the light-shaded and
hatched histograms are $\pi$ $\to$ $\pi$ (vector$\to$ vector) and vector $\to$ $\pi$
($\pi$ $\to$ vector) cross-feed, respectively (charge final states can
feed neutral and vice-versa). The histogram normalizations
are from the nominal fit. The arrows indicate the momentum cuts. 
}
\end{figure}

The third method uses exclusive decays.
CLEO has measured the decay 
rates for the exclusive final states $\pi\ell\nu$ and 
$\rho\ell\nu$  (Alexander 1996). The data are shown in Figure~\ref{pilnxx}.
The model of K\"orner and 
Schuler (KS) is ruled out by the measured ratio of $\rho/\pi$. Other models include those of (Isgur 1995)
(Isgur 1989) (Wirbel 1985) (Bauer 1989) (Korner 1988) (Melikhov 1996)
(Altarelli 1982) (Ramierz 1990). CLEO has 
presented an updated analysis for $\rho\ell\nu$ where
they have used several different models to evaluate
their efficiencies and extract $V_{ub}$. These
theoretical approaches include quark models, light cone sum
rules (LCRS), and lattice QCD. The CLEO values are shown
in Table~\ref{tab:Vub}.

\begin{table}[th]
\vspace{-2mm}
\begin{center}

\vskip 0.1 in
\caption{Values of $|V_{ub}|$ using 
$B\to \rho\ell^-\overline{\nu}$ and some theoretical models \label{tab:Vub}}
\begin{tabular}{lc}\hline
Model & $V_{ub}$ $(\times 10^{-3})$\\
\hline
ISGW2 (Isgur 1989)& $3.23\pm 0.14^{+0.22}_{-0.29}$ \\
Beyer/Melikhov (Beyer 1998) &$3.32\pm 0.15^{+0.21}_{-0.30}$  \\
Ligeti/Wise (Ligeti 1996) &$2.92\pm 0.13^{+0.19}_{-0.26}$  \\
LCSR (Ball 1998) &$3.45\pm 0.15^{+0.22}_{-0.31}$  \\
UKQCD (Debbio 1998) &$3.32\pm 0.14^{+0.21}_{-0.30}$  \\
\hline\hline
\end{tabular}
\end{center}
\end{table}

The uncertainties in the quark model calculations (first three in the table)
 are guessed to be
25-50\% in the rate. The Ligeti/Wise model uses charm data and
SU(3) symmetry to reduce the model dependent errors. The other models 
estimate their errors
at about 30-50\% in the rate, leading conservatively to a 25\% error in $|V_{ub}|$. 
Note that the models differ by 18\%, but it would be incorrect to assume
that this spread allows us to take a smaller error. It may be that the models
share common assumptions, e.g. the shape of the form-factors. At this time it is
prudent to assign a 25\% model dependent error realizing that the errors
in the models cannot be averaged. The fact that the models do not differ
much allows us to comfortably assign a central value 
$|V_{ub}|=(3.25\pm 0.14^{+0.22}_{-0.29}\pm 0.80)\times 10^{-3}$, and
a derived value $|V_{ub}/V_{cb}|=0.08 \pm 0.02$~.
 
\begin{figure}[hbtp]
\centerline{\epsfig{figure=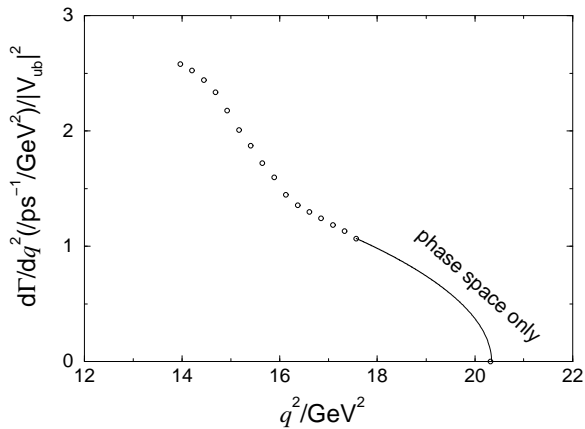,width=3.in}}
\caption{\label{integ} The UKQCD lattice calculation
for $B\to\rho\ell\nu$ shown as circles. The line is an estimate.}
\end{figure}

Lattice QCD has predicted form-factors and resulting rates for the exclusive
semileptonic final states $\pi\ell\nu$ and $\rho\ell\nu$ (Sachrajda 1999)
in the quenched approximation.
These calculations require the momentum of the final-state light meson to be
small in order to avoid discretization errors. This means we only obtain results
at large values of the invariant four-momentum transfer squared, $q^2$. 
Figure~\ref{integ} shows the predictions of the $B\to\rho\ell\nu$ width as
a function of $q^2$. Note that the horizontal scale is highly zero suppressed.
The region marked ``phase space only" is not calculated but estimated using a
phase space extrapolation from the last lattice point. 

The integral over the region for $q^2$ $>$ 14 GeV$^2$ gives a rate of
$\Delta\Gamma = 8.3\left|V_{ub}\right|^2$ ps$^{-1}$-GeV$^2$. The CLEO measurement
in the same interval gives $(7.1\pm 2.4)\times 10^{-5}$ps$^{-1}$-GeV$^2$, which yields a
value for $V_{ub} = (2.9\pm 0.5)\times 10^{-3}$ (Sachrajda 1999).
Ultimately unquenched lattice calculations when coupled with more precise
data will yield a much better value for $V_{ub}$.
 

\section{Facilities for {\boldmath $b$} Studies}

\subsection{{\boldmath $b$} Production Mechanisms}
Although most of what is known about $b$ physics
presently has been obtained from $e^+e^-$ colliders operating either at the
$\Upsilon(4S)$ or at LEP, interesting information
is now appearing from the hadron collider experiments, CDF and D0, 
which were designed to look
for considerably higher energy phenomena. The appeal of hadron
colliders arises mainly from the large measured $b$ cross-sections. At the FNAL
collider, 1.8 TeV in the $p\bar{p}$ center-of-mass, the cross-section has been
measured as $\sim$100 $\mu$b, while it is expected to be about five times higher at the
LHC (Artuso 1994).

\begin{figure}[hbtp]
\centerline{\epsfig{figure=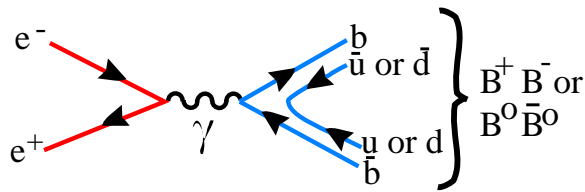,width=3in}}
\caption{\label{epemtoBB}$B$ production at the $\Upsilon(4S)$.}
\end{figure}

\begin{figure}[hbtp]
\centerline{\epsfig{figure=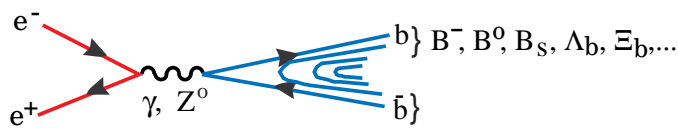,width=3.5in}}
\caption{\label{epemtobbX}$b$ production in the continuum at $e^+e^-$ colliders.}
\end{figure}

\begin{figure}[hbtp]
\centerline{\epsfig{figure=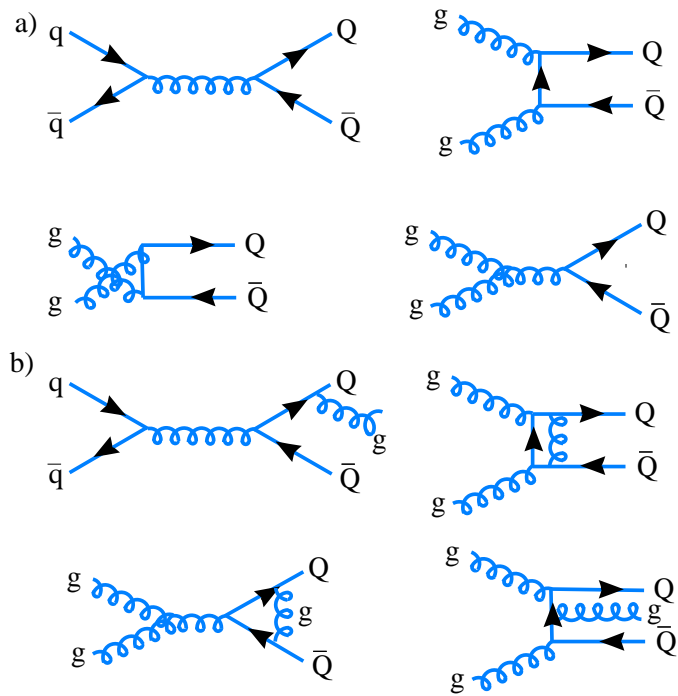,width=3.5in}}
\caption{\label{hadron_b_prod}Heavy quark, $Q$, production at hadron colliders.
(a) Second order in $\alpha_s$, while (b) is third order.}
\end{figure}

The different production mechanisms of $b$ quarks at various accelerators leads
to dissimilar methods of measurements. Figure~\ref{epemtoBB} shows the production
of $B^-$ and $B^o$ mesons at the $\Upsilon(4S)$, while Figure~\ref{epemtobbX}
shows the production mechanism of the different $b$ species at a higher energy
$e^+e^-$ collider such as LEP. Figure~\ref{hadron_b_prod} shows the production
mechanisms for a heavy $b$ or $c$ quark. 

At the $\Upsilon(4S)$ the total $b$ production cross-section is 1.05 nb. In hadron
colliders the measured cross-section is about 100$\mu$b. The third order
diagrams appear about as important as the second order diagrams and the overall
theoretical calculation gives about 1/2 of the measured value.

\subsection{Accelerators for {\boldmath $b$} Physics}

Experiments on $b$ decays started with CLEO and ARGUS using $e^+e^-$ colliders
operating at the $\Upsilon(4S)$ and were quickly joined by the PEP and PETRA
machines operating around 30 GeV. 
Table~\ref{tab:oldfac} lists some of the machines used to study $b$ quarks 
in the last century (Artuso 1994).
\begin{table}[th]
\vspace{-2mm}
\begin{center}
\caption{Machines used for $b$ physics in the 20th century. The total
number of $\overline{b}b$ pairs accumulated per experiment is also listed
when known.
\label{tab:oldfac}}
\vspace*{2mm}
\begin{tabular}{lcccccc}\hline\hline
Machine  & Beams & Energy & $\sigma(b)$ &
$b$ fraction & ${\cal{L}}$  & Total \#  \\
        &              & (GeV) & && cm$^{-2}$s$^{-1}$ &$b$-pairs \\
\hline
CESR &  $e^+e^-$ & 10.8 & 1.05 nb & 0.25 & 1.3$\times 10^{33}$ & 
$9.8\times 10^6$\\
DORIS &  $e^+e^-$ & 10.8 & 1.05 nb & 0.23 & $\sim 10^{31}$ & 
$0.4\times 10^6$\\
PEP &    $e^+e^-$ & 29  & 0.4 nb & 0.09 & 3.2$\times 10^{31}$ & 
\\
PETRA &   $e^+e^-$ & 35  & 0.3 nb & 0.09 & 1.7$\times 10^{31}$ & 
\\
LEP &  $e^+e^-$ & 91 & 9.2 nb & 0.22 & 2.4$\times 10^{31}$ & 
$1.8\times 10^6$\\
SLC &  $e^+e^-$ & 91 & 9.2 nb & 0.22 & 3.0$\times 10^{30}$ & 
$8.8\times 10^4$\\
TEVATRON & $\overline{p}p$ & 1800 & 100 $\mu$b &0.002 &3$\times 10^{31}$  & \\
\hline\hline
\end{tabular}
\end{center}
\end{table}

In the year 2000 the PEP II and KEK-B storage ring accelerators began
operation. These machines have separate $e^-$ and $e^+$ magnet rings so they
can operate at asymmetric energies; PEP II has beam energies of 9.0 GeV and
3.1 GeV, while KEK-B has energies of 8.0 GeV and 3.5 GeV. This allows the
$B$ meson to move with
velocity $\beta\sim0.6$, which turns out to be very important in measurements
of CP violation, since time integrated CP violation via mixing must be exactly zero due to the C odd nature of the $\Upsilon (4S)$.
 These machines also make very high luminosities. 
Current and future machines for $B$ physics are listed in Table~\ref{tab:fut}.
The CDF and D0 experiments will continue at the Tevatron with higher
luminosities. CDF has already made significant contributions including
studies of $b$ production, lifetimes and the discovery of the $B_c$ meson
(Abe 1998). BTeV and LHCb will go into operation around 2007 with much
larger event rates. The CMS and ATLAS experiments at the LHC will also
contribute to $b$ physics especially in the early stages when the luminosity
will be relatively low; at design luminosity these experiments have an average
of 23 interactions per crossing making finding of detached vertices difficult.

\begin{table}[thb]
\vspace{-2mm}
\begin{center}
\caption{Machines in use or approved for dedicated $b$ physics
experiments.
\label{tab:fut}}
\vspace*{2mm}
\begin{tabular}{llcccccc}\hline\hline
Machine & Exp. & Beam & Energy & $\sigma(b)$ &
$b$  & ${\cal{L}}$(Design)  & Interactions \\
        &    & &(GeV) & &fraction& cm$^{-2}$s$^{-1}$ &per crossing \\
\hline
PEP II &BABAR &  $e^+e^-$ & 10.8 & 1.05 nb & 1/4 & 3$\times 10^{33}(\dagger)$ & 
$\ll 1$\\
KEK-B &BELLE &  $e^+e^-$ & 10.8 & 1.05 nb & 1/4 &   $10^{34}$ & 
$\ll 1$\\
HERA &HERA-b&$pN$ & 800  & 10 nb & $2\cdot 10^{-6}$ &  & 4 \\
Tevatron & BTeV& $\overline{p}p$ & 2000 & 100 $\mu$b &1/500 &
2$\times 10^{32}$  & 2 \\
LHC & LHCb& $\overline{p}p$ & 14000 & 500 $\mu$b &1/160 &
2$\times 10^{32}$  & 0.6 \\
\hline\hline
\multicolumn{8}{l}{$\dagger$ Machine has already exceeded design luminosity.}
\end{tabular}
\end{center}
\end{table}

\subsection{{\boldmath $e^+e^-$} Detectors}
\begin{figure}[hbtp]
\centerline{\epsfig{figure=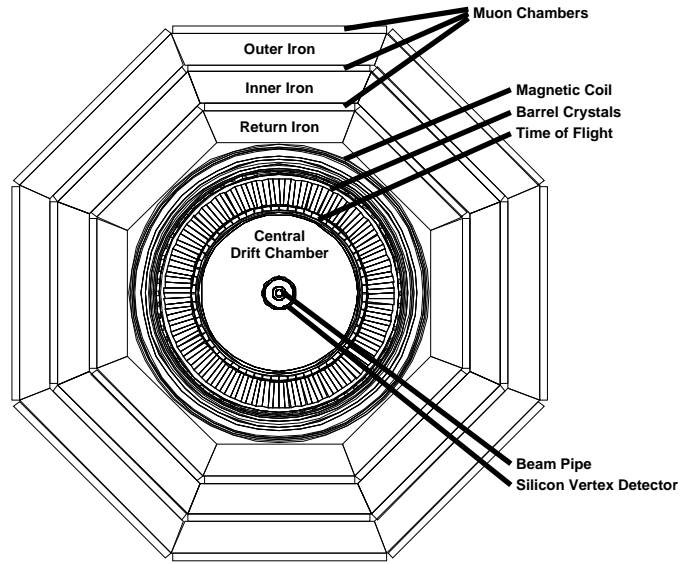,width=3.5in}}
\caption{\label{cleoII}Electrons view of the CLEO II detector.}
\end{figure}

Most experiments at $e^+e^-$ storage rings look quite similar. CLEO II,
shown in Figure~\ref{cleoII}, was the
first detector to have both an excellent tracking system and an excellent
electromagnetic calorimeter. 
Starting from the inside there is a thin beryllium beam pipe surrounded by
a silicon vertex detector; this detector measures positions very accurately
on the order of 10 $\mu$m. Then there is a wire drift chamber whose main
function is to measure the curving trajectories of particles in the 1.5T
solenoidal magnetic field. The next device radially outward is time-of-flight
system to distinguish pions, kaons and protons. This system only works for
lower momenta.
The most important advance in the CLEO III, BELLE and BABAR
detectors is
much better charged hadron identification. Each experiment uses
different techniques based on Cherenkov radiation to
extend $\pi/K$ separation up to
the limit from $B$ decays. The next device is an electromagnetic calorimeter
that uses Thallium doped CsI crystals; indeed this was the most important new
technical implementation done in CLEO II and has also been adopted by
BABAR and BELLE. Afterwards there is segmented iron
that serves as both a magnetic flux return and a filter for muon
identification. 

\begin{figure}[hbtp]
\centerline{\epsfig{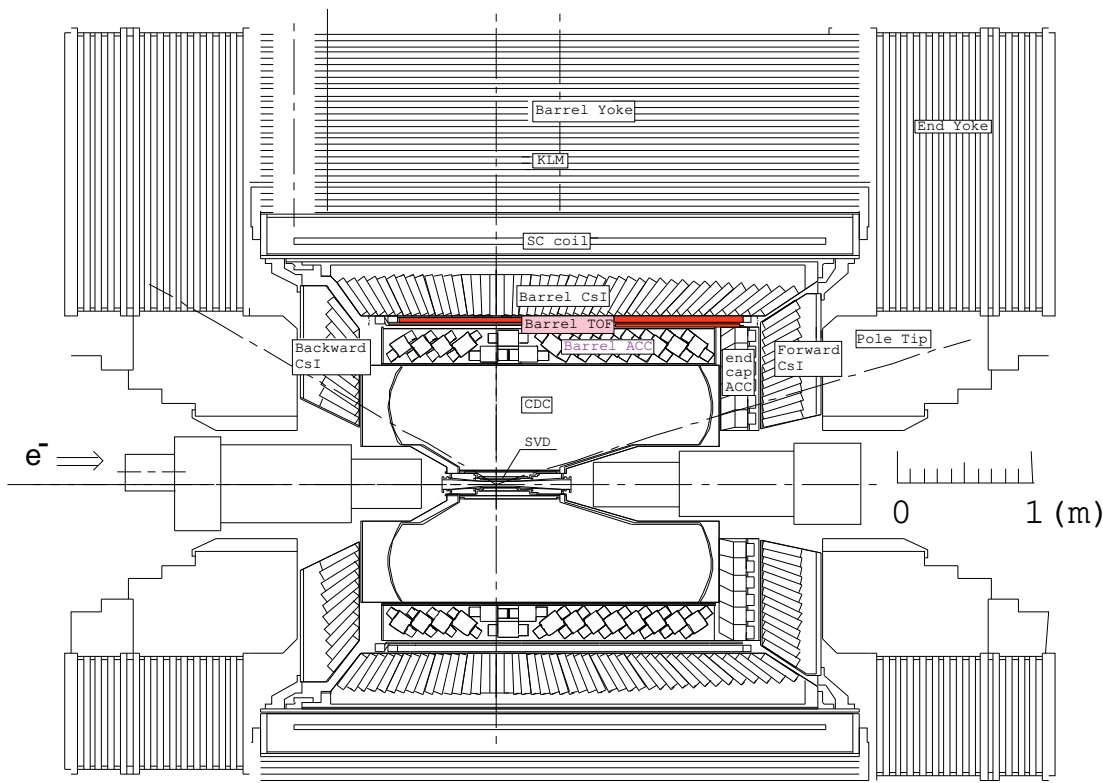}}
\caption{\label{belle_det} through the BELLE detector.}
\end{figure}

Figure~\ref{belle_det} shows a view of the BELLE detector parallel to the beam.

\subsection{{\boldmath $b$}  Production Characteristics at Hadron Colliders}

To make precision measurements, large samples of $b$'s are necessary.
Fortunately, these are available. With the Fermilab Main
Injector, the Tevatron collider will produce $\approx 4\times 10^{11}$  $b$ 
hadrons/$10^7$ s at a luminosity of $2\times 10^{32}$cm$^{-2}$s$^{-1}$. 
These rates compare very favorably to
e$^+$e$^-$ machines operating at the $\Upsilon$(4S). At a luminosity of 
$10^{34}$cm$^{-2}$s$^{-1}$ they would produce $2\times 10^8$ B's/$10^7$ s. Furthermore
$B_s$, $\Lambda_b$ and other $b$-flavored hadrons are accessible for study at
hadron colliders. The LHC has about a five times larger $b$ 
production cross-section. Also important are the large
charm rates, $\sim$10 times larger than the $b$ rate.

\begin{figure}[bht]
\centerline{\epsfig{figure=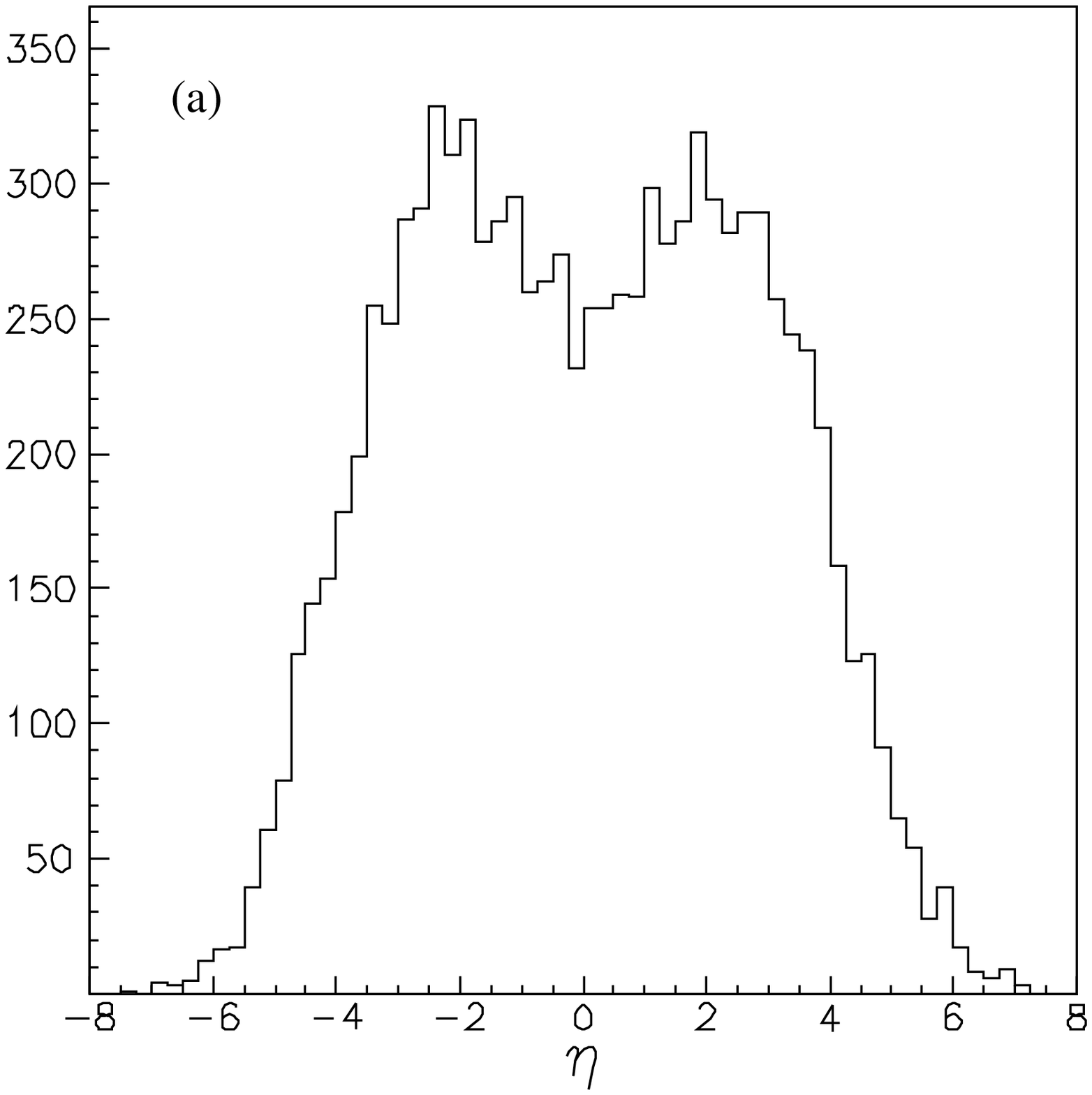,width=1.75in}
\epsfig{figure=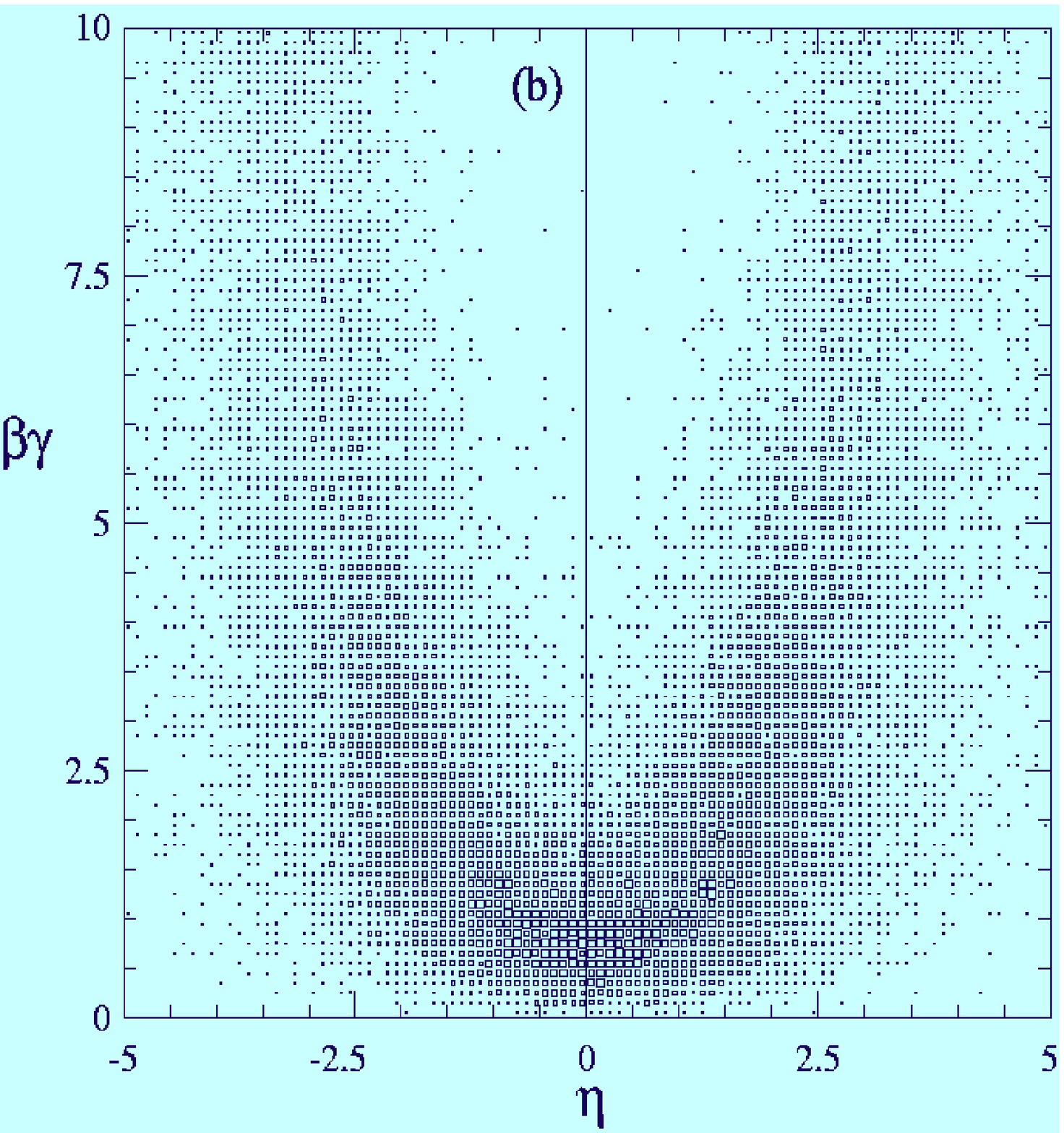,width=1.70in}
\epsfig{figure=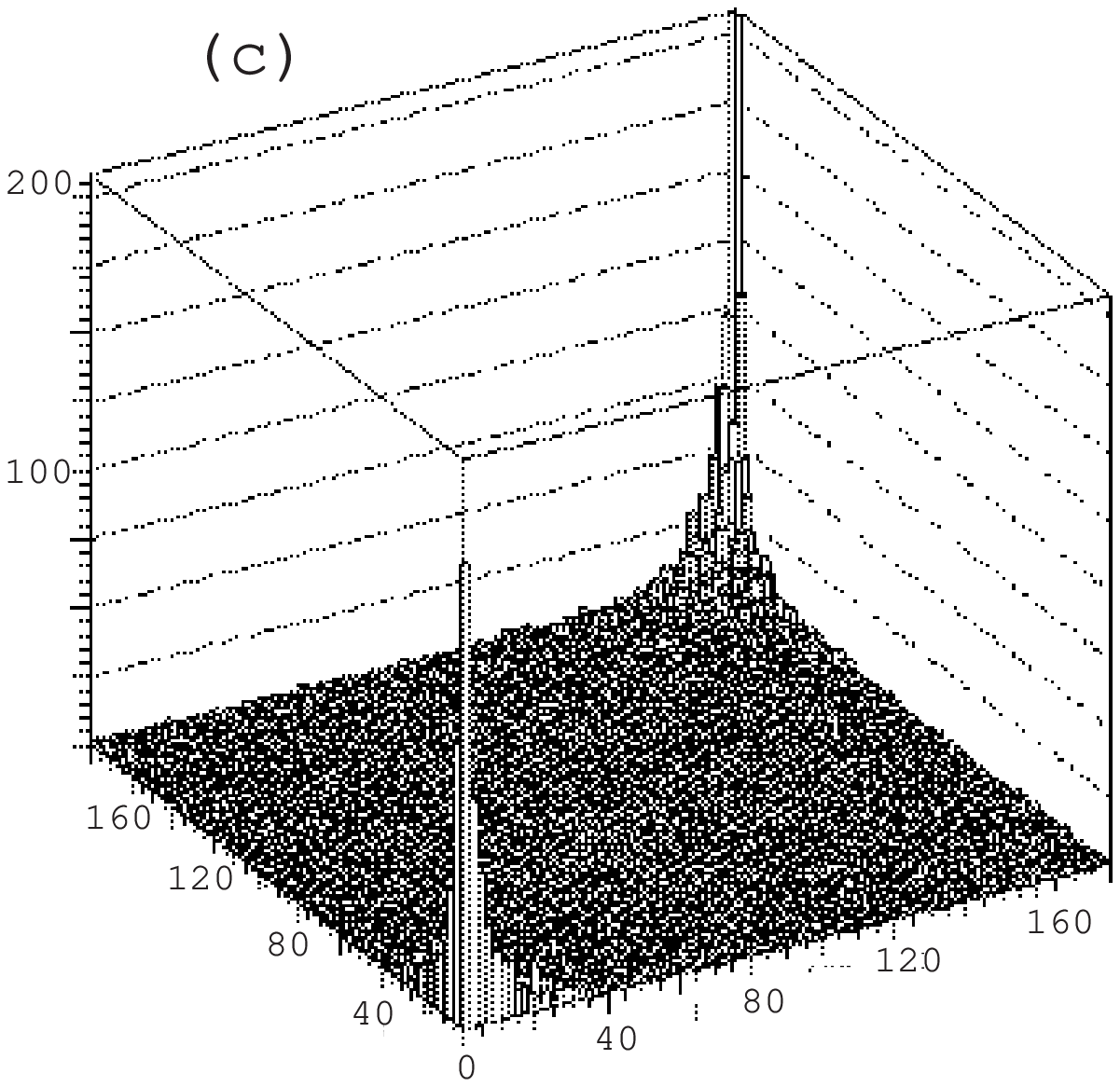,height=1.8in}}
\vspace{-.2cm}
\caption{\label{n_vs_eta} (a) The $B$ yield versus $\eta$. 
(b) $\beta\gamma$ of the  $B$  versus $\eta$. (c) The production angle (in degrees) for the hadron
containing a $b$ quark plotted versus the production angle for a hadron
containing $\bar{b}$ quark.}
\end{figure}

In order to understand the detector design it is useful to examine
the characteristics of $b$ quark production at $p\overline{p}$ collider.
It is often
customary to characterize heavy quark production in hadron collisions with the
two variables $p_t$ and $\eta$, where  $\eta =
-ln\left(\tan\left({\theta/2}\right)\right),$ and $\theta$ is the angle of the
particle with respect to the beam direction. According to QCD based
calculations of $b$ quark production, the $B$'s are produced ``uniformly" in
$\eta$ and have a truncated transverse momentum, $p_t$, spectrum characterized
by a mean value approximately equal to the $B$ mass (Artuso 1994). The
distribution in $\eta$ is shown in Figure~\ref{n_vs_eta}(a). Note that at larger
values of $|\eta|$, the $B$ boost, $\beta\gamma$, increases rapidly (b).

The ``flat" $\eta$ distribution hides an important correlation of
$b\bar{b}$ production at hadronic colliders. In Figure~\ref{n_vs_eta}(c) the production
angles of the hadron containing the $b$ quark is plotted versus the production
angle of the hadron containing the $\bar{b}$ quark according to the Pythia
generator. Many important measurements require the reconstruction of a $b$ decay and 
the determination of the flavor of the other $\bar{b}$, thus requiring both $b$'s to be 
observed in the detector. There is a very strong
correlation in the forward (and backward) direction: when the $B$ is forward
the $\overline{B}$ is also forward. This correlation is not present in the
central region (near 90$^{\circ}$). By instrumenting a relative small region of
angular phase space, a large number of $b\bar{b}$ pairs can be detected. 
Furthermore the $B$'s populating the forward and backward regions have large
values of $\beta\gamma$. 

BTeV, a dedicated heavy flavor experiment approved to run at the Fermilab
Tevatron collider, uses two forward spectrometers (along both the $p$ and $\overline{p}$
directions) that utilize the boost of the
$B$'s at large rapidities. This is of crucial importance because the main way
to distinguish $b$ decays is by the separation of decay vertices from the main
interaction. LHCb, approved for operation at the LHC,  needs a larger detector
to analyze the higher momentum
decay products, and thus has only one arm.

\subsubsection{The BTeV Detector Description}
I will describe BTeV here though LHCb shares many of the same considerations.
There are difficulties that heavy quark experiments at hadron colliders must 
overcome. First of all, the huge $b$ rate is accompanied by an even larger 
rate of uninteresting interactions. At the Tevatron the $b$-fraction is only 
1/500. In searching for rare processes, 
at the level of parts per million, the background from $b$ events is dominant.
(Of course all $b$ experiments have this problem.) The large data rate of $b$'s
must be handled. For example, BTeV, has 1 kHz of $b$'s into the detector, and these
events must be selected and written out. The electromagnetic calorimeter must
be robust enough to deal with the particles from the underlying event and still
have useful efficiency. Furthermore, radiation damage can destroy detector
elements.

\begin{figure}[hbt]
\centerline{\epsfig{figure=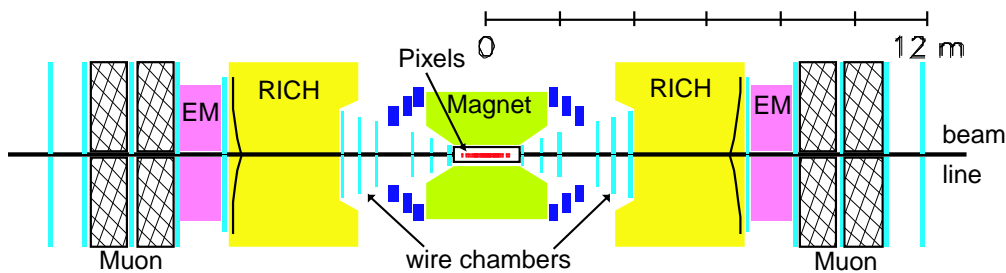,height=1.4in}}
\caption{\label{btev_det}Schematic of the BTeV detector.}
\end{figure}

\begin{figure}[htb]
\centerline{\epsfig{figure=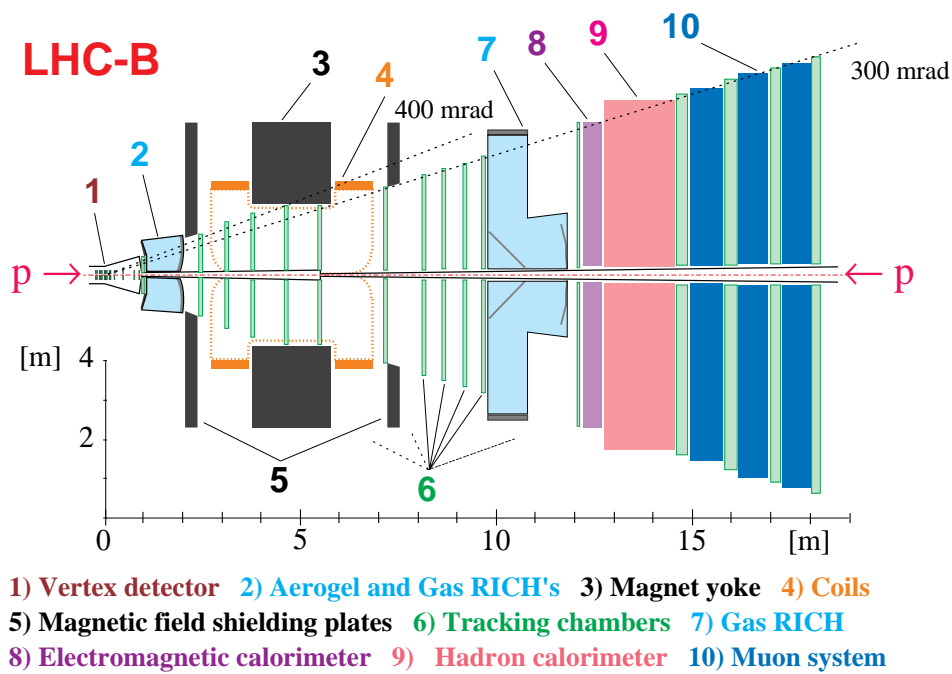,height=4.in}}
\vspace{-1.1cm}
\caption{\label{LHC_B}A schematic diagram of the LHCb detector.}
\end{figure}

\begin{figure}[hbt]
\centerline{\epsfig{figure=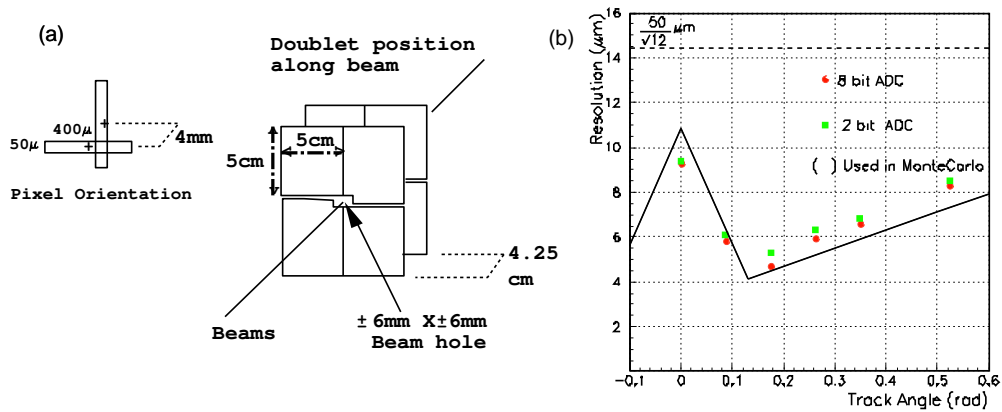,width=5.2in}}
\vspace{-.2cm}
\caption{\label{pixel_schm} (a) Pixel detector geometry in BTeV. The detector is inside the beam pipe. (b) The spatial resolution
as a function of the incident track angle for both 2-bit and 8-bit ADC's as
measured in an 800 GeV/c pion beam. The straight lines are piecewise fits to
the data used in the Monte Carlo simulation. The dashed line near the top
indicates the resolution obtainable without using pulse height information.}
\end{figure}

The BTeV Detector is shown in Figure~\ref{btev_det} (Skwarnicki 2001) and the LHCb detector
in Figure~\ref{LHC_B} (Muheim 2001). The central part of the BTeV
detector has a silicon pixel detector inside a 1.5 T dipole magnet. The LHCb  experiment uses silicon strips. The BTeV pixel detector
provides precision space points for use in both the offline analysis and the
trigger. The pixel geometry is sketched in Figure~\ref{pixel_schm}(a). Pulse
heights are measured on each pixel. Prototype detectors were tested in
a beam at Fermilab; excellent resolutions were obtained, especially
when reading out pulse heights (Appel 2001)
(see Figure~\ref{pixel_schm}(b)). The final design uses a 3-bit ADC. 

The pixel tracker
provides excellent vertex resolution, good enough to trigger on events with
detached vertices characteristic of $b$ or $c$ decays. BTeV shows a rejection
of 100:1 for minimum bias events in the first trigger level while accepting
about 50\% of the usable $b$ decays. A good explanation of the trigger
algorithm can be found at\newline
http://www-btev.fnal.gov/public\_documents/animations/Animated\_Trigger/~~.
Further trigger levels reduce the
background by about a factor of twenty while decreasing the $b$ sample by
only 10\%. The trigger system stores data in a pipeline that is long enough to ensure no deadtime. The data acquisition system has sufficient throughput to accommodate an output of 1
kHz of $b$'s, 1 kHz of $c$'s and 2 kHz of junk.
Tracking is accomplished using straw tube wire chambers with silicon strip chambers in the high track density region near the beam.

Charged particle identification is done using a Ring Imaging CHerenkov detector.
A gaseous C$_4$F$_{10}$ radiator is used with a large mirror that focuses light
on  plane of photon detectors; these currently are Hybrid Photo-Diodes. They have
a photocathode and a 20 KV potential difference between the photocathode
and a silicon diode that is segmented into 163 individual pads. The
photoelectron is accelerated and focused onto the
diode yielding position information for the initial photon.
The system will provide four standard-deviation kaon/pion 
separation between 3-70 GeV/c, electron/pion separation up to 22 GeV/c and 
pion/muon separation up to 15 GeV/c. Because protons don't radiate until 9
GeV/c they can't be distinguished from kaons below this momenta. BTeV is
considering an additional liquid C$_6$F$_{14}$ radiator, 1 cm thick, in the
front of the gas along with a proximity focused phototube array adjacent to the sides
of the gas volume, to resolve this ambiguity.

In addition, BTeV has an excellent Electromagnetic 
calorimeter made from PbWO$_4$ crystals, based on the design of CMS. Finally,
the Muon system is used to both identify muons and
provide an independent trigger on dimuons (BTeV 2000).

\section{{\boldmath $B^o-\overline{B}^o$} Mixing}
\subsection{Introduction}
\begin{figure}[thb]
\centerline{\epsfig{figure=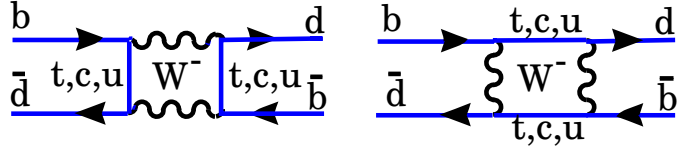,width=3.5in}}
\caption{\label{bmix}The two diagrams for $B_d$ mixing.} 
\end{figure}

Neutral $B$ mesons can transform to 
their anti-particles before they decay. The 
diagrams for this process are shown in Figure~\ref{bmix} for the $B_d$. There
is a similar diagram for the $B_s$. Although $u$, $c$ and
$t$ quark exchanges are all shown, the $t$ quark plays a dominant role mainly
due to its mass, as the amplitude of this process is proportional to the
mass of the exchanged fermion.

Under the weak interactions the eigenstates of flavor, degenerate in pure QCD
can mix. Let the quantum mechanical basis vectors be $\{|1\rangle,|2\rangle\}
\equiv \{|B^o\rangle,|\overline{B}^o\rangle\}$. Then the Hamiltonian is

\begin{equation}
{\cal H}=M-{i\over 2}\Gamma=\left(\begin{array}{cc}
M & M_{12}\\
M_{12}^* & M 
\end{array}
\right) -{i\over 2}\left(\begin{array}{cc}
\Gamma & \Gamma_{12} \\
\Gamma_{12}^* & \Gamma
\end{array}\right)  .
\end{equation}
Diagonalizing we have
\begin{equation}
\Delta m = m_{B_H}-m_{B_L}=2\left|M_{12}\right|~~~.
\end{equation}
Here $H$ refers to the heavier and $L$ the lighter of the two weak
eigenstates.

\begin{figure}[thb]
\centerline{\epsfig{figure=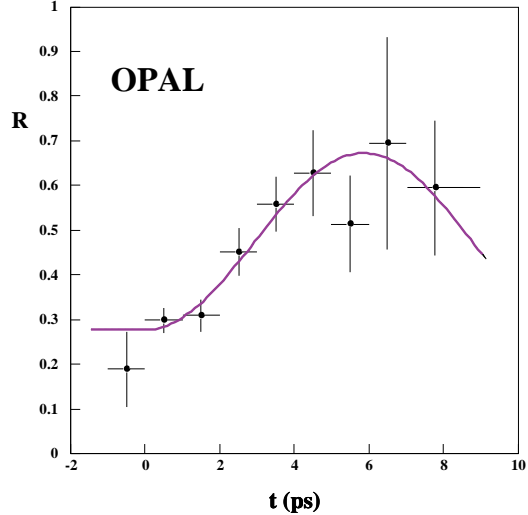,width=2.7in}}
\caption{\label{opal_mix}
The ratio, R, of like-sign to total events as a function of proper decay
time, for selected $B\to D^{*+}X\ell^-\bar{\nu}$ events. The jet charge in the
opposite hemisphere is used to determine the sign correlation. The curve is the result of 
a fit to the mixing parameter.}
\end{figure}

$B_d$ mixing was first discovered by the ARGUS 
experiment (Albrecht 1983) (There was a previous measurement by UA1 indicating
mixing for a mixture of $B_d^o$ and $B_s^o$ (Albajar 1987) At the time it was quite a surprise, since $m_t$ was thought to be in
the 30 GeV range.
It is usual to define  $R$ as probability for a $B^o$ to materialize as a $\overline{B}^o$
divided by the probability it decays as a $B^o$.   The OPAL data for $R$
(Akers 1995) are shown in Figure~\ref{opal_mix}. 

\begin{figure}[thb]
\centerline{\epsfig{figure=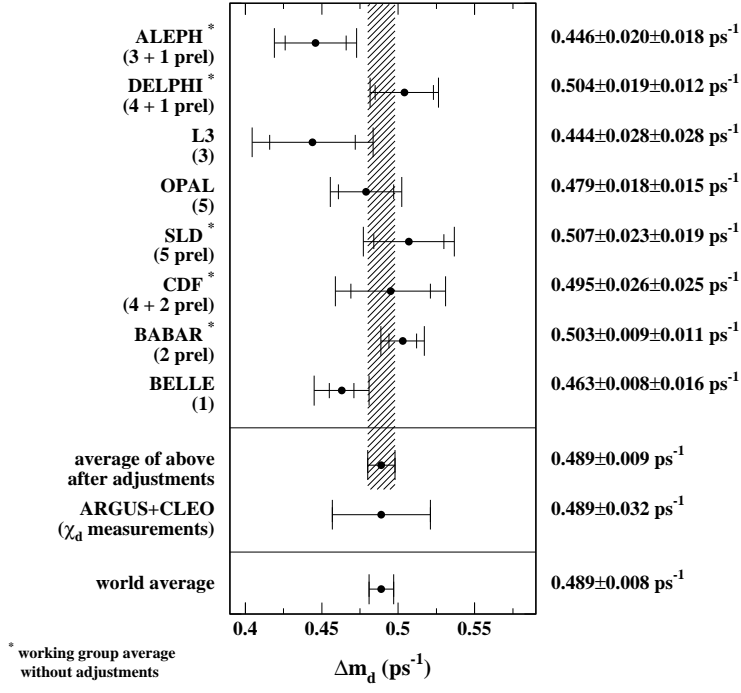,width=4in}}
\caption{\label{Bd_mix_sum}
Values of the $B_d$ mixing parameter $\Delta m_d$ for each experiment.}
\end{figure}

Data from many experiments has been combined by ``The LEP Working Group,"
to obtain an average value $\Delta m_d = (0.489\pm0.0008)\times
10^{12}\hbar$s$^{-1}$. Values from individual experiments are listed in
Figure~\ref{Bd_mix_sum}.

 The probability of mixing is given 
by (Gaillard 1974) (Bigi 2000)
\begin{equation}
x\equiv \frac{\Delta m}{\Gamma}={G_F^2\over 6\pi^2}B_Bf_B^2m_B\tau_B|V^*_{tb}V_{td}|^2m_t^2
F{\left(m^2_t\over M^2_W\right)}\eta_{QCD},
\label{eq:Bdmix}
\end{equation}
where $B_B$ is a parameter related to the probability of the $d$ and $\bar{b}$ 
quarks 
forming a hadron and must be estimated theoretically, $F$ is a known function 
which 
increases approximately as $m^2_t$, and $\eta_{QCD}$ is a QCD
correction, with value about 0.8. By far the largest uncertainty arises from the 
unknown decay constant, $f_B$. In principle $f_B$ can be measured. The decay
rate of the 
annihilation process $B^-\to\ell^-\overline{\nu}$ is proportional to the
product of $f_B^2|V_{ub}|^2$. Even if $V_{ub}$ were well known this is a very
difficult process to measure. Our current best hope is to rely on unquenched
lattice QCD which can use the measurements of the analogous $D^+\to\mu^+\nu$
decay as check. This will take the construction of a ``$\tau$-charm factory." 

Now we relate the mixing measurement to the CKM parameters. Since
\begin{equation} 
 |V^*_{tb}V_{td}|^2\propto |(1-\rho-i\eta)|^2=(\rho-1)^2+\eta^2,
\label{eq:mixrhoeta}
\end{equation}
measuring mixing gives a circle centered at (1,0) in the $\rho - \eta$ plane.
This could in principle be a very powerful constraint. Unfortunately, the
parameter $B_B$ is not experimentally accessible and $f_B$, although in 
principle measurable, has not been and may not be for a very long time, so it
too must be calculated. The errors on the calculations are quite large.

\subsection{{\boldmath $B_s$} Mixing in the Standard Model}

$B^o_s$ mesons can mix in a similar fashion to $B^o_d$ mesons. The diagrams
in Figure~\ref{bmix} are modified by substituting $s$ quarks for $d$ quarks,
thereby changing the relevant CKM matrix element from $V_{td}$ 
to $V_{ts}$. The time dependent mixing fraction is
\begin{equation}
x_s\equiv \frac{\Delta m_s}{\Gamma_s}={G_F^2\over 6\pi^2}
B_{B_s}f_{B_s}^2m_{B_s}\tau_{B_s}|V^*_{tb}V_{ts}|^2m_t^2
F{\left(m^2_t\over M^2_W\right)}\eta_{QCD},
\end{equation} 
which differs from equation~(\ref{eq:Bdmix}) by having parameters relevant for the
$B_s$ rather than the $B_d$.

Measuring $x_s$ allows us to use ratio of $x_d/x_s$ to provide constraints
on the CKM parameters $\rho$ and $\eta$. We still obtain
a circle in the ($\rho ,\eta$) plane centered at (1,0): 
\begin{eqnarray}
\left|V_{td}\right|^2&=&A^2\lambda^4\left[(1-\rho)^2+\eta^2\right] \\
{\left|V_{td}\right|^2 \over \left|V_{ts}\right|^2}&=&(1-\rho)^2+\eta^2
\nonumber ~~.
\end{eqnarray} 
Now however we must calculate only the SU(3) broken ratios $B_{B_d}/B_{B_s}$
and $f_{B_d}/f_{B_s}$. 

$B^o_s$ mixing has been searched for at LEP, the Tevatron, and the SLC. A combined analysis
has been performed. The probability, ${\cal{P}}(t)$ for a $B_s$ to oscillate into a
$\overline{B}_s$ is given as
\begin{equation}
{\cal{P}}(t)\left(B_s\to\overline{B}_s\right)={1\over 2}\Gamma_s e^{-\Gamma_s t}
\left[1+\cos\left(\Delta m_s t\right)\right]~~,
\end{equation}
where $t$ is the proper time.

\begin{figure}[thb]
\centerline{\epsfig{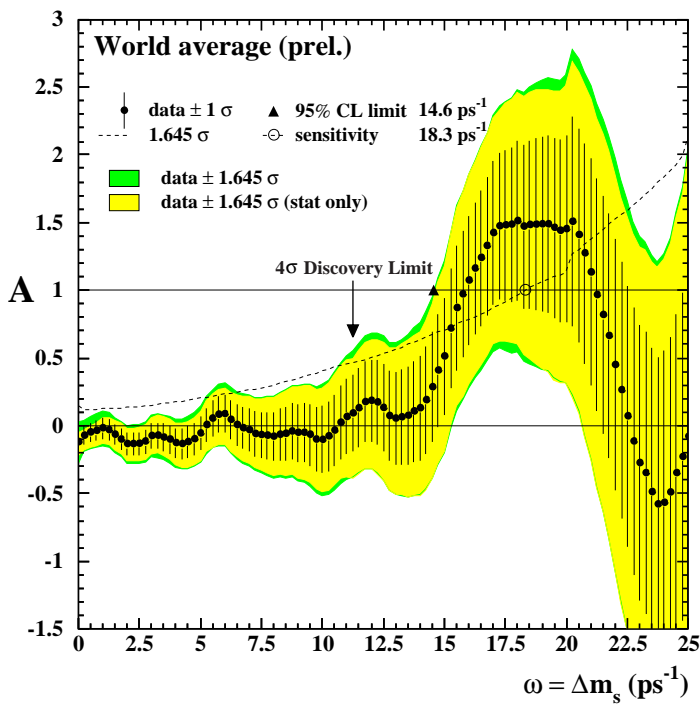}}
\caption{\label{Bs_mix_sum}
Combined experimental values of the amplitude $A$ versus the test  frequency
$\omega = \Delta m_d$ as defined in equation~\ref{eq:Bs}. The inner (outer)
envelopes give the 95\% confidence levels using statistical (statistical
and systematic) errors. The ``sensitivity" shown at 18.3 ps$^{-1}$ is the
likely place a 95\% c.l. upper limit could be set. Also indicated is the
maximum value, 11 ps$^{-1}$, where a 4$\sigma$ discovery would be possible.}
\end{figure}

To combine different experiments a framework has been established where each
experiment finds a amplitude $A$ for each test frequency $\omega$, defined
as
\begin{equation}
{\cal{P}}(t)={1\over 2}\Gamma_s e^{-\Gamma_s t}
\left[1+A\cos\left(\omega t\right)\right]~~.
\label{eq:Bs}
\end{equation}
Figure~\ref{Bs_mix_sum} shows the world average measured amplitude $A$ as a
function of the test frequency $\omega=\Delta m_s$ (Leroy 2001). For each frequency the
expected result is either zero for no mixing or one for mixing. No other value
is physical, although allowing for measurement errors other values are possible.
The data do indeed cross one at a $\Delta m_s$ of 16 ps$^{-1}$, however here
the error on $A$ is about 0.6, precluding a statistically significant
discovery. The quoted upper limit at 95\% confidence level is 14.6 
ps$^{-1}$. This is the point where the value of $A$ plus 1.645 times the error
on $A$ reach one. Also indicated on the figure is the point where the error
bar is small enough that a 4$\sigma$ discovery would be possible. This is at
$\Delta m_s = 11$ ps$^{-1}$. Also, one should be aware that
all the points are strongly correlated. 

The upper limit on $\Delta m_s$ translates to an upper limit on $x_s$ $< $21.6
also at 95\% confidence level. CDF plans to  probe higher sensitivity and
eventually LHCb and BTeV can reach values of $\sim$80.

\section{Rare {\boldmath $b$} Decays}
\subsection{Introduction}
\begin{figure}[htb]
\vspace{-.3cm}
\centerline{\epsfig{figure=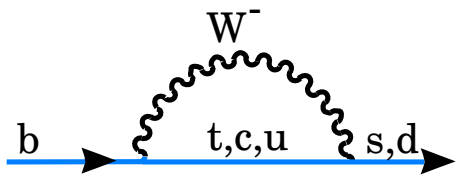,height=1.0in}}
\caption{\label{loop}Loop or ``Penguin" diagram for a $b\to s$ or $b\to d$
transition.}
\vspace{-1mm}
\end{figure}

These processes proceed
through higher order weak interactions involving loops, which are often called
``Penguin" processes, for unscientific reasons (Lingel 1998). A Feynman
loop diagram is shown in Figure~\ref{loop} that describes the transition of a $b$
quark into a charged -1/3 $s$ or $d$ quark, which is effectively a
neutral current transition. The dominant charged current
decays change the $b$ quark into a charged +2/3 quark, either $c$ or $u$.

The intermediate quark inside the loop can be any charge +2/3 quark. The relative
size of the different contributions arises from different quark masses
and CKM elements. In terms of the Cabibbo angle ($\lambda$=0.22), we have
for $t$:$c$:$u$ - $\lambda^2$:$\lambda^2$:$\lambda^4$. The mass dependence 
favors the $t$ loop, but the amplitude for $c$ processes can be 
quite large $\approx$30\%. Moreover, as
pointed out by Bander, Silverman and Soni (Bander 1979), interference can occur
between $t$, $c$ and $u$ diagrams and lead to CP violation. In the Standard Model it is
not expected to occur when $b\to s$, due to the lack of a CKM phase difference,
but could occur when $b \to d$. In any case, it is always worth looking for
this effect; all that needs to be done, for example, is to compare the number
of $K^{*-}\gamma$ events with the number of $K^{*+}\gamma$ events.

There are 
other ways for physics beyond the Standard Model to appear. 
For example, the $W^-$
in the loop can be replaced by some other charged object such as a Higgs; it is
also possible for a new object to replace the $t$.

\subsection{Standard Model Theory}

In the Standard Model the effective Hamiltonian for the intermediate $t$ quark is given
by (Desphande 1994)
\begin{equation}
H_{eff}=-{{4G_F}\over
\sqrt{2}}V_{tb}V_{ts}^*\sum_{i=1}^{10}C_i(\mu)O_i(\mu)~~.
\label{eq:ham}
\end{equation}
Some of the more important operators are
\begin{equation}
O_1=\overline{s}_L^i\gamma_{\mu}b^j_L\overline{c}^j_L\gamma^{\mu}c^j_L,
~~~~O_7={e \over 16\pi^2}m_b \overline{s}_L^i 
\sigma_{\mu\nu}b_R^jF^{\mu\mu}~~.
\label{eq:ops}
\end{equation}

The matrix elements are evaluated at the scale $\mu=M_W$ and then evolved to
the $b$ mass scale using renormalization group equations, which mixes the
operators:
\begin{equation}
C_i(\mu)=\sum_j U_{ij}(\mu,M_W)C_j(M_W)~~.
\label{eq:mixeq}
\end{equation}

\subsection{\boldmath $b \to s\gamma$}

This process occurs when any of the charged particles in Figure~\ref{loop} emits
a photon. The only operator which enters into the calculation is $C_7(\mu)$.
We have for the inclusive decay
\begin{eqnarray}
H_{eff}&=&{4G_F\over \sqrt{2}}\left(V_{tb}V_{ts}\right|^2C_7(m_b)O_7  \\
O_7&=&{e\over 16\pi^2}m_b\overline{s}_L\sigma_{\mu\nu}b_RF^{\mu\nu} \\
\Gamma(b\to s\gamma)&=&{{G_F^2\alpha m_b^5}\over 32\pi^4}\left|C_7\right|^2
\left|V_{tb}V_{ts}^*\right|^2~~.
\end{eqnarray}

Its far more difficult to calculate the exclusive radiative decay rates, but
they are much easier to measure.
Note that the reaction $B\to K\gamma$ would violate angular momentum conservation,
so the simplest exclusive final states are $B\to K^*\gamma$. 

\begin{figure}[htb]
\centerline{\epsfig{figure=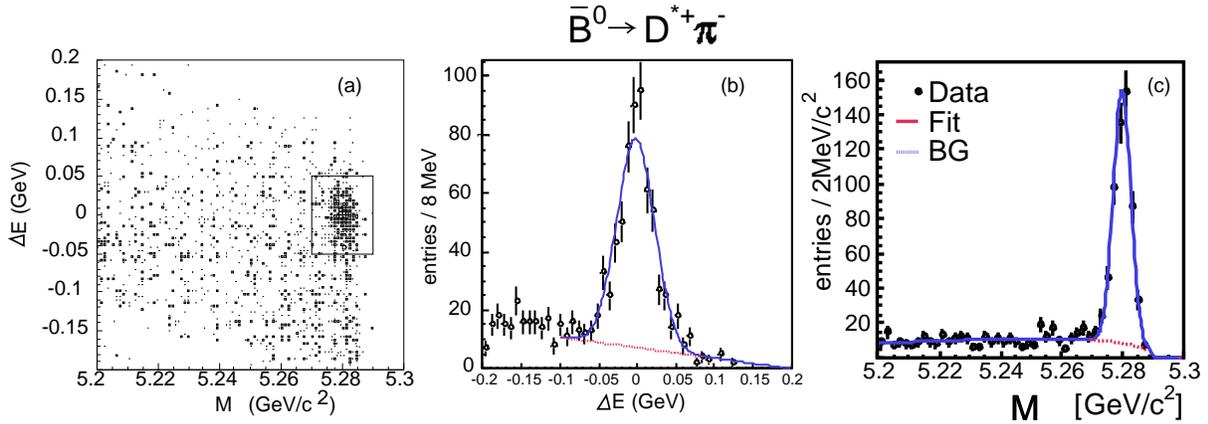,height=2.3in}}
\caption{\label{tajimaDspi}BELLE data for the reaction $\overline{B}^o
\to D^{*+}\pi^-$. (a) The correlation between $\Delta E$ and $M$. The box
shows the signal region. (b) The projection in $\Delta E$ for events
in the $M$ signal region; the line shows a fit to the background. (c)
The $M$ distribution for $\Delta E$ in the signal region; the line shows
a fit to the background.}
\end{figure}

Different techniques are used for reconstructing exclusive and inclusive
Decays and unique methods are invoked for exclusive decays on the
$\Upsilon(4S)$. At other machines the decay products, $i$, from an exclusive
$B$ decays are used to reconstruct an ``invariant mass" via 
$M^2=\sum_iE_i^2-\sum_i\overrightarrow{p}_{\!i}^2$. At the $\Upsilon(4S)$ its
done a bit differently, the decay products are first tested to see if the
sum of their energies is close to the beam energy, $E_{beam}$. If this is true
then the ``beam constrained" invariant mass is calculated as
$M^2=E_{beam}^2-\sum_i\overrightarrow{p}_{\!i}^2$. These methods are used for
all exclusive $B$ decays, in combination with other requirements.
Figure~\ref{tajimaDspi} shows the BELLE data for the reaction $\overline{B}^o
\to D^{*+}\pi^-$, where the $D^{*+}\to \pi^+ D^o$. BELLE first selects events
with candidate $D^o$'s. Then they require an additional $\pi^+$ where the 
measured mass difference between the $D^o\pi^+$ minus the $D^o$ candidate is
consistent with the known mass difference. Selecting the $D^{*+}$ candidates
they combine them with candidate $\pi^-$. In Figure~\ref{tajimaDspi}(a) the
correlation between difference in measured energy $\Delta E=\sum_i E_i
-E_{beam}$ versus the beam-constrained invariant mass is shown. In
Figure~\ref{tajimaDspi}(b) $\Delta E$ is shown after selecting the signal region
in $M$ and in Figure~\ref{tajimaDspi}(c) $M$ is shown after selecting on $\Delta
E$. These plots show how clean signals can be selected.

\begin{figure}[htb]
\centerline{\epsfig{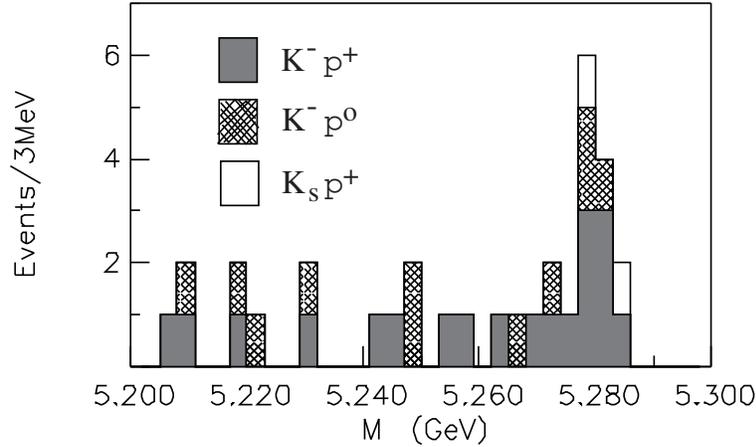}}
\caption{\label{Ksg_cleo} First published CLEO data for the reaction 
${B}\to K^{*}\gamma$ showing the
$M$ distribution for $\Delta E$ in the signal region.}
\end{figure}

CLEO first measured the inclusive rate (Alam 1995) as well as the exclusive rate into
$K^*(890)\gamma$ (Ammar 1993) shown in Figure~\ref{Ksg_cleo}. Here several different decay modes
of the $K^*(890)$ are used. The current world average value for 
${\cal{B}}(B\to K^*\gamma)=(4.2\pm 0.8\pm 0.6)\times 10^{-5}$. 

To find inclusive decays two techniques are used.
The first one, which provides the cleanest signals, is to sum the exclusive
decays for the final states $Kn\pi\gamma$, where $n\le 4$ and only one of the
pions is a $\pi^o$. These requirements are necessary or the backgrounds become
extremely large. (Both charged and neutral kaons are used.) Of course, imposing
these restrictions leads to a model dependence of the result that must be
carefully evaluated. This is why having an independent technique is useful.
That is provided by detecting only the high energy photon. The technique used
is to form a neutral network to discriminate between continuum and
$\Upsilon(4S)$ data using shape variables.

\begin{figure}[htb]
\centerline{\epsfig{figure=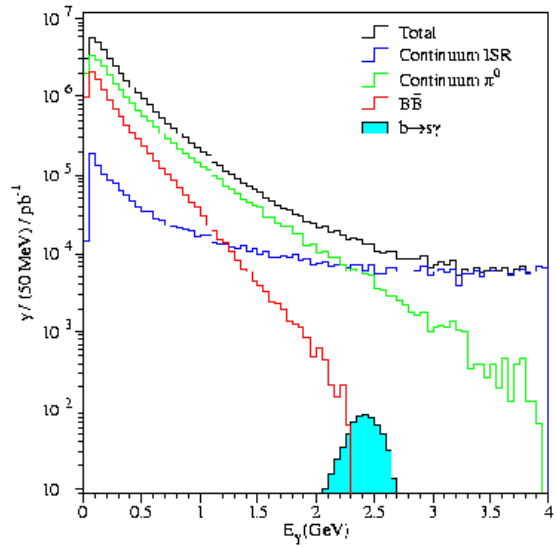,height=2.9in}}
\vspace{-.1cm}
\caption{\label{bsg_exp}Levels of inclusive photons from various background
processes at the $\Upsilon (4S)$ labeled largest to smallest at 2.5 GeV/c. Also shown is the
expected signal from $b\to s\gamma$.}
\end{figure}

The momentum spectrum of the $\gamma$ peaks close to its maximum value at half
the $B$ mass. If we had data with only $B$ mesons, it would be easy to pick out
$b \to s \gamma$. We have, however, a large background from other processes. At
the $\Upsilon (4S)$, the $\gamma$ spectrum from the different background
processes is shown. The largest is $\pi^o$ production from continuum $e^+e^-$
collisions, but another large source is initial state radiation (ISR), where
one of the beam electrons radiates a hard photon before annihilation. The
backgrounds and the expected signal are illustrated in Figure~\ref{bsg_exp}.
Similar backgrounds exist at LEP.

\begin{figure}[htb]
\vspace{-.4cm}
\centerline{\epsfig{figure=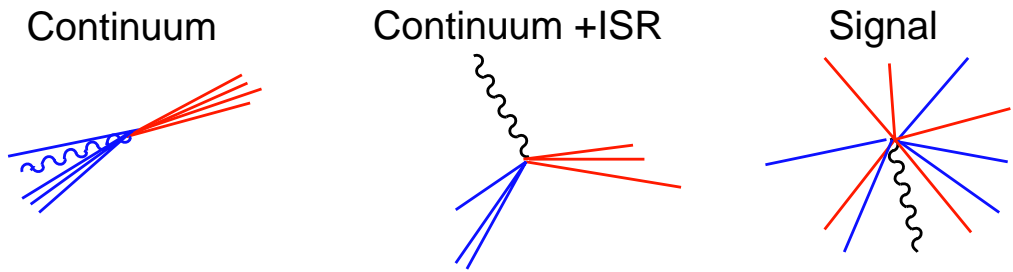,height=1.4in}}
\vspace{-.4cm}
\caption{\label{shape0}Examples of idealized event shapes. The straight lines
indicate hadrons and the wavy lines photons.}
\end{figure}

To remove background CLEO used two techniques originally, one based on ``event
shapes" and the other on summing exclusively reconstructed $B$ samples.
Examples of idealized events are shown in Figure~\ref{shape0}. CLEO uses eight
different shape variables described in Ref. [3], and defines a
variable $r$ using a neural network to distinguish signal from  background. The
idea of the $B$ reconstruction analysis is to find the inclusive branching
ratio by summing over exclusive modes. The allowed hadronic system is comprised
of either a $K_s\to\pi^+\pi^-$ candidate or a $K^{\mp}$ combined with 1-4
pions, only one of which can be neutral. The restriction on the number and kind
of pions maximizes efficiency while minimizing background. It does however lead
to a model dependent error. For all combinations CLEO evaluates 
\begin{equation}
\chi^2_B = \left({M_B-5.279}\over{\sigma_M}\right)^2 +  
\left({E_B-E_{beam}}\over{\sigma_E}\right)^2, \label{eq:chisq}
\end{equation}
where $M_B$ is the measured $B$ mass for that hypothesis and $E_B$ is its
energy.  $\chi^2_B$ is required to be $<$ 20. If any particular event has more
than one hypothesis, the solution which minimizes $\chi^2_B$ is chosen.
For events with a reconstructed $B$ candidate CLEO
also considers the angle between the direction of the $B$ and the thrust axis
of event with the $B$ candidate removed, $\cos(\theta_t)$. This is highly
effective in removing continuum background.

\begin{figure}[htb]
\vspace{-.2cm}
\centerline{\epsfig{figure=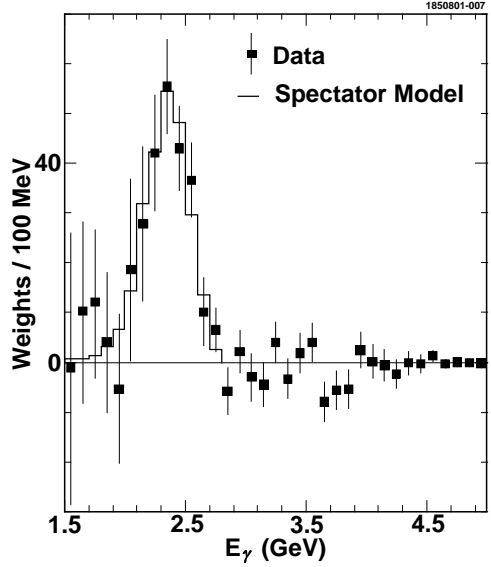,height=3in}}
\caption{\label{bsg_van}The background subtracted photon energy spectrum from
CLEO. The spectrum is not corrected for resolution or efficiency. The 
solid lines show the spectrum from a simulation of the Ali-Greub spectator
model with the $b$ quark mass set to 4.690 GeV and the Fermi momentum
set to 410 MeV/c.}
\end{figure}

A neural network is used to combine $r$, $\chi^2_B$, $\cos(\theta_t)$
into a new variable $r_c$ and events are then weighted according to their
value of $r_c$. This method maximizes the statistical potential of the
data.
Figure~\ref{bsg_van} shows the photon energy spectrum of the inclusive
signal from CLEO combining both reconstruction techniques. The signal is
compared to a theoretical prediction based on the model of Ali and Greub
(Ali 1991). A fit to the model 
over the photon energy range from 2.0 to 2.7 GeV/c gives the branching 
ratio result shown in Table~\ref{tab:btosgresults}, where the first error is
statistical, the second systematic and the third dependence on the theoretical
model (Chen 2001). 

ALEPH reduces the backgrounds by weighting candidate decay
tracks in a $b \to s\gamma$ event by a combination of their momentum, impact
parameter with respect to the main vertex and rapidity with respect to the
$b$-hadron direction (Barate 1998).

Current results are also shown in
Table~\ref{tab:btosgresults}.  The data are in
agreement with the Standard Model theoretical prediction to next to leading
order, including quark mass effects of
$(3.73\pm0.30)\times 10^{-4}$ (Hurth 2001). A deviation here would show physics beyond the
Standard Model. More precise data and better theory are needed to further
limit the parameter space of new physics models, or show an effect. 

\begin{table}[htb]
\vspace{-2mm}
\begin{center}
\caption{${\cal{B}}(b\to s\gamma)$.
\label{tab:btosgresults}}
\vspace*{2mm}
\begin{tabular}{lc}\hline\hline
Experiment & ${\cal{B}}\times 10^{-4}$\\
\hline
CLEO & $3.21\pm 0.43\pm 0.27^{+0.18}_{-0.10}$ \\
ALEPH & $3.11\pm 0.80\pm 0.72$ \\
BELLE & $3.36\pm 0.53\pm 0.44^{+0.50}_{-0.54}$\\\hline
Average & $3.23\pm 0.42$ \\
\hline\hline
\end{tabular}
\end{center}
\end{table}

\subsubsection{{\boldmath $|V_{cb}|$} Using Moments of the Photon Energy
Spectrum}
\label{section:Vcbsg}
In section(\ref{section:Vcbinc}) we found a value of $V_{cb}$ using the first and
second hadronic mass moments. Here we use the first moment of the photon energy
distribution in $b\to s\gamma$. The values
found for the moments and $\overline{\Lambda}$ which is directly proportional
to $\langle E_{\gamma}\rangle$ are (Chen 2001)
\begin{eqnarray}
\langle E_{\gamma}\rangle &=& 2.346\pm 0.032\pm 0.011{~\rm GeV} \\
\langle E_{\gamma}^2\rangle - \langle E_{\gamma}\rangle^2&=&
 0.0226\pm 0.0066\pm 0.0020{~\rm GeV^2} \\
\overline{\Lambda} &=& 0.35\pm 0.08\pm 0.10 {~\rm GeV}.
\end{eqnarray}

\begin{figure}[htb]
\centerline{\epsfig{figure=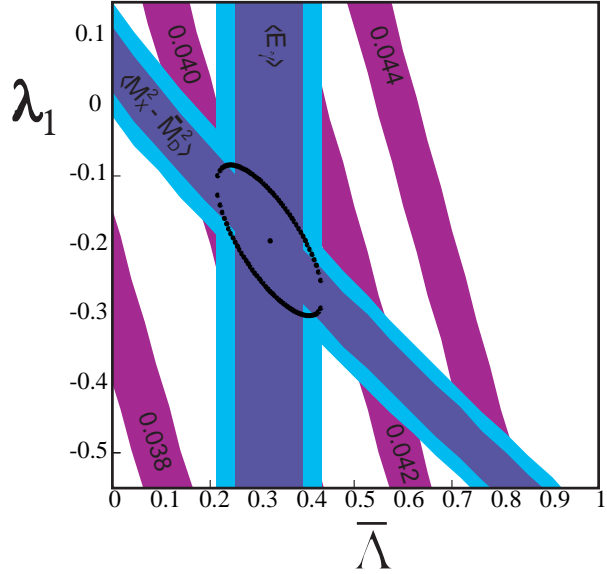,height=3in}}
\vspace{-.3cm}
\caption{\label{sgamma_moment}Correlation between $\lambda_1$,
$\overline{\Lambda}$ and $V_{cb}$ derived from $\langle E_{\gamma}\rangle$
and $\langle M_X^2-\overline{M}_D^2\rangle$. The lighter bounds include
both experimental and theoretical errors.}
\end{figure}

In Figure~\ref{sgamma_moment} we show the combination of first moments from
photon energy in $b\to s\gamma$ and hadron moments in $b\to c\ell\nu$.
This implies a value of $V_{cb}$ around 0.0406.

\subsection{Rare Hadronic Decays}
\subsubsection{{\boldmath $B\to \pi\pi$} and {\boldmath $B\to K\pi$}}
\begin{figure}[htb]
\centerline{\epsfig{figure=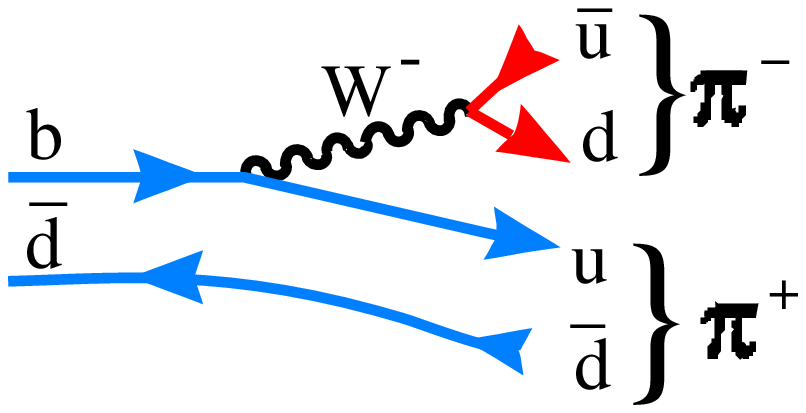,height=1.5in}
\epsfig{figure=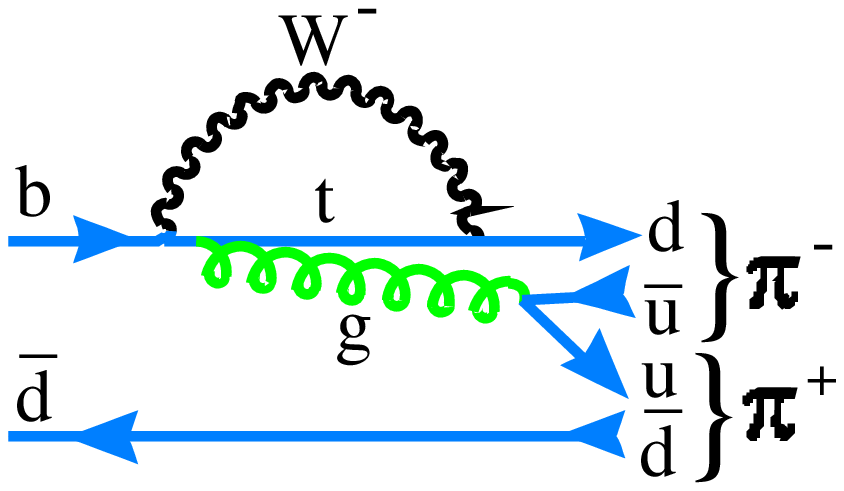,height=1.5in}}
\vspace{-.3cm}
\caption{\label{pippim}Decay diagrams for $\overline{B}^o\to\pi^+\pi^-$. (left) Via
tree level $V_{ub}$ moderated decay. (right) Via a Penguin process.}
\end{figure}

The decays $\overline{B}^o\to\pi^+\pi^-$ and $\overline{B}^o\to K^-\pi^+$
 do not contain any charm
quarks in the final state so must proceed via either the tree level $V_{ub}$
process shown in Figure~\ref{pippim}(left) or via the Penguin process shown
on the right side.

These diagrams can interfere and they can also interfere through $B^o$ mixing,
thus complicating any weak phase extraction. The same diagrams are applicable
for $\overline{B}^o\to K^-\pi^+$ by replacing $W^-\to \bar{u}d$ in the 
tree level diagram by $W^-\to \bar{u} s$ and replacing the $td$ coupling
in the Penguin by a $ts$ coupling.

\begin{figure}[htbp]
\centerline{\epsfig{figure=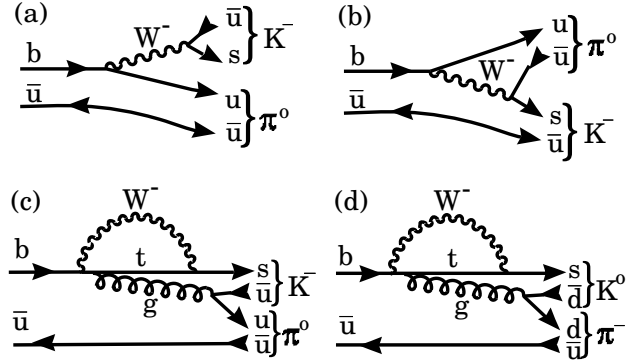,width=3.3in}}
\caption{\label{kpi} Diagrams for $B^-\to K^-\pi^o$, (a) and (b) are tree level
 diagrams where (b) is color suppressed; (c) is a  penguin diagram. (d) shows
  $B^-\to K^o\pi^-$, which cannot be produced via a tree diagram.}
\end{figure}

Other diagrams for producing $K\pi$ final states are shown in Figure~\ref{kpi}.
In section~\ref{section:CPV} it will be shown that CP violation can result
from the interference between two distinct decay amplitudes leading to the
same final state.
Consider the possibility of observing CP violation 
by measuring a rate difference between $B^-\to K^-\pi^o$ and $B^+\to K^+\pi^o$.
The $K^-\pi^o$ final state can be reached either by tree or penguin diagrams. 
The tree diagram has an imaginary part coming from the $V_{ub}$ coupling, while 
the penguin term does not, thus insuring a weak phase difference. This type of CP 
violation is called ``direct." Note also that the process  $B^-\to K^o\pi^-$
can only be produced by the penguin diagram in Figure~\ref{kpi}(d). Therefore,
we do not expect a rate difference between  $B^-\to K^o\pi^-$ and
$B^+\to K^o\pi^+$. 

\begin{figure}[htbp]
\centerline{\epsfig{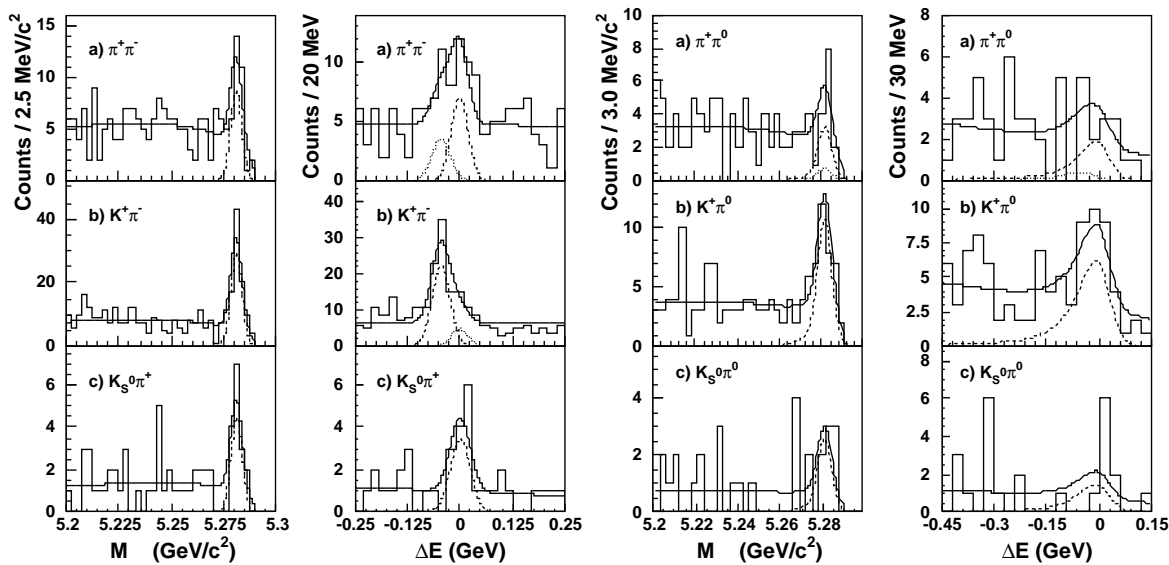}}
\caption{\label{kp-pipi-belle}Signals in $M$ and $\Delta E$ for
two-body decay modes from BELLE. The data result from projections of
a Likelihood fit that takes into account event shape and particle identification
information. The dashed lines are the signal projections. The dotted lines
in the $\Delta E$ distributions are projections of the background component
from the $\pi\Leftrightarrow K$ substitution.}
\end{figure}

Measurements of these rates have been by several groups. Recent data from BELLE
are shown in Figure~\ref{kp-pipi-belle} (Abe 2001a). Table~\ref{tab:kp-pipi} lists the
currently measured branching ratios.

\begin{table}[htb]
\vspace{-2mm}
\begin{center}
\caption{Branching Ratios for $B\to K\pi$ and $B\to \pi\pi$ in units
of $10^{-6}$.
\label{tab:kp-pipi}}
\vspace*{2mm}
\begin{tabular}{lcccc}\hline\hline
Mode & CLEO & BABAR & BELLE & Average \\
\hline
$\pi^+\pi^-$ & $4.7^{+1.8}_{-1.5}\pm 0.6$ & $4.1\pm 1.0\pm 0.7$ &
$5.6^{+2.3}_{-2.0}\pm 0.4$  & $4.5^{+0.9}_{-0.8}$\\
$\pi^+\pi^o$  & $<$12 & $<$9.6 &$<$13.4 &\\
$K^{\pm}\pi^{\mp}$ & $18.8^{+2.8}_{-2.6}\pm 1.3$ & $16.7\pm 1.6 \pm 1.3$ &
$19.3^{+3.4+1.5}_{-3.2-.06}$ & $17.7^{+1.6}_{-1.5}$\\
$K^+\pi^o$& $12.1^{+3.0+2.1}_{-2.8-1.4}$ & $10.8^{+2.1}_{-1.9}\pm 1.6$ &
$16.3^{+3.5+1.6}_{-3.3-1.8}$ & $12.1^{+1.7}_{-1.6}$ \\
$K^o\pi^-$& $18.2^{+4.6}_{-4.0}\pm 1.6$ & $18.2^{+3.3}_{-3.0}\pm 2.0$ &
$13.7^{+5.7+1.9}_{-4.8-1.8}$ & $17.3^{+2.7}_{-2.4}$ \\
$K^o\pi^o$& $14.8^{+5.9+2.4}_{-5.1-3.3}$ & $8.2^{+3.1}_{-2.7}\pm 1.2$ &
$16.0^{+7.2+2.5}_{-5.9-2.7}$ & $10.4^{+2.7}_{-2.5}$ \\
\hline\hline
\end{tabular}
\end{center}
\end{table}

\section{Hadronic Decays}
\subsection{Introduction}
Mark Wise in his talk at the 2001 Lepton Photon conference (Wise 2001) gave some
advice to theorists: ``If you drink the nonlep tonic your physics
career will be ruined and you will end up face down and in the gutter."
Presumably Mark's statement was inspired by the difficulty in predicting
hadronic decays. Here we have lots of gluon exchange with low energy gluons,
while perturbation theory works well when the energies are large compared with
$\Lambda_{QCD}$ $\sim$200 MeV. Furthermore, multibody decays are currently
impossible to predict, so we will consider only two-body decays.

\subsection{Two-Body Decays into a Charm or Charmonium}

\begin{figure}[htb]
\centerline{\epsfig{figure=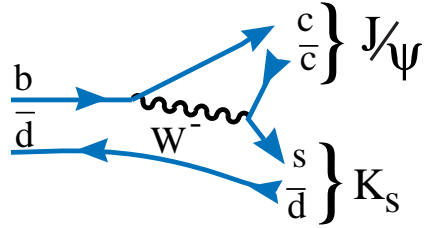,height=1.3in}}
\caption{\label{psi_ks}Color suppressed spectator decay diagram
for $B^o\to J/\psi K_s$.}
\end{figure}

We start by considering two-body decays into a charmed hadron (Neubert 1997).
Figure~\ref{DSPI} shows the processes for both $B^-$ and $\overline{B}^o$ decays
into a $D$ and a $\pi^-$. There is only one possible process for the
$\overline{B}^o$, the simple spectator process (left), while the color
suppressed spectator (right) is also allowed for the $B^-$. We call decays with
only the simple spectator diagram allowed ``class I" and decays where both the
simple and color suppressed diagrams are allowed ``class III." Note, that
because the colors of all the outgoing quarks must be the same in the color
suppressed case, naively the amplitude is only $\sim$1/3 that of the simple
spectator case where the $W^-$ can transform into quarks of all three colors.
``Class II" decays are processes than can only be reached by the color
suppressed spectator diagram, for example the $B^o\to J/\psi K_s$ decay shown
in Figure~\ref{psi_ks}.

The effective Hamiltonian consists of local 4-quark operators renormalized at
the scale $\mu$ and the Wilson coefficients (from the Operator Product
Expansion) $c_i(\mu)$. We have
\begin{eqnarray}
H_{eff}&=&{G_F\over \sqrt{2}}\left\{V_{cb}\left[c_1(\mu)Q_1^{cb}+
c_2\mu)Q_2{cb}\right]+V_{ub}...\right\} \\
Q_1^{cb}&=&\left[(\overline{d}'u)_{V-A}+(\overline{s}'c)_{V-A}\right]
(\overline{c}'b)_{V-A} \nonumber \\
Q_2^{cb}&=&(\overline{c}u)_{V-A}(\overline{d}'b)_{V-A}+(\overline{c}c)_{V-A}
(\overline{s}'b)_{V-A}~~, \nonumber
\end{eqnarray}
where the notation $(\overline{q_1}q_2)_{V-A}\equiv 
\overline{q_1}\gamma_{\mu}(1-\gamma_5)q_2$.
Without QCD corrections $c_1(\mu)=1$ and $c_2(\mu)=0$. For non-leading order
correction using the renormalization group equations, we have $c_1(\mu)=1.132$ 
and $c_2(\mu)=-0.249$.

We can factorize the amplitude by assuming that the current producing the
$\pi^-$ is independent of the one producing the charmed hadron. Lets consider a
class I case, $\overline{B}^o\to D^+\pi^-$. The amplitude can be written as
\begin{equation}
A_{fact}=-{G_F\over \sqrt{2}}V_{cb}V_{ud}^*{{a_1}}\langle\pi^-\left|
(\overline{d}u)_A\right|0\rangle
\langle D^+\left|(\overline{c}b)_V\right|\overline{B}^o\rangle
\end{equation}
The part of the amplitude dealing with the $\pi^-$ is known from pion decay.
We have $\langle\pi^-\left|\overline{d}\gamma_{\mu}\gamma_5 u\right|0\rangle
=if_{\pi}p_{\mu}$, where the axial vector structure is made explict, $p_{\mu}$
is the pion four-vector and $f_{\pi}$ is given by measuring the decay width
for $\pi^-\to\mu^-\nu$. The term ${ a_1}$ is defined as
\begin{equation}
{ a_1}=c_1(\mu_f)+\xi c_2(\mu_f)~~,
\end{equation}
where $\xi$ is equal to the number of colors and the scale $\mu_f$ is on the
order of the $b$ quark mass. Then
\begin{equation}
A_{fact}=-{G_F\over \sqrt{2}}V_{cb}V_{ud}^*{{a_1}}f_{\pi}(m_B^2-m_D^2)
F_0^{B\to D}(m_{\pi}^2)~~.
\end{equation}
The $F_0$ form factor can either be calculated or measured, for example in
semileptonic decays. 

Let us also consider a class II process $B\to J/\psi K$. In this case
\begin{equation}
A_{fact}=-{G_F\over \sqrt{2}}V_{cb}V_{cs}^*{{a_2}}\langle J/\psi\left|
(\overline{c}c)_V\right|0\rangle
\langle K\left|(\overline{s}b)_V\right|{B}\rangle~~,\label{eq:fact}
\end{equation}
where ${ a_2}=c_2(\mu_f)+\xi c_1(\mu_f).$

A class III decay has a term ${ a_1}+x{ a_2}$ in the amplitude, where
$x$ equals one from flavor symmetry. However the actual values of ${ a_1}$,
$x$, and ${ a_2}$ are not well predicted from theory but we can obtain them
from the data.

\begin{figure}[htb]
\centerline{\epsfig{figure=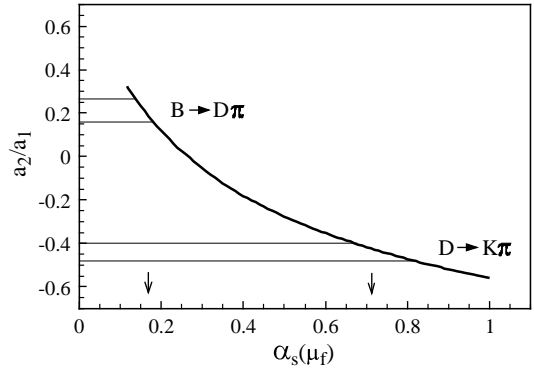,height=2.0in}}
\caption{\label{stech} Calculated values of $a_2/a_1$ versus $\alpha_s$. The
arrows indicate the ratios chosen for $B$ and $D$ decays.}
\end{figure}

One method is to use the class I decays to obtain ${ a_1}=1.08\pm0.04$. It
is possible to calculate $a_2/a_1$ as shown in Figure~\ref{stech} (Neubert
1997). Using these values the measured branching ratios are compared with the
predicted ones in Table~\ref{tab:haddec}. Here $x=+1$ is used, taken from the
data. This is opposite to the interference in $D$ decays but is expected from
the calculation shown in Figure~\ref{stech}.

\begin{table}[htb]
\vspace{-2mm}
\begin{center}
\caption{Predicted Branching Ratios ($10^{-4}$) (Neubert 1997) Compared to
Measurement For Two-Body Hadronic Decays.
\label{tab:haddec}}
\vspace*{2mm}
\begin{tabular}{lcclcclcc}\hline\hline
\multicolumn{3}{c}{Class I}&\multicolumn{3}{c}{Class II}&
\multicolumn{3}{c}{Class III}\\\hline
Mode & Model & Data &Mode & Model & Data &Mode & Model & Data\\\hline
$D^+\pi^-$ & 30 & 30$\pm$4 & $J/\psi K^-$ & 11 & 10$\pm$1 
& $D^o\pi^-$ & 48 & 53$\pm$5\\
$D^{*+}\pi^-$ &30 & 28$\pm$2 & $J/\psi K^*$ & 17 & 15$\pm$2
&$D^{*o}\pi^-$ & 49 & 46$\pm$4 \\
$D^+\rho^-$ & 70 & 79$\pm$14 & $D^o\pi^o$ & 0.7 & 2.8$\pm$0.4
&$D^o\rho^-$ & 110 & 134$\pm$18 \\
$D^{*+}\rho^-$ & 85& 73$\pm$15 & $D^{*o}\pi^o$ &1.0& 2.1$\pm$0.9
&$D^{*o}\rho^-$ & 119 & 155$\pm$31 \\

\hline\hline
\end{tabular}
\end{center}
\end{table}
The agreement is rather good except for the newly measured $D^{o(*)}\pi^o$
modes where it is rather miserable. CLEO and BELLE both have rates for $D^o\pi^o$ of $(2.7\pm 0.3\pm 0.5)
\times 10^{-4}$ and $(2.9^{+0.4}_{-0.3}\pm 0.6)\times 10^{-4}$, respectively,
while CLEO alone has measured $D^{*o}\pi^o$ as $(2.1\pm 0.5\pm 0.8)
\times 10^{-4}$.  

\subsubsection{Isospin Analysis of the {\boldmath $B\to D\pi$} System}
\begin{figure}[htb]
\centerline{\epsfig{figure=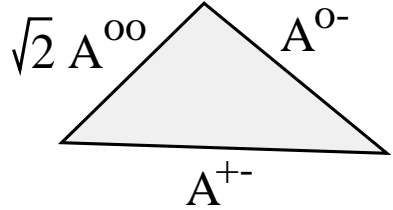,height=1.2in}}
\caption{\label{isospin_amp} The isospin amplitude triangle for $B\to D\pi^-$ decays.}
\end{figure}

All the decay rates for $B\to D\pi$ have now been measured. The four-quark
operator $(\overline{d}u)(\overline{c}b)$ has isospin I=1 and I$_3$=+1. It
transforms the $\overline{B}^o$ into final states $D^+\pi^-$ and
$D^o\pi^o$ with I=1/2 or I=3/2. The $B^-$ decays into $D^o\pi^-$ with
I=3/2 only. It is thought that the isospin amplitudes cannot be modified by
final state interactions, so we can look for evidence of final state phase
shifts by doing an ``isospin analysis." The decay amplitudes form a triangle as
shown in Figure~\ref{isospin_amp}.

The relationship among $B$ decay amplitudes and isospin amplitudes is given by
\begin{eqnarray}
A(\overline{B}^o\to D^+\pi^-)&=&\sqrt{1\over 3} A_{3/2}+\sqrt{2\over 3}A_{1/2} \\
A(\overline{B}^o\to D^o\pi^o)&=&\sqrt{2\over 3} A_{3/2}-\sqrt{1\over 3}A_{1/2}
\nonumber \\
A({B}^-\to D^o\pi^-)&=&\sqrt{3} A_{3/2}~~. \nonumber
\end{eqnarray}

These equations may be solved for the isopsin amplitudes and the relative phase
shift between the two amplitudes. The solution is
\begin{eqnarray}
\left|A_{1/2}\right|^2&=&\left|A(\overline{B}^o\to D^+\pi^-)\right|^2
+\left|A(\overline{B}^o\to D^o\pi^o)\right|^2-{1\over 3}
\left|A({B}^-\to D^o\pi^-)\right|^2  \\
\left|A_{1/2}\right|^2&=&{1\over 3}\left|A({B}^-\to D^o\pi^-)\right|^2
\nonumber  \\
\cos\delta&=&\cos\left(\delta_{3/2}-\delta_{1/2}\right) =
{{3\left|A({B}^-\to D^o\pi^-)\right|^2-2\left|A_{1/2}\right|^2
-\left|A_{3/2}\right|^2}\over{2\sqrt{2}\left|A_{1/2}\right|\left|A_{3/2}\right|}}
\nonumber
\end{eqnarray}

Solving these equations for the $D\pi$ final states gives 
$\cos\delta=0.88\pm 0.05$ which indicates a
phase shift of about $28\pm 8$ degrees, but is not statistically
significant.

\subsubsection{Factorization Tests Using Semileptonic Decays}

The factorized amplitude for $D\pi^-$ decays in equation~\ref{eq:fact} is the
product of two hadronic currents, one for $W^-\to\pi^-$ and the
other for $B\to D$. In semileptonic decay (Figure~\ref{Bdecaymech}) we
have the product of the known lepton current and the pion current.
At $q^2=m_{\pi}^2$ the $B\to D$ should be the same in both decays, at least
for class I.
The comparision for the general case of any hadron $h^-$ is
\begin{equation}
\left.\Gamma\left(\overline{B}\to D^{(*)}h\right)=6\pi^2a_1^2f_h^2
\left|V_{ud}\right|^2{d\Gamma\over dq^2}
\left(\overline{B}\to D^{(*)}\ell^-\overline{\nu}\right)\right|_{q^2=m_h^2}~~.
\end{equation}
Tests of this equation for $D^{*+}$ and a $\pi^-$, $\rho^-$ or $a^-_1$ are satisfied at about 15\% accuracy (Bortoletto 1990) (Browder 1996).

Another test compares the polarization of the $D^*$ in both hadronic and
semileptonic cases:
\begin{equation}
\left.{\Gamma_L\over \Gamma}\left(\overline{B}\to D^*h\right)=
{\Gamma_L\over \Gamma}\left(\overline{B}\to D^*h\right)\right|_{q^2=m_h^2}~~,
\end{equation}
where $\Gamma_L$ denotes the longitudinally polarized fraction of the decay
width. Comparisons with data will be shown in section~\ref{sec:rhop}.

There are also more modern approaches to factorization (Beneke 2001) (Bauer
2001). However these approaches predict 
\begin{equation}
{{\Gamma(B^-\to D^o\pi^-)}\over {\Gamma(\overline{B}^o\to D^+\pi^-)}}
=1+{\cal O}(\Lambda_{QCD}/m_b)~~
\end{equation}
which seems to contradict current observations.

\subsection{Observation of the $\rho'$ in $B$ Decays}
\label{sec:rhop}

\begin{figure}[bht]
\centerline{\epsfig{figure=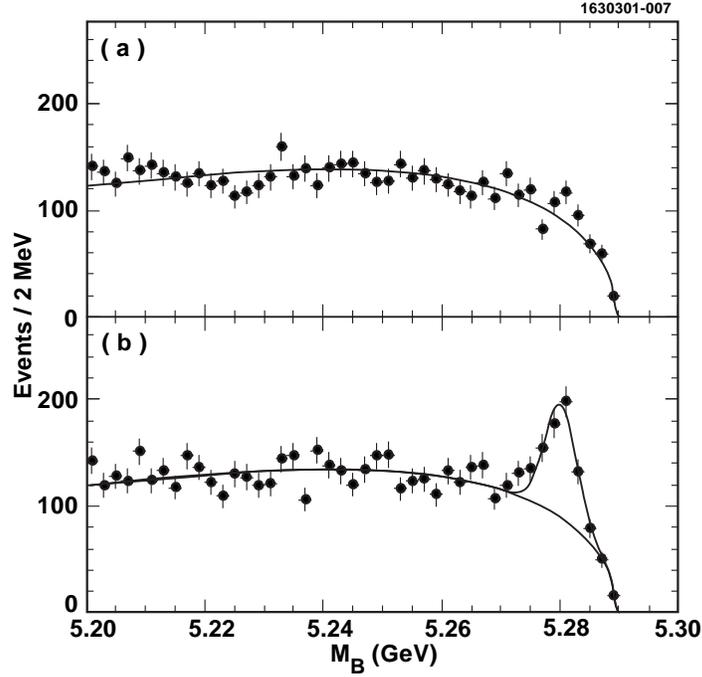,height=3.6in}}
\caption{\label{007}The $\overline{B}$ candidate mass spectra for the
final state
$D^{*+}\pi^+\pi^-\pi^-\pi^0$, with $D^0\to K^-\pi^+$ (a) for 
$\Delta E$
sidebands and (b) for $\Delta E$ consistent with zero. The curve 
in (a) is a
fit to the background distribution described in the text, while in 
(b) the
shape from (a) is used with the normalization allowed to float and 
a signal
Gaussian of width 2.7 MeV is added.}
\end{figure}

CLEO made the first statistically significant observations of     
six hadronic $B$ decays shown in Table~\ref{table:brs} that result from
studying the reactions $B\to D^{(*)}\pi^+\pi^-\pi^-\pi^o$ (Alexander 2001).
The signal in one of these final states $\overline{B}^o\to
D^{(*+)}\pi^+\pi^-\pi^-\pi^o$, where $D^{*+}\to \pi^+ D^o$ and
$D^o\to K^-\pi^+$ is shown in Figure~\ref{007}. 

\begin{figure}[hbt]
\vspace{-0.3cm}
\centerline{\epsfig{figure=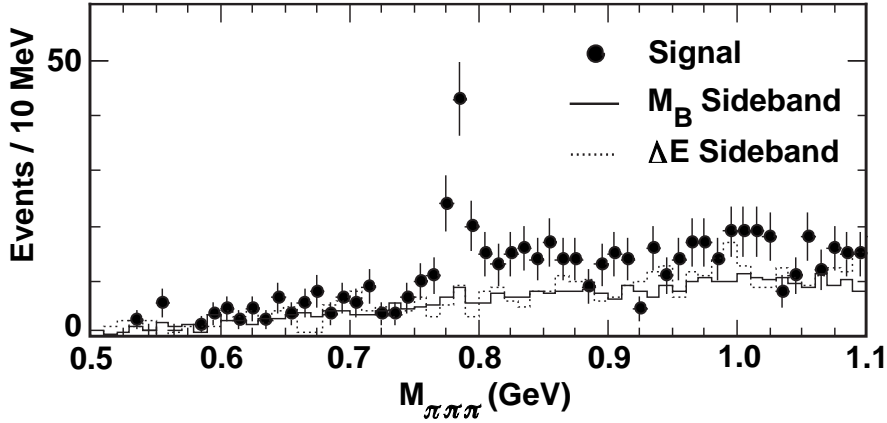,height=2.2in}}
\caption{\label{019}The invariant mass spectra of 
$\pi^+\pi^-\pi^0$ for the final state
$D^{*+}\pi^+\pi^-\pi^-\pi^0$ for all three $D^o$ decay modes $K^-\pi^+$,
$K^-\pi^+\pi^+\pi^-$ and $K^-\pi^+\pi^o$.
The solid histogram is the background
estimate from the $M_B$ lower sideband and the dashed histogram is 
from the
$\Delta E$ sidebands; both are normalized to the fitted number of 
background events. (There is an additional cut selecting the center of the
Dalitz plot.)}
\end{figure}

\begin{figure}[htb]   
\centerline{\epsfig{figure=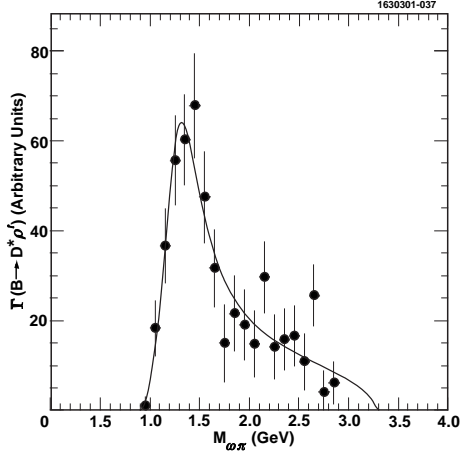, height=2.4in}} 
\caption{\label{omega_pi}(left) The background subtracted efficiency
corrected $\omega\pi^-$ mass spectrum from 
$\overline{B}^o\to (D^{*+}+D^o+D^+)\omega\pi^-$ decays fit to a Breit-Wigner
shape.} 
\end{figure}

\begin{figure}[bth]    
\centerline{\epsfig{figure=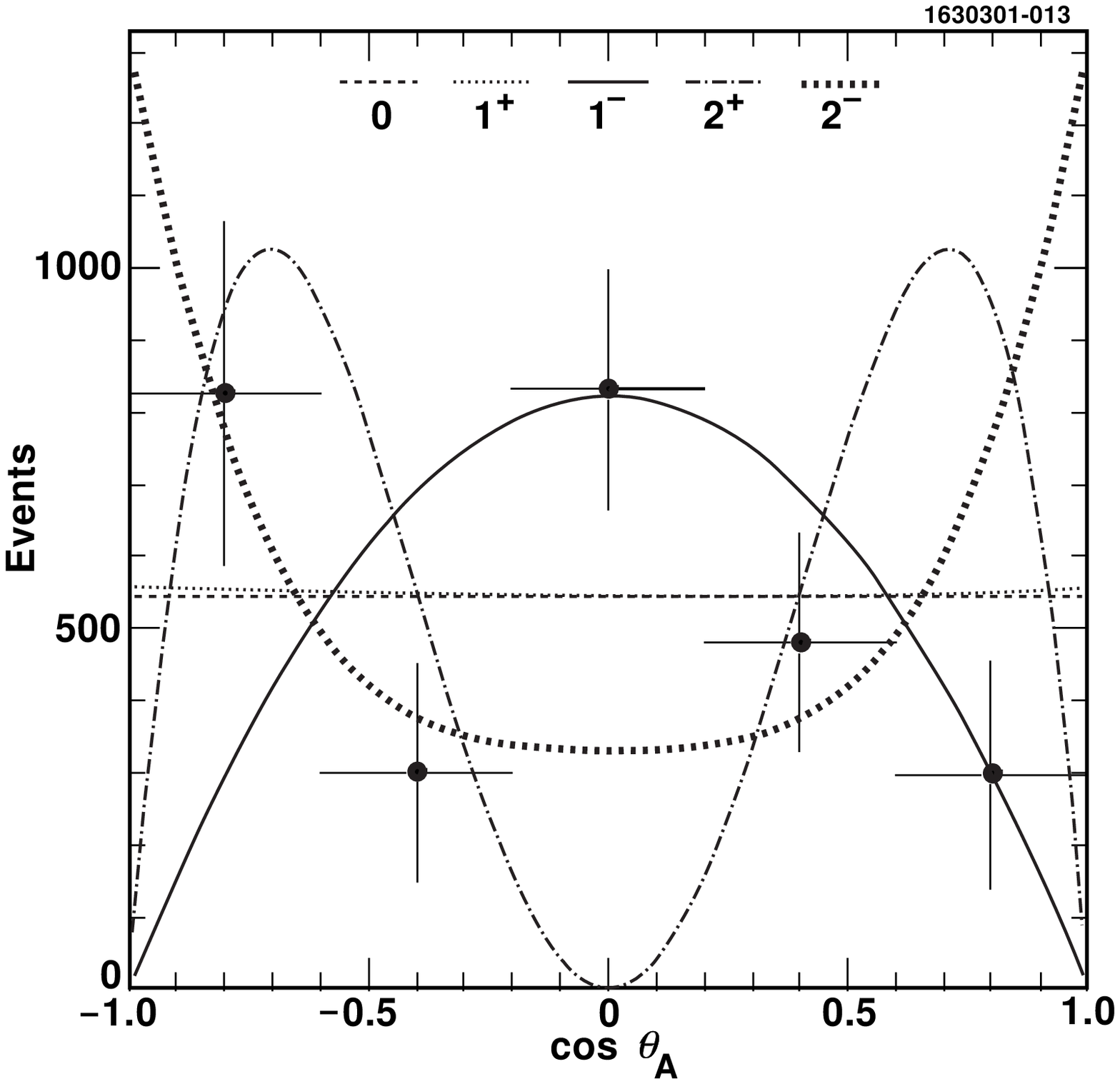,height=1.9in}
\epsfig{figure=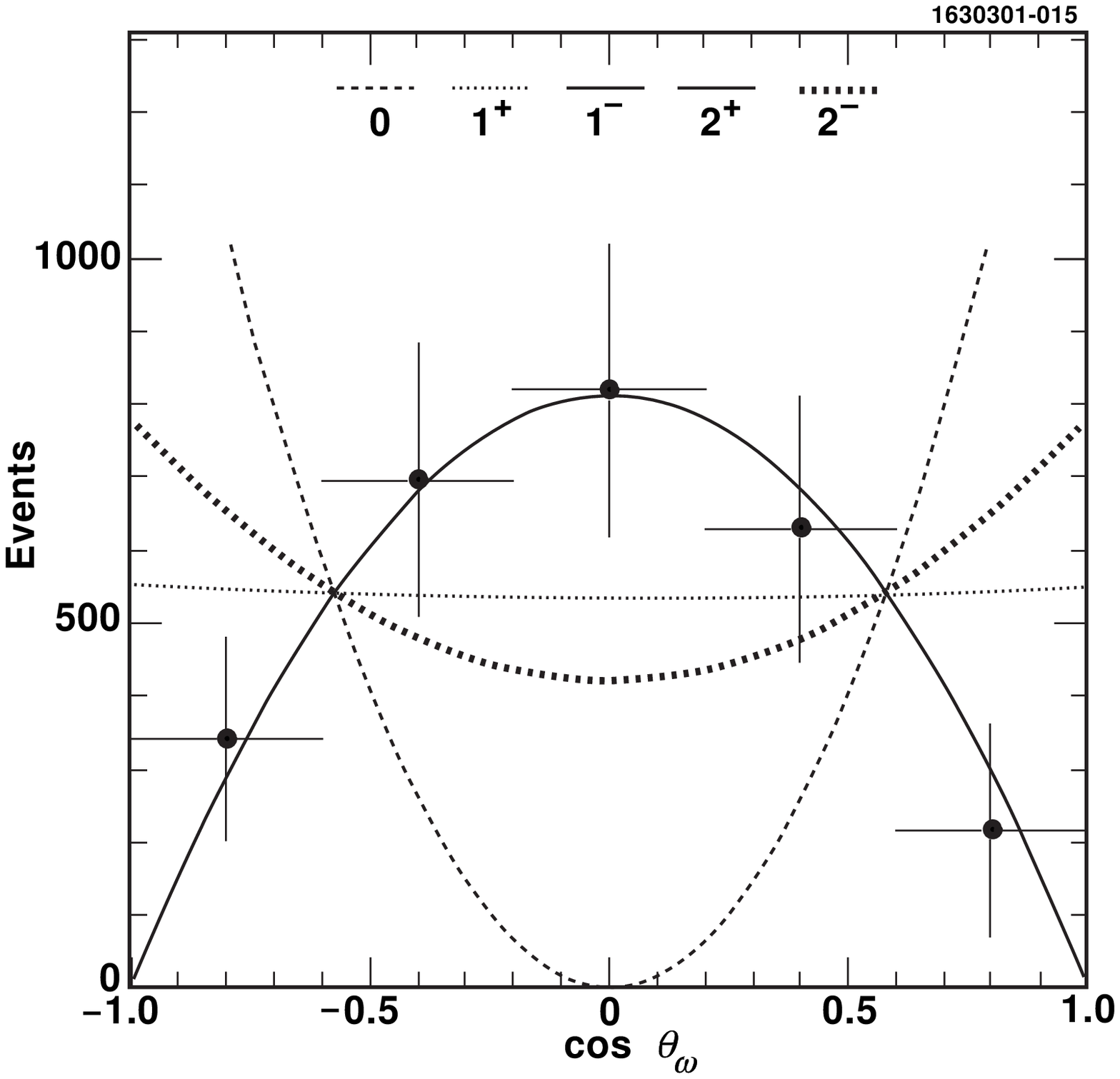,height=1.9in}
\epsfig{figure=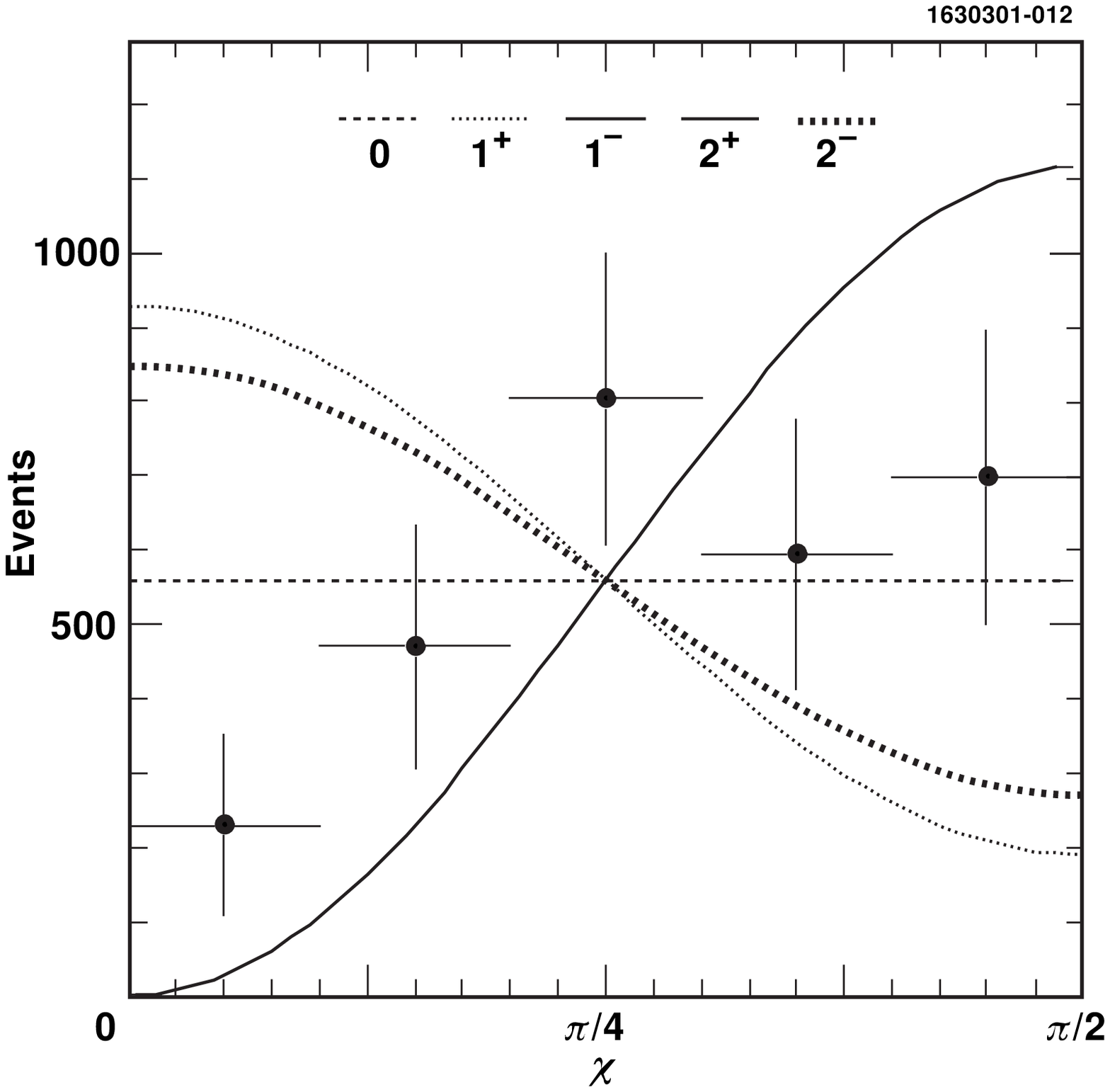,height=1.9in}
}
\vspace{.2cm}  
\caption{\label{helicity_angles}The angular distribution of $\theta_A$ 
(top-left),$\theta_\omega$ (top-right) and $\chi$ (bottom).
The curves show the best fits to the data for for different $J^P$ assignments. 
The $0^-$ and $1^+$ are almost indistinguishable in $\cos\theta_A$, while the 
$1^-$ and $2^+$ are indistinguishable in $\cos\theta_{\omega}$ and $\chi$.
The vertical axis gives efficiency corrected events, 104 events are used.}
\end{figure}

In examining the substructure of the four-pions, a clear $\omega$ signal
was observed in the $\pi^+\pi^-\pi^o$ mass as can be seen in Figure~\ref{019},
leading to 
a significant amount of  $D^{(*)}\omega\pi^-$. Furthermore,    
there is a low-mass resonant substructure in the $\omega\pi^-$     
mass. (See Figure~\ref{omega_pi}).
\begin{table} [htb]   
\begin{center}    
\caption{Measured Branching Ratios}    
\label{table:brs}    
\begin{tabular}{lc}\hline\hline    
\raisebox{0pt}[12pt][6pt]{Mode} & $\cal{B}$ (\%) \\\hline    
\raisebox{0pt}[12pt][6pt]{$\overline{B}^o\to D^{*+}\pi^+\pi^-\pi^-\pi^o$} &     
1.72$\pm$0.14$\pm$0.24    \\    
\raisebox{0pt}[12pt][6pt]{$\overline{B}^o\to D^{*+}\omega\pi^-$} & 0.29$\pm$0.03$\pm$0.04     
\\   
\raisebox{0pt}[12pt][6pt]{$\overline{B}^o\to D^{+}\omega\pi^-$} & 0.28$\pm$0.05$\pm$0.03       
  \\   
\raisebox{0pt}[12pt][6pt]{${B}^-\to D^{*o}\pi^+\pi^-\pi^-\pi^o$}&1.80$\pm$0.24$\pm$0.25           
   \\ 
\raisebox{0pt}[12pt][6pt]{${B}^-\to D^{*o}\omega\pi^-$}  & 0.45$\pm$0.10$\pm$0.07                 
\\    
\raisebox{0pt}[12pt][6pt]{${B}^-\to D^{o}\omega\pi^-$}  &0.41$\pm$0.07$\pm$0.04        
     \\    
\hline\hline    
\end{tabular}    
\end{center}    
\end{table}    
The spin and parity of the $\omega\pi^-$ resonance (denoted by $A$ temporarily)
is determined by considering the decay sequence 
$B\rightarrow A\ D$; $A\rightarrow\omega \pi$ and 
$\omega\rightarrow\pi^+\pi^-\pi^o$. The angular distributions are shown in 
Figure~\ref{helicity_angles}. Here $\theta_A$ is the angle between the 
$\omega$ direction in the $A$ rest frame and the $A$ direction in the $B$ rest
frame; $\theta _{\omega}$ is the orientation of the $\omega$ decay plane in 
the $\omega$ rest frame, and $\chi$ is the angle between the $A$ and $\omega$ 
decay planes.

The data are fit to the expectations for the various $J^P$ assignments. 
The $\omega$ polarization is very clearly transverse ($\sin^2\theta_{\omega}$)
and that infers a $1^-$ or $2^+$ assignment. The $\cos\theta_A$ distribution
prefers $1^-$, as does the fit to all three projections.  

This structure is identified with the 
$\rho'$ because it has the correct $J^P$ and is at approximately the right
mass. To determine the mass and width parameters, that are not well known,
we write the decay width as a function of $\omega\pi^-$ mass as
\begin{eqnarray}
d\Gamma(B\to D\omega\pi^-)&=&{1\over 2M_B}\left|A(B\to D\rho')\cdot
{\cal BW}(m_{\omega\pi})\cdot A(\rho'\to \omega\pi^-)\right|^2 \\
& &\times dP(B\to D\rho')\cdot dP(\rho'\to \omega\pi^-) 
{{dm^2_{\omega\pi}}\over 2\pi}~~, \nonumber 
\end{eqnarray}
where $dP$ indicates phase space and the Breit-Wigner amplitude is given by
\begin{equation}
{\cal BW}(m_{\omega\pi})={1\over {(m_{\omega\pi}^2-m_{\rho'}^2)
-im_{\omega\pi}\Gamma_{tot}(m_{\omega\pi})}}~~.
\end{equation}

The Breit-Wigner fit assuming a single resonance and no background gives a 
mass of 1349$\pm$25$^{+10}_{-5}$ MeV with an intrinsic width of 
547$\pm$86$^{+46}_{-45}$ MeV.
The fit shows that the $\omega\pi^-$ mass spectrum is consistent with being
entirely one resonance.
This state is likely to be the elusive $\rho'$ resonance (Clegg 1994).
These are by far the most accurate and least model dependent measurements of 
the $\rho'$ parameters. The $\rho'$ dominates the final state. 
(Thus the branching ratios for the $D^{(*)}\omega\pi^-$ apply also for 
$D^{(*)}\rho'^-$.)

Heavy quark symmetry predicts equal partial widths for $D^*\rho'$ and$D\rho'$. 
CLEO measures the relative rates to be 
$ {{\Gamma\left({B}\to D^{*}\rho'^-\right)}/    
{\Gamma\left({B}\to D\rho'^-\right)}} = 1.06 \pm 0.17 \pm 0.04$,       
consistent within the relatively large errors.     
    
\begin{figure}
\centerline{\epsfig{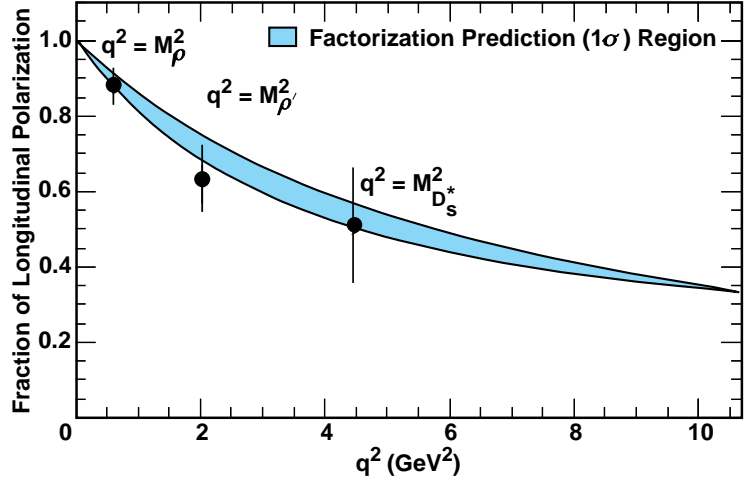}}
\caption{Measured $D^{*+}$ polarization versus semileptonic model predictions.}
\label{polar_bw}
\end{figure}

Factorization predicts that the fraction of longitudinal polarization of the 
$D^{*+}$ is the same as in the related semileptonic decay 
$B\to D^*\ell^-\bar{\nu}$ at four-momentum transfer $q^2$ equal to the 
mass-squared of the $\rho'$    
\begin{equation}    
{{\Gamma_L\left({B}\to D^{*+}\rho'^-\right)}\over     
{\Gamma\left({B}\to D^{*+}\rho'^-\right)}} =     
{{\Gamma_L\left({B}\to D^{*}\ell^-\bar{\nu}\right)}\over     
{\Gamma\left({B}\to D^{*}\ell^-    
\bar{\nu}\right)}}\left|_{q^2=m^2_{\rho'}}\right.~~.    
\end{equation}    
CLEO's measurement of the $D^{*+}$ polarization is (63$\pm$9)\%.     
The model predictions in semileptonic decays for a $q^2$ of 2 GeV$^2$, 
are between 66.9 and 72.6\% (Isgur 1990) (Wirbel 1985) (Neubert 1991).
Figure~\ref{polar_bw} shows the 
measured polarizations for the $D^{*+}\rho'^-$, the $D^{*+}\rho^-$,
(Artuso 1999) and the $D^{*+}D_s^{*-}$ final states (Stone 2000). The latter is 
based on a new measurement using partial reconstruction of the $D^{*+}$
(Ahmed 2000). Thus this prediction of factorization is satisfied.     

\section{CP Violation}
\subsection{Introduction}
\label{section:CPV}
Recall that the operation of Charge Conjugation (C) takes particle to
anti-particle and Parity (P) takes a vector $\overrightarrow{r}$ to
$-\overrightarrow{r}$. CP violation can occur because of the imaginary term in
the CKM matrix, proportional to $\eta$ in the Wolfenstein representation (Bigi
2000).

Consider the case of a process $B\to f$ that goes via two amplitudes \cal{A}
and \cal{B} each of which has a strong part e. g. $s_{\cal{A}}$ and a weak part
$w_{\cal{A}}$. Then we have
\begin{eqnarray}
\Gamma(B\to f)&=&\left(\left|{\cal{A}}\right|e^{i(s_{\cal{A}}+w_{\cal{A}})}
+\left|{\cal{B}}\right|e^{i(s_{\cal{B}}+w_{\cal{B}})}\right)^2 \\
\Gamma(\overline{B}\to \overline{f})&=&
\left(\left|{\cal{A}}\right|e^{i(s_{\cal{A}}-w_{\cal{A}})}
+\left|{\cal{B}}\right|e^{i(s_{\cal{B}}-w_{\cal{B}})}\right)^2 \\
\Gamma(B\to f)-\Gamma(\overline{B}\to \overline{f})&=&2\left|{\cal{AB}}\right|
\sin(s_{\cal{A}}-s_{\cal{B}})\sin(w_{\cal{A}}-w_{\cal{B}})~~.
\end{eqnarray}

Any two amplitudes will do, though its better that they be of approximately
equal size. Thus charged $B$ decays can exhibit CP violation as well as
neutral $B$ decays. In some cases, we will see that
it is possible to guarantee that $\left|\sin(s_{\cal{A}}-s_{\cal{B}})\right|$
is unity, so we can get information on the weak phases. 
In the case of neutral $B$ decays, mixing can be the second amplitude.

\subsection{Unitarity Triangles}
The unitarity of the CKM matrix\footnote{Unitarity implies that any pair of 
rows or 
columns are orthogonal.} allows us to construct six relationships. The most useful 
turns out to be
\begin{equation}
V_{ud}V_{td}^*+V_{us}V_{ts}^*+V_{ub}V_{tb}^*=0~~.
\end{equation}
To a good approximation
\begin{equation}
V_{ud} \approx V_{tb}^*\approx 1 {\rm ~~ and~~}V_{ts}^*\approx -V_{cb},
\end{equation}
then
\begin{equation}
{V_{ub}\over V_{cb}} + {V_{td}^*\over V_{cb}} - V_{us} = 0~~.
\end{equation}
Since $V_{us}=\lambda$, we can define a triangle with sides
\begin{eqnarray}
1 & & \\ 
\left|{V_{td}\over A\lambda^3 }\right| &=&{1\over \lambda}\sqrt{\left(\rho-
1\right)^2+\eta^2}
={1\over \lambda} \left|{V_{td}\over V_{ts}}\right|\\
\left|{V_{ub}\over A\lambda^3}\right|  &=&{1\over \lambda}\sqrt{\rho^2+\eta^2}
={1\over \lambda} \left|{V_{ub}\over V_{cb}}\right|.
\end{eqnarray}

\begin{figure}[hbtp]
\vspace{-6mm}
\centerline{\epsfig{figure=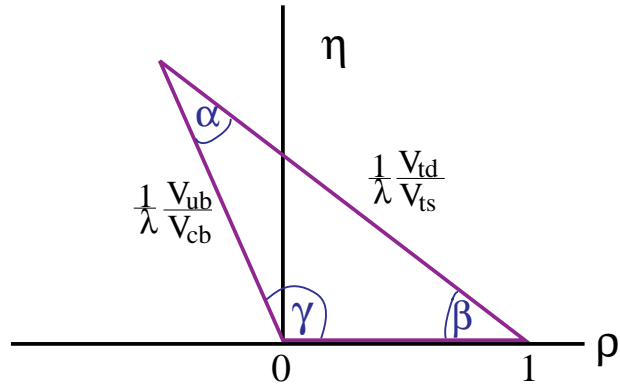,width=3.5in}}
\vspace{-4mm}
\caption{The unitarity triangle shown in the $\rho-\eta$ plane. The
left side is determined by measurements of $b\to u/b\to c$ and the right side can be
determined using mixing measurements in the $B_s$  and $B_d$ systems. The 
angles can be found by making measurements of CP violating asymmetries in
hadronic $B$ decays.
\label{ckm_tri}}
\end{figure}

The CKM triangle is depicted in Figure~\ref{ckm_tri}. 
We know the length of two sides already:
the base is  defined as unity and the left side is determined by the
measurements of  $|V_{ub}/V_{cb}|$, but the error is still quite
substantial. The right side can be determined using
mixing measurements in the neutral $B$ systems. 
 Figure~\ref{ckm_tri} also shows the angles
as $\alpha,~\beta$, and $\gamma$. These angles can be determined by measuring
CP violation in the $B$ system. 

\begin{figure}[bht]
\centerline{\epsfig{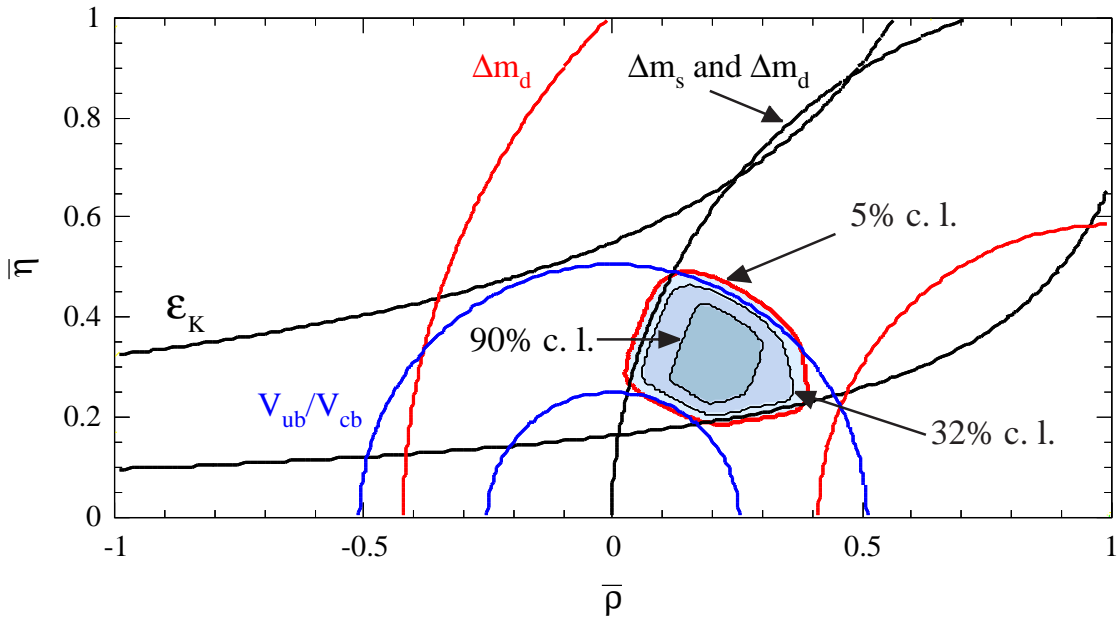}}
\caption{\label{ckm_fig}The regions in $\overline{\rho}-\overline{\eta}$ space
(shaded), where $\overline{\rho}=\rho(1-\lambda^2/2)$ and 
$\overline{\eta}=\eta(1-\lambda^2/2)$, consistent
with measurements of CP violation in $K_L^o$ decay ($\epsilon$), $V_{ub}/V_{cb}$
in semileptonic $B$ decay, $B_d^o$ mixing, and the excluded region from
limits on $B_s^o$ mixing. The allowed region is defined by a fit from
(Hocker 2001). The large width of the $B_d$ mixing band is
dominated by the uncertainty in  $B_Bf_B^2$. The lines that are not
specified are at 5\% confidence level.} 
\end{figure}

Another constraint on $\rho$ and $\eta$ is given by the $K_L^o$ CP
violation  measurement ($\epsilon$) (Buras 1995):
\begin{equation}
\eta\left[(1-\rho)A^2(1.4\pm 0.2)+0.35\right]A^2{B_K \over 0.75}=(0.30\pm 0.06),
\end{equation}
where $B_K$ is parameter that cannot be measured and thus must be calculated.
 A reasonable range is $0.9>B_k>0.6$, given by an assortment
of theoretical calculations (Buras 1995); this number is one of the largest sources of
uncertainty. Other constraints come from current measurements
on $V_{ub}/V_{cb}$, $B_d$  mixing  and $B_s$ mixing.  
The current status of constraints on $\rho$ and $\eta$ is shown in 
Figure~\ref{ckm_fig} (Hocker 2001). The width of both of these
bands are also dominated by theoretical errors.  This shows that the data are
consistent with the standard model.

It is crucial to check if measurements of the sides and angles are consistent,
i.e., whether or not they actually form a triangle. The standard model is
incomplete. It has many parameters including the four CKM numbers, six quark
masses, gauge boson masses and coupling constants. Perhaps measurements of the
angles and sides of the unitarity triangle will bring us beyond the standard
model and show us how these paramenters are related, or what is missing.

Furthermore, new physics can also be observed by measuring branching
ratios which violate standard model predictions. Especially important are
 ``rare decay," processes such as $B\to K\mu^+\mu^-$ or $D\to \pi\mu^+\mu^-$.
 These processes occur only through loops, and are an important class of Penguin decays.

\subsection{Formalism of CP Violation in Neutral $B$ Decays}
Consider the operations of Charge Conjugation, C,  and Parity, P:
\begin{eqnarray}
&C|B(\overrightarrow{p})\big>=|\overline{B}(\overrightarrow{p})\big>,~~~~~~~~~
&C|\overline{B}(\overrightarrow{p})\big>=|{B}(\overrightarrow{p})\big> \\
&P|B(\overrightarrow{p})\big>=-|{B}(-\overrightarrow{p})\big>,~~~~~
&P|\overline{B}(\overrightarrow{p})\big>=-|\overline{B}(-\overrightarrow{p})\big> \\
&CP|B(\overrightarrow{p})\big>=-|\overline{B}(-\overrightarrow{p})\big>,~~~
&CP|\overline{B}(\overrightarrow{p})\big>=-|{B}(-\overrightarrow{p})\big> ~~.
\end{eqnarray}
For neutral mesons we can construct the CP eigenstates
\begin{eqnarray}
\big|B^o_1\big>&=&{1\over \sqrt{2}}\left(\big|B^o\big>-
\big|\overline{B}^o\big>\right)~~,\\
\big|B^o_2\big>&=&{1\over 
\sqrt{2}}\left(\big|B^o\big>+\big|\overline{B}^o\big>\right)~~,
\end{eqnarray}
where 
\begin{eqnarray}
CP\big|B^o_1\big>&=&\big|B^o_1\big>~~, \\
CP\big|B^o_2\big>&=&-\big|B^o_2\big>~~.
\end{eqnarray}
Since $B^o$ and $\overline{B}^o$ can mix, the mass eigenstates are a superposition of
$a\big|B^o\big> + b\big|\overline{B}^o\big>$ which obey the Schrodinger equation
\begin{equation}
i{d\over dt}\left(\begin{array}{c}a\\b\end{array}\right)=
{\cal H}\left(\begin{array}{c}a\\b\end{array}\right)=
\left(M-{i\over 2}\Gamma\right)\left(\begin{array}{c}a\\b\end{array}\right).
\label{eq:schrod}
\end{equation}
If CP is not conserved then the eigenvectors, the mass eigenstates $\big|B_L\big>  $ 
and  $\big|B_H\big>$, are not the CP eigenstates but are 
\begin{equation}
\big|B_L\big> = p\big|B^o\big>+q\big|\overline{B}^o\big>,~~\big|B_H\big> = 
p\big|B^o\big>-q\big|\overline{B}^o\big>,
\end{equation}
where
\begin{equation}
p={1\over \sqrt{2}}{{1+\epsilon_B}\over {\sqrt{1+|\epsilon_B|^2}}},~~
q={1\over \sqrt{2}}{{1-\epsilon_B}\over {\sqrt{1+|\epsilon_B|^2}}}.
\end{equation}
CP is violated if $\epsilon_B\neq 0$, which occurs if $|q/p|\neq 1$.

The time dependence of the mass eigenstates is 
\begin{eqnarray}
\big|B_L(t)\big> &= &e^{-\Gamma_Lt/2}e^{im_Lt/2} \big|B_L(0)\big> \\
 \big|B_H(t)\big> &= &e^{-\Gamma_Ht/2}e^{im_Ht/2} \big|B_H(0)\big>,
\end{eqnarray}
leading to the time evolution of the flavor eigenstates as
\begin{eqnarray}
\big|B^o(t)\big>&=&e^{-\left(im+{\Gamma\over 2}\right)t}
\left(\cos{\Delta mt\over 2}\big|B^o(0)\big>+i{q\over p}\sin{\Delta mt\over 
2}\big|\overline{B}^o(0)\big>\right) \\
\big|\overline{B}^o(t)\big>&=&e^{-\left(im+{\Gamma\over 2}\right)t}
\left(i{p\over q}\sin{\Delta mt\over 2}\big|B^o(0)\big>+
\cos{\Delta mt\over 2}\big|\overline{B}^o(0)\big>\right),
\end{eqnarray}
where $m=(m_L+m_H)/2$, $\Delta m=m_H-m_L$ and 
$\Gamma=\Gamma_L\approx \Gamma_H$, and $t$ is the decay time in the $B^o$
rest frame, the so called proper time. Note that the probability of a 
$B^o$ decay as a function of $t$ is given by $\big<B^o(t)\big|B^o(t)\big>^*$, and is 
a pure exponential, $e^{-\Gamma t/2}$, in the absence of CP violation.

\subsubsection{CP violation for $B$ via interference of mixing and decays}

\begin{figure}[htb]
\centerline{\epsfig{figure=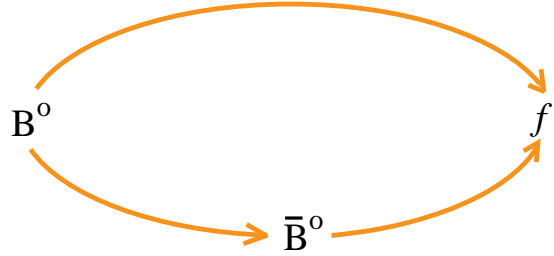,height=1.5in}}
\caption{\label{eigen_CP}Two interfering ways for a $B^o$ to decay into a final
state $f$.} 
\end{figure}

Here we choose a final state $f$ which is accessible to both $B^o$ and
$\overline{B}^o$  decays (Bigi 2000). The second amplitude necessary for
interference is provided by mixing.  Figure~\ref{eigen_CP} shows the decay into
$f$ either directly or indirectly via  mixing. 
It is  necessary only that $f$ be
accessible directly from either state; however if $f$ is a CP  eigenstate the
situation is far simpler. For  CP eigenstates \begin{equation}
CP\big|f_{CP}\big>=\pm\big|f_{CP}\big>. \end{equation}

It is useful to define the amplitudes
\begin{equation}
A=\big<f_{CP}\big|{\cal H}\big|B^o\big>,~~
\overline{A}=\big<f_{CP}\big|{\cal H}\big|\overline{B}^o\big>.
\end{equation}
If $\left|{\overline{A}\over A}\right|\neq 1$, then we have ``direct" CP violation in the 
decay 
amplitude, which we will discuss in detail later. Here CP can be violated by having
\begin{equation}
\lambda = {q\over p}\cdot {\overline{A}\over A}\neq 1,
\end{equation}
which requires only that $\lambda$  acquire a non-zero phase, i.e. $|\lambda|$ 
could be 
unity and CP violation can occur. 

A comment on neutral $B$ production at $e^+e^-$ colliders is in order. At the 
$\Upsilon 
(4S)$ resonance there is coherent production of $B^o\bar{B}^o$ pairs. This puts the 
$B$'s in a $C=-1$ state. In hadron colliders, or at $e^+e^-$ machines operating 
at the $Z^o$, the $B$'s are produced incoherently. For the rest
of this article I will 
assume incoherent production except where explicitly noted.

The asymmetry, in this case, is defined as
\begin{equation}
a_{f_{CP}}={{\Gamma\left(B^o(t)\to f_{CP}\right)- 
\Gamma\left(\overline{B}^o(t)\to 
f_{CP}\right)}\over
{\Gamma\left(B^o(t)\to f_{CP}\right)+ \Gamma\left(\overline{B}^o(t)\to 
f_{CP}\right)}},
\end{equation}
which for $|q/p|=1$ gives
\begin{equation}
a_{f_{CP}}={{\left(1-|\lambda|^2\right)\cos\left(\Delta mt\right)-2{\rm Im}\lambda 
\sin(\Delta mt)}\over {1+|\lambda|^2}}.
\end{equation}
For the cases where there is only one decay amplitude $A$, $|\lambda |$ equals 1, 
and we have
\begin{equation}
a_{f_{CP}}=-{\rm Im}\lambda \sin(\Delta mt).
\end{equation}
Only the amplitude, ${\rm -Im}\lambda$ contains information about the level of CP 
violation, 
the sine term is determined only by $B_d$ mixing. In fact, the time integrated 
asymmetry is given by
\begin{equation}
a_{f_{CP}}=-{x \over {1+x^2}}{\rm Im}\lambda = -0.48 {\rm Im}\lambda ~~. \label{eq:aint}
\end{equation}
This is quite lucky as the maximum size of the coefficient for any $x$ is $-0.5$.

Let us now find out how ${\rm Im}\lambda$ relates to the CKM parameters. Recall 
$\lambda={q\over p}\cdot {\overline{A}\over A}$. The first term is the part that comes 
from mixing:
\begin{equation}
{q\over p}={{\left(V_{tb}^*V_{td}\right)^2}\over {\left|V_{tb}V_{td}\right|^2}}
={{\left(1-\rho-i\eta\right)^2}\over {\left(1-\rho+i\eta\right)\left(1-\rho-
i\eta\right)}}
=e^{-2i\beta}{\rm~~and}
\end{equation}
\begin{equation}
{\rm Im}{q\over p}= -{{2(1-\rho)\eta}\over {\left(1-\rho\right)^2+\eta^2}}=\sin(2\beta).
\end{equation}

To evaluate the decay part we need to consider specific final states. For example, 
consider 
$f\equiv\pi^+\pi^-$. The simple spectator decay diagram is shown in 
Figure~\ref{pippim} (left).
For the moment we will assume that this is the only diagram which contributes,
which we know is 
not true. For this $b\to u\bar{u}d$ process we have
\begin{equation}
{\overline{A}\over A}={{\left(V_{ud}^*V_{ub}\right)^2}\over 
{\left|V_{ud}V_{ub}\right|^2}}={{(\rho-i\eta)^2}\over
 {(\rho-i\eta)(\rho+i\eta)}}=e^{-2i\gamma},
\end{equation}
and 
\begin{equation}
{\rm Im}(\lambda)={\rm Im}(e^{-2i\beta}e^{-2i\gamma})=
{\rm Im}(e^{2i\alpha})=-\sin(2\alpha).
\end{equation}

For our next example let's consider the final state $J/\psi K_s$. The decay diagram is 
shown in Figure~\ref{psi_ks}. In this case we do not get a phase from the decay part because
\begin{equation}
{\overline{A}\over A} = {{\left(V_{cb}V_{cs}^*\right)^2}\over 
{\left|V_{cb}V_{cs}\right|^2}}
\end{equation}
is real to order $1/\lambda^4$.

In this case the final state is a state of negative $CP$, i.e. 
$CP\big|J/\psi K_s\big>=-\big|J/\psi K_s\big>$. This introduces an additional
minus sign in the result for ${\rm Im}\lambda$.
Before finishing discussion of this final state we need to consider in more detail 
the presence of the $K_s$ in the final state. Since neutral kaons can mix, we pick 
up another mixing phase (similar diagrams as for $B^o$, see Figure~\ref{bmix}). 
This term creates a phase given by
\begin{equation}
\left({q\over p}\right)_K={{\left(V_{cd}^*V_{cs}\right)^2}\over 
{\left|V_{cd}V_{cs}\right|^2}},
\end{equation}
which is real to order $\lambda^4$. It necessary to include this term, however,
since there are other  formulations of the CKM matrix than Wolfenstein, which
have the phase in a  different location. It is important that the physics
predictions not depend on the  CKM convention.\footnote{Here we don't include
CP violation in the neutral kaon  since it is much smaller than what is
expected in the $B$ decay. The term of order $\lambda^4$ in $V_{cs}$ is
necessary to explain $K^o$ CP violation.}

In summary, for the case of $f=J/\psi K_s$, ${\rm Im}\lambda=-\sin(2\beta)$.
\subsubsection{Comment on Penguin Amplitude}

In principle all processes can have penguin components. One such diagram is
shown  in Figure~\ref{pippim}(right). The $\pi^+\pi^-$ final state is expected to
have a rather large penguin  amplitude $\sim$10\% of the tree amplitude. Then
$|\lambda |\neq 1$ and  $a_{\pi\pi}(t)$ develops a $\cos(\Delta mt)$ term. It
turns out that $\sin(2\alpha)$ can be extracted using isospin considerations
and  measurements of the branching ratios for $B^+\to \pi^+\pi^o$ and $B^o\to 
\pi^o\pi^o$, or other methods the easiest of which appears to be the
study of $B^o\to\rho\pi$.

In the $J/\psi K_s$ case, the penguin amplitude is expected to be small since a 
$c\bar{c}$ pair must be ``popped" from the vacuum. Even if the penguin decay 
amplitude were of significant size, the decay phase is the same as the tree level 
process, namely zero.

\subsection{Results on {\boldmath $\sin{2\beta}$}}

For years observation of large CP violation in the $B$ system was considered to
be one of the corner stone predictions of the Standard Model. Yet it took
a very long time to come up with definitive evidence. 
The first statistically significant measurements of CP violation in the $B$
system were made recently by BABAR and BELLE. This enormous achievement
was accomplished using an asymmetric $e^+e^-$ collider on the $\Upsilon (4S)$
which was first suggested by Pierre Oddone. The measurements are listed in
Table~\ref{tab:s2b}, along with other previous indications (Groom 2001).

\begin{table}[htb]
\vspace{-2mm}
\begin{center}
\caption{Measurements of $\sin 2\beta$.
\label{tab:s2b}}
\vspace*{2mm}
\begin{tabular}{lc}\hline\hline
Experiment & $\sin 2\beta$ \\
\hline
BABAR & $0.59\pm 0.14 \pm 0.05 $ \\
BELLE & $0.99\pm 0.14 \pm 0.06 $ \\\hline
Average & 0.79$\pm$0.11 \\\hline
CDF & 0.79$^{+.41}_{-0.44}$ \\
ALEPH & 0.84$^{+0.82}_{-1.04}\pm 0.16$\\
OPAL & 3.2$^{+1.8}_{-2.0}\pm 0.5$\\
\hline\hline
\end{tabular}
\end{center}
\end{table}

\begin{figure}[htb]
\centerline{\epsfig{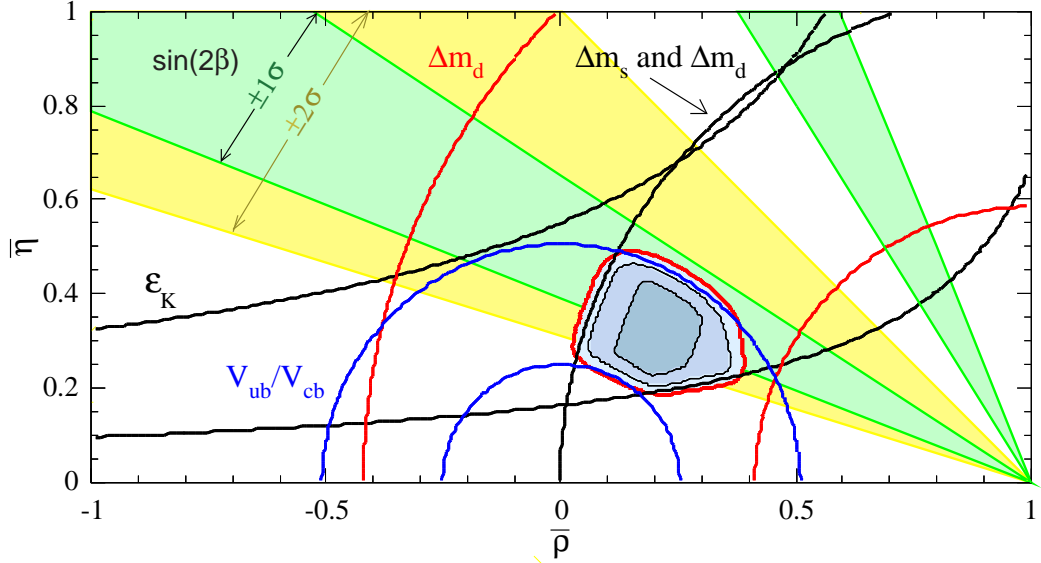}}
\caption{\label{rhoeta+ss_2beta} Constraints from $\sin2\beta$ measurement
overlaid with other constraints (Hocker 2001). The inner band is at 1$\sigma$
while the outer band, shown on one band only, is at $2\sigma$.}
\end{figure}

The average value of $0.79\pm 0.11$ is taken from BABAR and BELLE only. These
two measurements do differ by a sizeable amount, but the confidence level
that they correctly represent the same value is 6\%.
This value is consistent with what is expected from the other known constraints
on $\rho$ and $\eta$. We have
\begin{equation}
\overline{\eta}=(1-\overline{\rho}){{1\pm\sqrt{1-\sin^2{2\beta}}}\over
\sin{2\beta}}~~.
\end{equation}
There is a four fold ambiguity in the translation between $\sin{2\beta}$
and the linear constraints in the $\rho-\eta$ plane. These occur
at $\beta$, $\pi/2-\beta$, $\pi+\beta$ and $3\pi/2-\beta.$ Two of these constraints
are shown in Figure~\ref{rhoeta+ss_2beta}. The other two can be viewed by
extending these to negative $\overline{\eta}$. We think $\eta >0$ based only
on measurement of $\epsilon'$ in the neutral kaon system.

\subsection{Remarks on Global Fits to CKM parameters}
The fits shown in this paper (Hocker 2001) for $\rho$ and $\eta$ have been done
by others with a somewhat different statistical framework (Ciuchini 2001)
(Mele 1999). The
latter group uses ``Bayesian" statistics which means that they use apriori
knowledge of the probability distribution functions. The former are termed
``frequentists" (Groom 2001), almost for the lack of a better term. The
frequentists are more conservative than the Bayesians. 

The crux of the issue is
how to treat theoretically predicted parameters that are used to translate
measurements into quantities such as $V_{ub}$ or $\epsilon_K$ that form
constraints in the
$\rho-\eta$ plane. This of course is a problem because it is difficult to
estimate the uncertainties in the theoretical predictions. Both groups treat
the experimental measurements as Gaussian distributions with the $1\sigma$
width derived from both the statistical and systematic errors combined. Note,
that the systematic errors are also difficult sometimes to quantify and are not
necessarily Gaussian, but they
judged to be sufficiently well known as to not cause a problem. 

Hocker et al. (Hocker 2001) have decided to use a method which restricts the theoretical
quantities to a 95\% confidence interval where the parameter in question is
just as likely to lie anywhere in the interval. They call this the ``Rfit"
method. Of course assigning the 95\% confidence interval is a matter of
judgment which they fully admit. Ciuchini et al. (Ciuchini 2001) treat the theoretical errors
in the same manner as the experimental errors. They call theirs ``the standard
method" with just a bit of hubris. They argue that QCD is mature enough to
trust its predictions, that they know the sign and rough magnitude of corrections and
they can assign reasonable errors, so it would be wrong to throw away
information.

Hocker et al. point out an extreme interesting but generally unknown danger with
the Bayesian approach, which is that in multi-dimension problems the Bayesian
treatment unfairly predicts a narrowing of possible results. The following
discussion will demonstrate this.

Let $x_i$ denote N theoretical parameters over the identical ranges $[-\Delta,
+\Delta]$; then the theoretical prediction is 
\begin{equation}
T_P^{(N)}=\prod_i^N x_i~~.
\end{equation}
In the
95\% scan scheme $[T^{(N)}_P]=[-\Delta^N,+\Delta^N]$ while in the Bayesian
approach the convoluted Probability Density Function (PDF) is
\begin{equation}
\rho(T)=\int_{-\infty}^{+\infty}...\int_{-\infty}^{+\infty}
\prod_i^N dx_iG(x_i)\delta(T-T_P^{(N)})~~,
\end{equation}
where the $G(x_i)$ are  PDF's for each individual variable taken to be equal here. This function
has a singularity in $\rho(T)$ that goes as $(-ln T)^{N-1}$, so it increases as
$N$ increases. 

Now suppose $G(x_i)$ is flat, then for $N=1$ both approaches are the same, but
for $N\geq 2$, the Bayesian approach gets a $\rho(T)$ that peaks at zero. In
effect, when the number of theoretical predictions entering the computation increases, and hence
our knowledge of the corresponding observable decrease the Bayesian approach
claims the converse. 

\begin{figure}[htb]
\vspace{0.4cm}
\centerline{\epsfig{figure=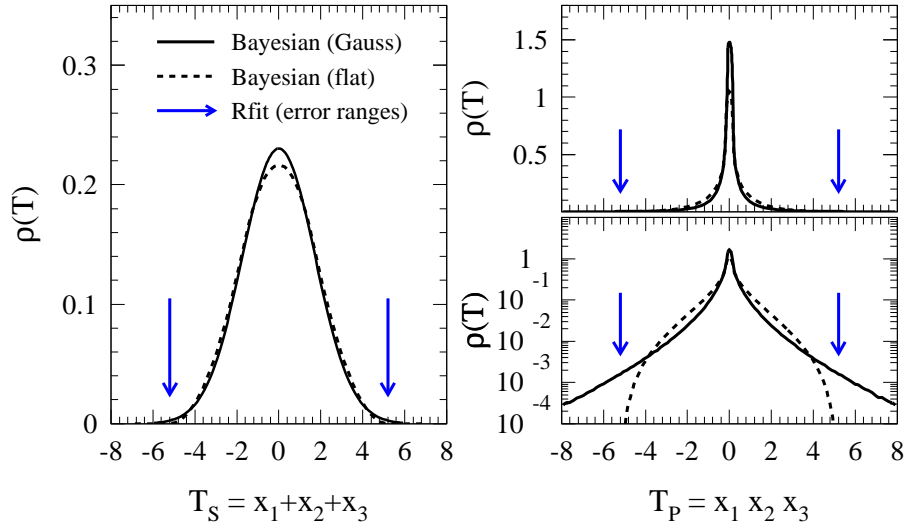,height=3in}}
\caption{\label{bayestest}Convolution of the sum $T_S=x_1+x_2+x_3$ for both
Rfit and Bayesian methods (left)
and the product $T_P=x_1x_2x_3$ (right) of 3 parameters for the Bayesian
method only. Plotted is the PDF
$\rho(T)$ obtained using for $G(x)$ a uniform (solid
lines, $\Delta=\sqrt{3}$) or a Gaussian (dashed lines, $\sigma=1$)
distribution. Both PDF's of $T_P$ have a singularity at the origin which is
not shown. The Rfit ranges of $T_S$ and $T_P$ are indicated by the arrows
located in both instances at $\pm 3\sqrt{3}$. From (Hocker 2001).}
\end{figure}

Lets look at a specific example: here $N=3$ and
$\Delta=\sqrt{3}$. Consider both the sum  $T_S=x_1+x_2+x_3$ and product
$T_P=x_1x_2x_3$ distributions. For Rfit the allowed ranges are identical
being $[-3\Delta,3\Delta]$. The left side of Figure~\ref{bayestest}
shows the probability density $\rho(T)$ for $T_S$, while the right side shows 
$\rho(T)$ for $T_P$ with $G(x)$ in the Bayesian case being either a Gaussian 
with $\sigma=1$ (solid lines) or a uniform distribution over the range 
$[-\sqrt{3},+\sqrt{3}]$ (dashed lines). The later distribution is closest to 
the Rfit method where the allowed range for either $T$ is $[-3\sqrt{3},+3\sqrt{3}]$
indicated by the arrows. 
   
In the Rfit scheme the two predictions for $T_P$ and $T_S$ are identical, while
in the Bayesian scheme there is large difference in
the PDF's and it really doesn't matter if a Gaussian or uniform distribution is
chosen. There is a clear distinction between the Rfit and Bayesian predictions
for $T_P$ in this case, and the Bayesian one is unreasonable because it
produces a very narrow PDF peaked very close to zero.

\begin{figure}[htb]
\centerline{\epsfig{figure=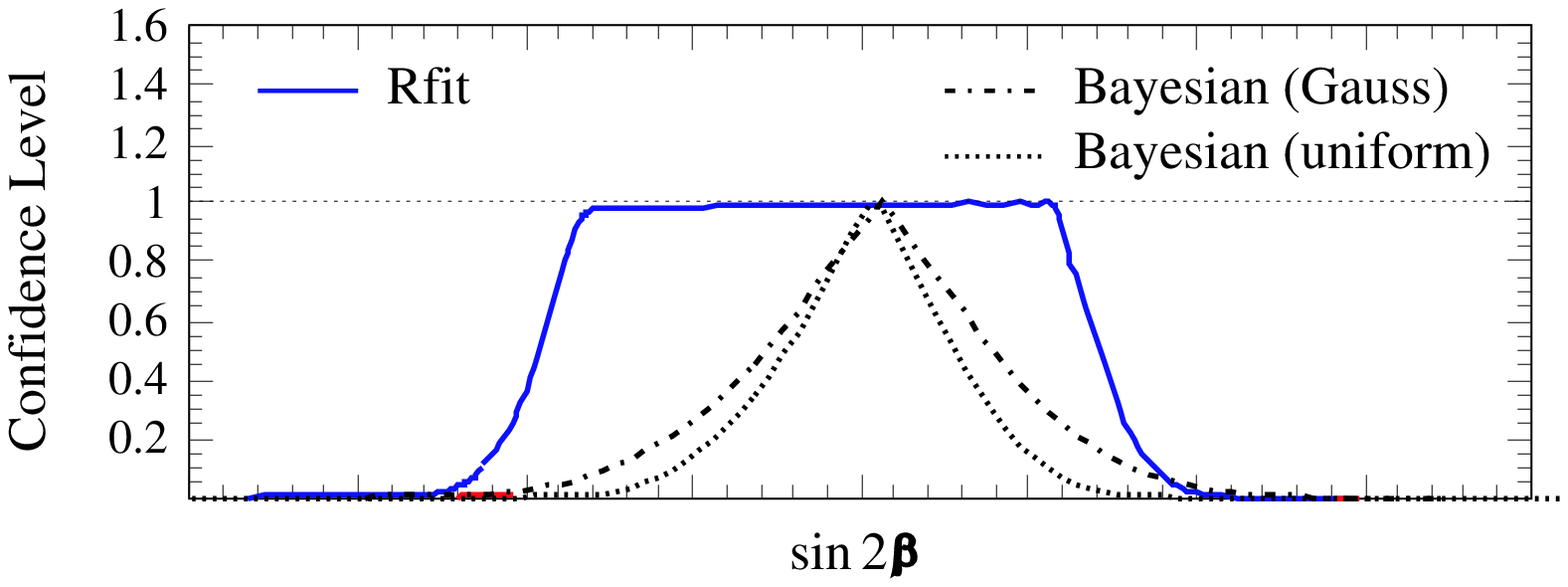,height=2.in}}
\vspace{.4cm}
\caption{\label{comprfitbayes}Comparison between Rfit (broad solid curve) and Bayesian fits for the indirect
CKM constraints on $\sin2\beta$. For the Bayesian fits: Gaussian (uniform)
theoretical PDF's are depicted as dashed-dotted (dotted) curves. (This example
 has been outdated by newer data.)}
\end{figure}

An example of how this can effect the results is shown on
Figure~\ref{comprfitbayes} where predictions of $\sin2\beta$ from the indirect
measurements are shown for the Rfit technique and either uniform or Gaussian
Bayesian PDF's. The predictions are quite different.

\section{New Physics Studies}
\subsection{Introduction}

There are many reasons why we believe that the Standard Model is incomplete and
there must be physics beyond. One is the plethora of ``fundamental parameters,"
for example quark masses, mixing angles, etc... The Standard Model cannot
explain the smallness of the weak scale compared to the GUT or Planck scales;
this is often called ``the hierarchy problem." In the Standard Model it is
believed that the CKM source of CP violation extensively discussed here is not
large enough to explain the baryon asymmetry of the Universe (Gavela 1993);
we can also take the view
that we will discover additional large unexpected effects in $b$ and/or
$c$ decays. Finally, gravity is not incorporated. John Ellis said ``My personal
interest in CP violation is driven by the search for physics beyond the
Standard Model" (Ellis 2000).

We must realize that {\it all} our current measurements are a combination of
Standard Model and New Physics; any proposed models must satisfy current
constraints. Since the Standard Model tree level diagrams are probably large,
lets consider them a background to New Physics. Therefore loop diagrams and CP
violation are the best places to see New Physics. 

The most important current constraints on New Physics models are
\begin{itemize}
\item The neutron electric dipole moment, $d_N~<6.3\times 10^{-26}$e-cm.
\item ${\cal{B}}(b\to s\gamma)=(3.23\pm 0.42)\times 10^{-4}$ and
${\cal{B}}(b\to s\ell^+\ell^-)<4.2\times 10^{-5}$.
\item CP violation in $K_L$ decay, $\epsilon_K =(2.271\pm 0.017)\times
10^{-3}$.
\item $B^o$ mixing parameter $\Delta m_d = (0.487\pm0.014)$ ps$^{-1}$.
\end{itemize}

\subsection{Generic Tests for New Physics}
We can look for New Physics either in the context of specific models or more
generically, for deviations from the Standard Model expectation.

One example is to examine the rare decays $B\to K\ell^+\ell^-$ and
$B\to K^*\ell^+\ell^-$ for branching ratios and polarizations.
According to Greub et al., ``Especially the decay into $K^*$ yields a wealth
of new information on the form of the new interactions since the Dalitz
plot is sensitive to subtle interference effects" (Greub 1995).

Another important tactic is to test for inconsistencies in Standard Model
predictions independent of specific non-standard models.
 
\begin{figure}[htb]
\vspace{0.4cm}
\centerline{\epsfig{figure=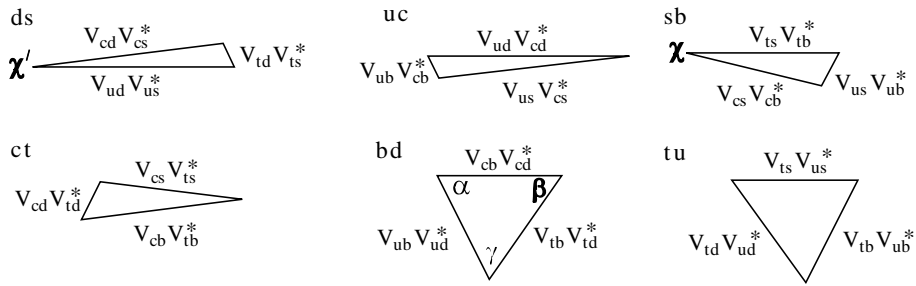,height=1.5in}}
\caption{\label{six_tri}The six CKM triangles. The bold labels, e.g. {\bf ds} 
refer to the rows or columns used in the unitarity relationship. The angles
defined in equation (\ref{eq:chi}) are also shown.}
\end{figure}

The unitarity of the CKM matrix allows us to construct six relationships.
These may be thought of as triangles in the complex plane  
shown in Figure~\ref{six_tri}. (The {\bf bd} triangle is the one depicted in 
Figure~\ref{ckm_tri}.)

All six of these triangles can be constructed knowing four and
only four independent angles (Silva 1997) (Aleksan 1994).
 These 
are defined as:
\begin{eqnarray} \label{eq:chi}
\beta=arg\left(-{V_{tb}V^*_{td}\over V_{cb}V^*_{cd}}\right),&~~~~~&
\gamma=arg\left(-{{V^*_{ub}V_{ud}}\over {V^*_{cb}V_{cd}}}\right), \\
\chi=arg\left(-{V^*_{cs}V_{cb}\over V^*_{ts}V_{tb}}\right),&~~~~~&
\chi'=arg\left(-{{V^*_{ud}V_{us}}\over {V^*_{cd}V_{cs}}}\right).\nonumber\\ 
\end{eqnarray}
($\alpha$ can be used instead of $\gamma$ or $\beta$.) Two of the phases $\beta$ and $\gamma$ are probably large while $\chi$ is
estimated to be small $\approx$0.02, but measurable, while $\chi'$ is likely
to be much smaller.

It has been pointed out by Silva and Wolfenstein (Silva 1997) that
measuring only angles may not be sufficient to detect
new physics. For example, suppose there is new physics that arises in 
$B^o-\overline{B}^o$ mixing. Let us assign a phase $\theta$ to this new
physics. If we then measure CP violation in $B^o\to J/\psi K_s$ and eliminate
any Penguin pollution problems in using $B^o\to\pi^+\pi^-$, then we actually
measure $2\beta' =2\beta + \theta$ and $2\alpha' = 2\alpha -\theta$. So while
there is new physics, we miss it, because
$2\beta' + 2\alpha' = 2\alpha +2\beta$ and $\alpha' + \beta' +\gamma
= 180^{\circ}$.

\subsubsection{A Critical Check Using $\chi$}

The angle $\chi$, defined in equation~\ref{eq:chi}, can be extracted by
measuring the time dependent CP violating asymmetry in the reaction
$B_s\to J/\psi \eta^{(}$$'^{)}$, or if one's detector is incapable of quality
photon detection, the $J/\psi\phi$ final state can be used.  However, in this
case there are
two vector particles in the final state, making this a state of mixed CP,
requiring a time-dependent angular analysis to extract $\chi$, that requires large
statistics.

Measurements of the magnitudes of 
CKM matrix elements all come with theoretical errors. Some of these are hard 
to estimate.
The best measured magnitude is that of $\lambda=|V_{us}/V_{ud}|=0.2205\pm 
0.0018$. 

Silva and 
Wolfenstein (Silva 1997) (Aleksan 1994)
show that the Standard Model 
can be checked in a profound manner by seeing if:
\begin{equation}
\sin\chi = \left|{V_{us}\over 
V_{ud}}\right|^2{{\sin\beta~\sin\gamma}\over{\sin(\beta+\gamma)}}~~.
\end{equation}
Here the precision of the check will be limited initially by the measurement of
$\sin\chi$, not of $\lambda$. This check can  reveal new physics, even 
if other measurements have not shown any anomalies. 

Other relationships to check include:
\begin{equation}
\sin\chi = \left|{V_{ub}\over 
V_{cb}}\right|^2{\sin\gamma~\sin(\beta+\gamma)\over{\sin\beta}}~~,
\end{equation}
\begin{equation}
\sin\chi = \left|{V_{td}\over 
V_{ts}}\right|^2{\sin\beta~\sin(\beta+\gamma)\over{\sin\gamma}}~~.
\end{equation}

The astute reader will have noticed that these two equations lead to the
non-trivial relationship:
\begin{equation}
\sin^2\beta\left|V_{td}\over V_{ts}\right|^2 = \sin^2\gamma\left|{V_{ub}\over 
V_{cb}}\right|^2 ~~.
\end{equation}
This constrains these two magnitudes in terms of two of the angles. 
Note, that it is in principle possible to determine the magnitudes of 
$|V_{ub}/V_{cb}|$ and $|V_{td}/V_{ts}|$ without model dependent errors
by measuring
$\beta$, $\gamma$ and $\chi$ accurately. Alternatively, $\beta$, $\gamma$
 and $\lambda$ can be used to give a much more precise value than is
possible at present with direct methods. For example, once $\beta$ and
$\gamma$ are known
\begin{equation}
\left|{V_{ub}\over V_{cb}}\right|^2 = \lambda^2
{{\sin^2\beta}\over {\sin^2(\beta +\gamma)}}~ ~~.
\end{equation}

Table~\ref{table:reqmeas} lists the most important physics quantities and the
decay modes that can be used to measure them. The necessary detector
capabilities include the ability to collect purely hadronic final states
labeled here as ``Hadron Trigger," the ability to identify
charged hadrons labeled as ``$K\pi$ sep," the ability to detect photons with
good efficiency and
resolution and excellent time resolution required to analyze rapid $B_s$
oscillations. Measurements of $\cos(2\phi)$ can eliminate 2 of the 4
ambiguities in $\phi$ that are present when $\sin(2\phi)$ is measured. 

\begin{table}[hbt]
\begin{center}
\label{table:reqmeas}
\begin{tabular}{|l|l|c|c|c|c|} \hline\hline
Physics & Decay Mode & Hadron & $K\pi$ & $\gamma$ & Decay \\
Quantity&            & Trigger & sep   & det & time $\sigma$ \\
\hline
$\sin(2\alpha)$ & $B^o\to\rho\pi\to\pi^+\pi^-\pi^o$ & $\surd$ & $\surd$& $\surd$ 
&\\
$\cos(2\alpha)$ & $B^o\to\rho\pi\to\pi^+\pi^-\pi^o$ & $\surd$ & $\surd$& 
$\surd$ &\\
sign$(\sin(2\alpha))$ & $B^o\to\rho\pi$ \& $B^o\to\pi^+\pi^-$ & 
$\surd$ & $\surd$ & $\surd$ & \\
$\sin(\gamma)$ & $B_s\to D_s^{\pm}K^{\mp}$ & $\surd$ & $\surd$ & & $\surd$\\
$\sin(\gamma)$ & $B^-\to \overline{D}^{0}K^{-}$ & $\surd$ & $\surd$ & & \\
$\sin(\gamma)$ & $B^o\to\pi^+\pi^-$ \& $B_s\to K^+K^-$ & $\surd$ & $\surd$& & 
$\surd$ \\
$\sin(2\chi)$ & $B_s\to J/\psi\eta',$ $J/\psi\eta$ & & &$\surd$ &$\surd$\\
$\sin(2\beta)$ & $B^o\to J/\psi K_s$ & & & & \\
$\cos(2\beta)$ &  $B^o\to J/\psi K^o$, $K^o\to \pi\ell\nu$  & &$\surd$ & & \\
$\cos(2\beta)$ &  $B^o\to J/\psi K^{*o}$ \& $B_s\to J/\psi\phi$  & & & 
&\\
$x_s$  & $B_s\to D_s^+\pi^-$ & $\surd$ & & &$\surd$\\
$\Delta\Gamma$ for $B_s$ & $B_s\to  J/\psi\eta'$, $ D_s^+\pi^-$, $K^+K^-$ &
$\surd$ & $\surd$ & $\surd$ & $\surd$ \\
\hline
\end{tabular}
\caption{Required CKM Measurements for $b$'s}
\end{center}
\end{table}

\subsubsection{Finding Inconsistencies}
Another interesting way of viewing the physics was given by Peskin
(Peskin 2000). Non-Standard Model physics would show up as discrepancies among
the values of $(\rho ,\eta)$ derived from independent determinations using CKM
magnitudes ($|V_{ub}/V_{cb}|$ and $|V_{td}/V_{ts}|$), or $B^o_d$ mixing ($\beta$ and
$\alpha$), or $B_s$ mixing ($\chi$ and $\gamma$).

\subsubsection{Required Measurements Involving $\beta$}

Besides a more precise measurement of $\sin 2\beta$ we need to 
resolve the ambiguities. There are two suggestions on how
this may be accomplished. Kayser (Kayser 1997) shows that time dependent
measurements of the final state
$J/\psi K^o$, where $K^o\to \pi \ell \nu$, give a direct measurement of
$\cos(2\beta)$ and can also be used for CPT tests. Another suggestion is to use
the final state $J/\psi K^{*o}$, $K^{*o}\to K_s\pi^o$, and to compare with
$B_s\to J/\psi\phi$ to extract the sign of the strong interaction phase shift
assuming SU(3) symmetry, and thus determine $\cos(2\beta)$ (Dighe 1998).

\subsubsection{Required Measurements Involving $\alpha$ and $\gamma$}
It is well known that $\sin (2\beta)$ can be measured without
problems caused by Penguin processes using the reaction $B^o\to J/\psi K_s$.
The simplest reaction that can be used to measure $\sin (2\alpha)$ is
$B^o\to \pi^+\pi^-$. This reaction can proceed via both the Tree and Penguin
diagrams shown in Figure~\ref{pippim}.

Current measurements shown in Table~\ref{tab:kp-pipi} show a large Penguin component.
The ratio of Penguin {\it amplitude} to Tree {\it amplitude} in the
$\pi^+\pi^-$ channel is about 15\% in magnitude.
Thus the effect of the Penguin must be
determined in order to extract $\alpha$. The only model independent way 
of doing this was suggested by Gronau and London, but requires the measurement
of $B^{\mp}\to\pi^{\mp}\pi^o$ and $B^o\to\pi^o\pi^o$, the latter being rather 
daunting.

There is however, a theoretically clean method to determine $\alpha$.
The interference between Tree and Penguin diagrams can be exploited by
 measuring the time dependent CP violating
 effects in the decays $B^o\to\rho\pi\to\pi^+\pi^-\pi^o$  
as shown by Snyder and Quinn (Snyder 1993).

\begin{figure}[htb]
\vspace{-0.4cm}
\centerline{\epsfig{figure=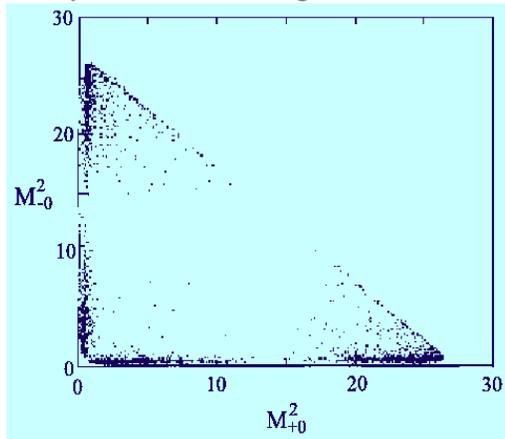,height=2.3in}}
\vspace{-.4cm}
\caption{\label{dalitz} The Dalitz plot for $B^o\to\rho\pi\to\pi^+\pi^-\pi^o$
from Snyder and Quinn.}
\end{figure}

The $\rho\pi$ final state has many advantages. First of all,
it has been seen with a relatively large rate. The 
branching ratio for the $\rho^o\pi^+$ final state as measured by CLEO is 
$(1.5\pm 0.5\pm 0.4)\times 10^{-5}$, and the rate for the neutral
 $B$ final state $\rho^{\pm}\pi^{\mp}$ is  
$(3.5^{+1.1}_{-1.0}\pm 0.5)\times 10^{-5}$, while the $\rho^o\pi^o$ final
state is limited at 90\% confidence level to $<5.1 \times 10^{-6}$
(Gao 1999). (BABAR (Bona 2001) measures
${\cal{B}}\left(B^o\to\rho^{\pm}\pi^{\mp}\right)$ as  
$(28.9\pm 5.4\pm 4.3)\times 10^{-6}$.) These
measurements are consistent with some theoretical expectations (Ali 1999).
Furthermore, the associated vector-pseudoscalar
Penguin decay modes have conquerable or smaller branching ratios. Secondly, since the 
$\rho$ is spin-1, the $\pi$ spin-0 and the initial $B$ also spinless, the $\rho$ 
is fully polarized in the (1,0) configuration, so it decays as $\cos^2\theta$, 
where $\theta$ is the angle of one of the $\rho$ decay products with the other
$\pi$ 
in the $\rho$ rest frame. This causes the periphery of the Dalitz plot to be 
heavily populated, especially the corners. A sample Dalitz plot is shown in 
Figure~\ref{dalitz}. This kind of distribution is good for maximizing the interferences, which 
helps minimize the error. Furthermore, little information is lost by excluding 
the Dalitz plot interior, a good way to reduce backgrounds.

To estimate the required number of events Snyder and 
Quinn preformed an idealized analysis that showed that a background-free,
flavor-tagged sample of 1000 to
2000 events was sufficient. The 1000 event sample usually yields good results 
for $\alpha$, but sometimes does not resolve the ambiguity. With the 2000 event sample, however, they always succeeded. 

This technique not only finds $\sin(2\alpha)$, it also determines 
 $\cos(2\alpha)$, thereby removing two of the remaining ambiguities. The final
 ambiguity can be removed using the CP asymmetry in $B^o\to\pi^+\pi^-$ and
 a theoretical assumption (Grossman 1997).

\begin{figure}[htb]
\centerline{\epsfig{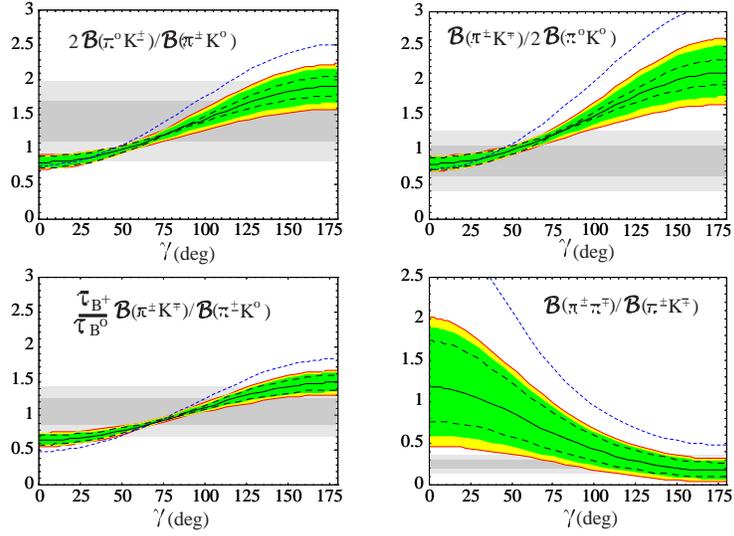}}
\caption{\label{beneke1} Model predictions from (Beneke 2001) ~as
a function of the indicated rate ratios. The dotted curve shows the predictions
from naive factorization. The curved bands show the total model uncertainties
where the inner band comes from theoretical input uncertainties, while
the outer band allows for errors to corrections on the theory. The specific
sensitivity to $|V_{ub}|$ is showed as the dashed curves. 
The gray bands show the current data
with a $1\sigma$ error while the lighter bands are at $2\sigma$.}
\end{figure}

Several model dependent methods using the light two-body pseudoscalar decay
rates have been suggested for measuring $\gamma$ The basic idea in all these
methods can be summarized as follows: $B^o\to\pi^+\pi^-$ has the weak decay
phase $\gamma$. In order to reproduce the observed suppression of the decay
rate for $\pi^+\pi^-$ relative to $K^{\pm}\pi^{\mp}$ we require a large
negative interference between the Tree and Penguin amplitudes. This puts
$\gamma$ in the range of 90$^{\circ}$. There is a great deal of theoretical
work required to understand rescattering, form-factors etc... We are left with
several ways of obtaining model dependent limits, due to Fleischer and Mannel 
(Fleischer 1998),
Neubert and Rosner (Neubert 1998),
Fleischer and Buras (Fleischer 2000), and Beneke \etal. (Beneke 2001). The latter 
make a sophisiticated model of QCD factorization and apply corrections.
Figure~\ref{beneke1} shows values of $\gamma$
that can be found in their framework, once better data are obtainable.

In fact, it may be easier to measure $\gamma$ than $\alpha$ in a model 
independent manner. There have been two methods suggested.

\begin{figure}[htb]
\centerline{\epsfig{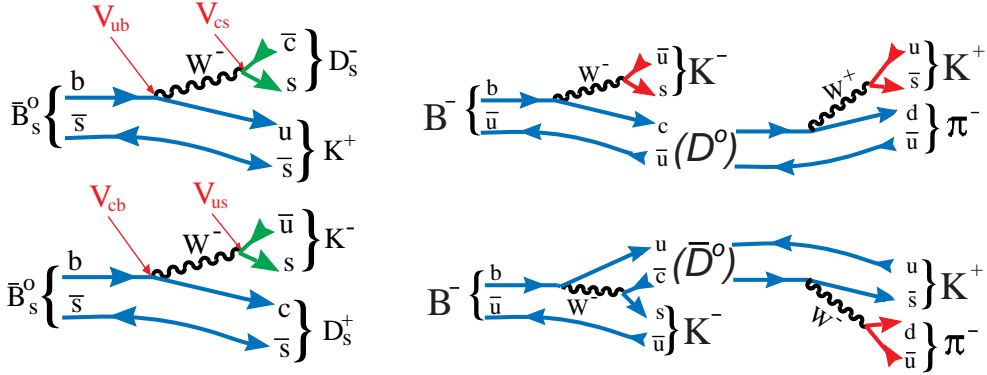}}
\caption{\label{DK_both} (left) The two diagram diagrams for $B_s\to
D_s^{\pm}K^{\mp}$ that interfere via $B_s$ mixing. (right) The two interfering
decay
diagrams for $B^-\to \overline{D}^o K^-$ where one is a $b\to u$ transition and
the other a doubly-Cabibbo suppressed decay.}
\end{figure}

(1) Time dependent flavor tagged analysis of $B_s\to D_s^{\pm}K^{\mp}$. This
is a direct model independent measurement (Du 1986) (Aleksan 1992) (Aleksan
1995). Here the Cabibbo suppressed $V_{ub}$ decay interferes with a somewhat
less suppressed $V_{cb}$ decay via $B_s$ mixing as illustrated in Figure~\ref{DK_both}
(left). Even though we are not dealing with CP eigenstates here there are no
hadronic uncertainties, though there are ambiguities.

(2) Measure the rate differences between $B^-\to \overline{D}^o K^-$ and
$B^+\to {D}^o K^+$ in two different $D^o$ decay modes such as $K^-\pi^+$
and $K^+ K^-$. This method makes use of the interference between the tree
and doubly-Cabibbo suppressed decays of the $D^o$, and does not depend
on any theoretical modeling (Atwood 1997) (Gronau 1991). See
Figure~\ref{DK_both}
(right).

\subsection{New Physics Tests in Specific Models}
\subsubsection{Supersymmetry}
Supersymmetry is a kind of super-model. The basic idea is that for every
fundamental fermion there is a companion boson and for every boson there
is a companion fermion. There are many different
implementations of couplings in this framework (Masiero 2000). In the most general case we pick up
80 new constants and 43 new phases. This is clearly too many to handle so
we can try to see things in terms of simpler implementations. In the minimum
model (MSSM) we have only two new fundamental phases. One $\theta_D$ would
arise in $B^o$ mixing and the other, $\theta_A,$ would appear in $B^o$ decay.
A combination would generate CP violation in $D^o$ mixing, call it $\phi_{K\pi}$
when the $D^o\to K^-\pi^+$ (Nir 1999). Table~\ref{tab:MSSM} shows the CP
asymmetry in three different processes in the Standard Model and the MSSM.

\begin{table}[htb]
\vspace{-2mm}
\begin{center}
\caption{CP Violating Asymmetries in the Standard Model and the MSSM.
\label{tab:MSSM}}
\vspace*{2mm}
\begin{tabular}{lcl}\hline\hline
Process & Standard Model & New Physics\\
\hline
$B^o\to J/\psi K_s$ & $\sin 2\beta$ & $\sin2(\beta+\theta_D)$\\
$B^o\to \phi K_s$ & $\sin 2\beta$ & $\sin2(\beta+\theta_D+\theta_A)$\\
$D^o\to K^-\pi^+$ & 0 & $\sim \sin\phi_{K\pi}$ \\
\hline\hline
\end{tabular}
\end{center}
\end{table}
Two direct effects of New Physics are clear here. First of all, the difference in
CP asymmetries between $B^o\to J/\psi K_s$ and $B^o\to \phi K_s$ would show
the phase $\phi_A$. Secondly, there would be finite CP violation in 
$D^o\to K^-\pi^+$ where none is expected in the Standard Model.

Manifestations of specific SUSY models lead to different patterns.
Table~\ref{tab:fNir} shows the expectations for some of these models in
terms of these variables and the neutron electric dipole moment $d_N$;
see (Nir 1999) for details.
\begin{table}[htb]
\vspace{-2mm}
\begin{center}
\caption{Some SUSY Predictions.
\label{tab:fNir}}
\vspace*{2mm}
\begin{tabular}{lcccc}\hline\hline
Model & $d_N\times 10^{-25}$ & $\theta_D $& $\theta_A$ &$\sin\phi_{K\pi}$ \\
\hline
Standard  Model & $\leq 10^{-6}$ & 0 & 0 & 0\\
Approx. Universality & $\geq 10^{-2}$  & $\cal O$(0.2) & $\cal O$(1)  & 0\\
Alignment  & $\geq 10^{-3}$  & $\cal O$(0.2) & $\cal O$(1)  & $\cal O$(1)\\
Heavy squarks  & $\sim 10^{-1}$  & $\cal O$(1) & $\cal O$(1)  & $\cal
O$($10^{-2}$)\\
Approx. CP & $\sim 10^{-1}$  & -$\beta$ & 0  & $\cal O$($10^{-3}$) \\
\hline\hline
\end{tabular}
\end{center}
\end{table}
Note, that ``Approximate CP" has already been ruled out by the measurements
of $\sin 2\beta$.

In the context of the MSSM there will be significant contributions to $B_s$
mixing, and the CP asymmetry in the charged decay $B^{\mp}\to \phi K^{\mp}$.
The contribution to $B_s$ mixing significantly enhances the CP violating
asymmetry in modes such as $B_s\to J/\psi \eta$. (Recall the CP asymmetry in
this mode is proportional to $\sin2\chi$ in the Standard Model.) 
The Standard Model and MSSM diagrams are shown in 
Figure~\ref{Bs_mssm}. The expected CP asymmetry in the MSSM is
$\approx \sin\phi_{\mu}\cos\phi_A\sin(\Delta m_s t)$, which is approximately
10 times the expected value in the Standard Model (Hinchliff 2001a).

\begin{figure}[htb]
\centerline{\epsfig{figure=Bs_mssm.eps,height=1.5in}}
\caption{\label{Bs_mssm} The Standard Model (left) and MSSM (right)
contributions to $B_s^o$ mixing.}
\end{figure} 

\begin{figure}[htb]
\centerline{\epsfig{figure=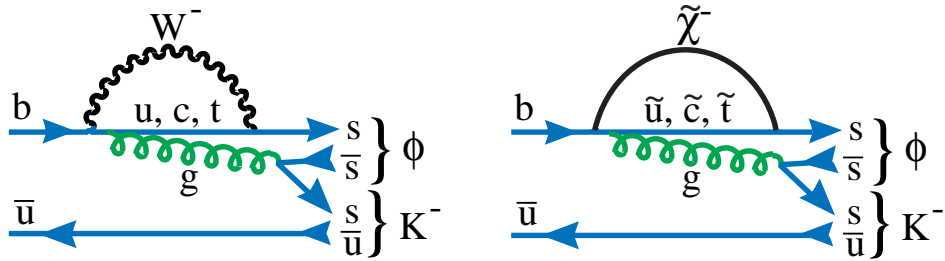,height=1.5in}}
\caption{\label{phi_K_mssm} The Standard Model (left) and MSSM (right)
contributions to $B^-\to \phi K^-$.}
\end{figure}

We observed that a difference between CP asymmetries in 
$B^o\to J/\psi K_s$ and $\phi K_s$ arises in the MSSM due to a CP asymmetry in
the decay phase. It is possible to observe this directly by looking for
a CP asymmetry in $B^{\mp}\to \phi K^{\mp}$. The Standard Model and MSSM
diagrams are shown in Figure~\ref{phi_K_mssm}. Here the interference of the two
diagrams provides the CP asymmetry. The predicted asymmetry is equal to
$\left(M_W/m_{squark}\right)^2\sin\phi_{\mu}$ in the MSSM, where $m_{squark}$
is the relevant squark mass (Hinchliff 2001a). 

\begin{figure}[htb]
\centerline{\epsfig{figure=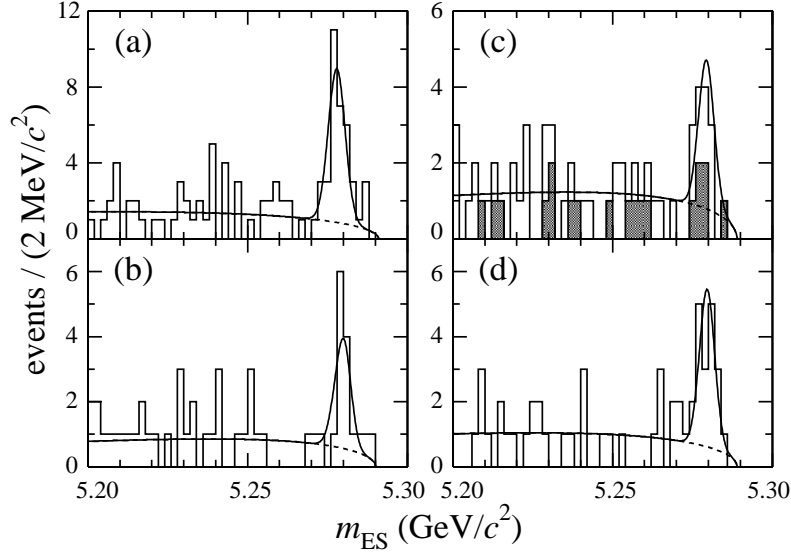,height=3in}}
\caption{\label{phi_K_BABAR} Projections of the maximum likelihood
fit onto the $B$ candidate mass $m_{ES}$ from BABAR. (a) $B^+\to \phi K^+$; (b)
$B^o\to \phi K^o$; (c) $B^+\to \phi K^{*+}$; (d) $B^o\to \phi K^{*o}$. In (c)
the histogram is the sum of the two contributing $K^*$ channels while the shaded
area is $K^o\pi^+$ alone (the other channel is $K^+\pi^o$). The solid line
shows the projection of the signal plus background fit while the dashed line shows the
projection of the background only.}
\end{figure}

The $\phi K$ and $\phi K^*$ final states have been observed, first by CLEO 
(Briere 2001) and
subsequently by BABAR (Aubert 2001). The BABAR data is shown in
Figure~\ref{phi_K_BABAR}. The average branching ratio is 
${\cal B}(B^-\to\phi K^-)=(6.8\pm 1.3)\times 10^{-6}$ showing that in principle
large samples can be acquired especially at hadronic machines. 

\subsubsection{Other New Physics Models}
There are many other specific models that predict New Physics in $b$ decays.
I list here a few of these with a woefully incomplete list of references, to
give a flavor of what these models predict. 
\begin{itemize}
\item {\it Two Higgs and Multi-Higgs Doublet Models-} They predict large
effects in $\epsilon_K$ and CP violation in $D^o\to K^-\pi^+$ with
only a few percent effect in $B^o$ (Nir 1999). Expect to see 1-10\% CP
violating effects in $b\to s\gamma$ (Wolfenstein 1994).
\item {\it Left-Right Symmetric Model-} Contributions compete with or even
dominate over Standard Model contributions to $B_d$ and $B_s$ mxing. This means
that CP asymmetries into CP eigenstates could be substantially different from
the Standard Model prediction (Nir 1999).
\item {\it Extra Down Singlet Quarks-} Dramatic deviations from Standard Model
predictions for CP asymmetries in $b$ decays are not unlikely (Nir 1999).
\item {\it FCNC Couplings of the $Z$ boson-} Both the sign and magnitude of the
decay leptons in $B\to K^*\ell^+\ell^-$ carry sensitive information on new
physics. Potential effects are on the of 10\% compared to an entirely
negligable Standard Model asymmetry of $\sim 10^{-3}$ (Buchalla 2000). These
models also predict a factor of 20 enhancement of $b\to d\ell^+\ell^-$ and 
could explain a low value of $\sin2\beta$ (Barenboim 2001a).
\item {\it Noncommutative Geometry-} If the geometry of space time is
noncommutative, i.e. $[x_{\mu},x_{\nu}]=i\theta_{\mu\nu}$, then CP violating
effects may be manifest a low energy. For a scale $<$2 TeV there are comparable
effects to the Standard Model (Hinchliffe 2001b).
\item {\it MSSM without new flavor structure-} Can lead to CP violation in
$b\to s\gamma$ of up to 5\% (Bartl 2001). Ali and London propose (Ali 1999)
that the Standard Model formulas are modified by Supersymmetry as
\begin{eqnarray}
\Delta m_d &=&\Delta m_d{\rm(SM)}\left[1+f\left(m_{\chi^{\pm}_2},m_{\tilde{t}_R},
m_{H^{\pm}},tan\beta\right)\right] \\
\Delta m_s &=&\Delta m_s{\rm(SM)}\left[1+f\left(m_{\chi^{\pm}_2},m_{\tilde{t}_R},
m_{H^{\pm}},tan\beta\right)\right] \\
\left|\epsilon_K\right|&=&{G_F^2f^2_KM_KM_W^2\over 6\sqrt{2}\pi^2\Delta M_K}
B_K(A^2\lambda^6\overline{\eta})\left[y_c\left(\eta_{ct}f_3(y_c,y_t)-\eta_{cc}
\right)\right. \nonumber \\
& &\left.+\eta_{tt}y_tf_s(y_t)\left[1+f\left(m_{\chi^{\pm}_2},m_{\tilde{t}_R},
m_{H^{\pm}},tan\beta\right)\right]A^2\lambda^4(1-\overline{\rho})\right]~~,
\end{eqnarray}
where $\Delta m (SM)$ refers to the Standard Model formula and the expression
for $\left|\epsilon_K\right|$ would be the Standard Model expression if $f$
were set equal to zero. Ali and London show that it is reasonable to expect
that $0.8>f>0.2$ so since the CP violating angles will not change
from the Standard Model, determining the value of $(\rho,~\eta)$ using the
magnitudes $\Delta m_s/\Delta m_d$ and $|\epsilon_K|$ will show an
inconsistency with values obtained using other magnitudes and angles. 
\item {\it Extra Dimensions-} We are beginning to see papers predicting $b$ decay
phenomena when the world has extra dimensions. See (Agashe 2001)
(Barenboim 2001b) (Branco 2001) (Chang 2001) and (Papavassiliou 2000).
\end{itemize}

I close this section with a quote from  Masiero and Vives (Masiero 2001):
``The relevance of SUSY searches in rare processes is not confined to the
usually quoted possibility that indirect searches can arrive `first' in signaling
the presence of SUSY. Even after the possible direct observation of SUSY
particles, the importance of FCNC and CP violation in testing SUSY remains of
utmost relevance. They are and will be complementary to the Tevatron and LHC
establishing low energy supersymmetry as the response to the electroweak
breaking puzzle."

I agree, except that I would replace ``SUSY" with ``New Physics."


\subsection{Expected Data Samples}
It is clear that precision studies of $b$ decays can bring a wealth of
information to bear on new physics, that probably will be crucial in sorting
out anything seen at the LHC. This is possible because we do expect to have
data samples large enough to test these ideas from
existing and approved experiments. In Table~\ref{tab:expectations} I show the expected rates in
BTeV for one year of running ($10^7$ s) and an $e^+e^-$ $B$-factory operating
at the $\Upsilon (4S)$ with a total accumulated sample of 500 fb${-1}$, about
what is expected around 2006. (LHCb numbers are the same order of magnitude
as the BTeV numbers for many of the modes.) 

\begin{table}[htb]
\vspace{-2mm}
\begin{center}
\caption{Comparisons of BTeV and $B$-factory Yields on Different Time Scales.
\label{tab:expectations}}
\vspace*{2mm}
\begin{tabular}{l|rrrrrr}\hline\hline
Mode&\multicolumn{3}{c}{BTeV $(10^7)s$}&\multicolumn{3}{c}{$B$-factory (500
fb$^{-1}$)}\\
  & Yield & Tagged$^{\dagger}$ & S/B & Yield & Tagged$^{\dagger}$ & S/B\\
\hline
$B_s\to J/\psi\eta^{(')}$ & 22000 & 2200 &$>$15 & &&\\
$B^-\to\phi K^-$ & 11000 & 11000 & $>$10 & 700 & 700 &4\\
$B^o\to\phi K_s$ & 2000 & 200 & 5.2 & 250 & 75 & 4\\
$B^o\to K^{*o}\mu^+\mu^-$ & 4400 &4400 & 11 & $\sim$50 & $\sim$50 & ?\\
$D^{*+}\to\pi^+ D^o$; $D^o\to K^-\pi^+$ & $\sim 10^8$ &$\sim 10^8$ & large &
$8\times 10^5$ & $8\times 10^5$ & large \\\hline\hline
\multicolumn{7}{l}{$\dagger$ Tagged here means that the intial flavor of the $B$
is determined.} 
\end{tabular}

\end{center}
\end{table}
\section{Conclusions}
I have attempted to cover the length and breadth of $b$ physics, and have only
scratched the surface. There is much  more to be said and much more to learn.
Why are there three families? What is the connection with neutrinos and that
mixing matrix? How do we explain the mystery of flavor? These and many more
unanswered questions I leave to the students.

\section*{Acknowledgments}

This research was supported by the US National Science Foundation.
I thank Ken Peach and Steve Playfer for a well organized and
interesting school that was very enjoyable to attend. The sunny day at
the Himalayas was memorable.
  
I thank Jon Rosner for several seminal discussions during the course of the
school. My colleagues at BTeV and CLEO contributed much to my understanding. In
particular I thank, M. Artuso, J. Butler and T. Skwarnicki. 

\section*{References}
\frenchspacing
\begin{small}

\reference{Abe F, et al., 1998,  
\textit{Phys. Rev. D} \textbf{58}, p.5513, [hep-ex/9804012].}

\reference{Abe K, et al., 2001a,  Measurement of Branching Fractions for $B\to\pi\pi$,
$K\pi$ and $KK$ Decays,
\textit{Phys. Rev. Lett.} \textbf{87}, p.101801, [hep-ex/0104030].}

\reference{Abbiendi G, et al., 2000, 
\textit{Phys. Lett. B} \textbf{482}, p.15 [hep-ex/0003013].}

\reference{Abreu P, et al., 2000, Determination of $|V_{ub}|/|V_{cb}|$ with
DELPHI at LEP,
\textit{Phys. Lett. B} \textbf{478}, p.14 [hep-ex/0105054].}

\reference{Abreu P, et al., 2001, 
\textit{Phys. Lett. B} \textbf{510}, p.55 [hep-ex/0104026].}

\reference{Acciarri M, et al., 1998,
\textit{Phys. Lett. B} \textbf{436}, p.174.}

\reference{Agashe K, Deshphande NG and Wu GH, 2001, Universal Extra Dimensions
and $b\to s\gamma$,
\textit{Phys. Lett. B} \textbf{514}, p.309 [hep-ph/0105084].}

\reference{Ahmed S, et al., 2000,
\textit{Phys. Rev. D},\textbf{62}, p.112003 [hep-ex/0008015].}

\reference{Akers R, et al., 1995,  
\textit{Z. Phys. C} \textbf{66}, p.555.}

\reference{Alam MS, et al., 1995,
\textit{Phys. Rev. Lett.}, \textbf{74}, p.2285.}

\reference{Albajar, et al., 1987,  
\textit{Phys. Lett. B} \textbf{186}, p.247, erratum-ibid B{\bf 197} p.565.}

\reference{Albrecht H, et al., 1983,  
\textit{Phys. Lett. B} \textbf{192}, p.245.}

\reference{Albrecht H, et al., 1990,  
\textit{Phys. Lett. B} \textbf{234}, p.16.}

\reference{Aleksan R, Dunietz I and Kayser B, 1992,  
\textit{Z. Phys. C} \textbf{54}, p.653.}

\reference{Aleksan R, Kayser B and London D, 1994,  
\textit{Phys. Rev. Lett.} \textbf{73}, p.18 [hep-ph/9403341].}

\reference{Aleksan R, et al., 1995,  
\textit{Z. Phys. C} \textbf{67}, p.251 [hep-ph/9407406.}

\reference{Alexander J, et al., 1996,  
\textit{Phys. Rev. Lett.} \textbf{77}, p.5000.}

\reference{Alexander J, et al., 2001,  
\textit{Phys. Rev. D} \textbf{64}, p.092001 [hep-ex/0103021].}

\reference{Ali A and Greub C, 1991 ,
\textit{Phys. Lett. B} \textbf{259}, p.182.}

\reference{Ali A, Kramer G and Lu CD,
1999,
\textit{Phys. Rev. D} \textbf{59} p.014005 [hep-ph/9805403].}

\reference{Altarelli G, 1982,
\textit{Nucl. Phys. B}, \textbf{208}, p.365.}

\reference{Ammar R, et al., 1993,
\textit{Phys. Rev. Lett.}, \textbf{71}, p.674.}

\reference{Appel JA, et al., 2001, Performance of Prototype BTeV Silicon Pixel
Detectors in a High Energy Pion Beam, [hep-ex/0108014].  }

\reference{Artuso M, 1994, Experimental Facilities for b-quark Physics, in
\textit{$B$ Decays 2nd Edition} ed. S. Stone, World Scientific, Singapore,
p.80.}

\reference{Artuso M and Barberio E, 2001,
\textit{Private communication}; also see \newline
http://www-cdf.lbl.gov/$\sim$weiming/vcb\_summary\_ckm\_workshop.ps~.}

\reference{Atwood D, Dunietz I and Soni A, 1997,
\textit{Phys. Rev. Lett.} \textbf{78}, p.3257.}

\reference{Atwood M and Jaros RA, 1994, Lifetimes,
\textit{$B$ Decays 2nd Edition} ed. S. Stone, World Scientific, Singapore,
p. 364.}

\reference{Aubert B, et al., 2001,
\textit{Phys. Rev. Lett.}, \textbf{87}, p.151801, [hep-ex/0105001].}

\reference{Ball P and Braun VM, 1998,
\textit{Phys. Rev. D} \textbf{58}, p.094016.}

\reference{Bander M, Silverman D and Soni A, 1979,
\textit{Phys. Rev. Lett.}, \textbf{43}, p.242.}

\reference{Barate R, et al., 1998,
\textit{Phys. Lett. B} \textbf{429}, p.169.}

\reference{Barate R, et al., 1999, 
\textit{Eur. Phys. J.} \textbf{C6}, p.555.}

\reference{Barenboim G, Botella FJ and Vives O, 2001a,
\textit{Phys. Rev. D} \textbf{64}, p.015007 [hep-ph/0012197].}

\reference{Barenboim G, Botella FJ and Vives O, 2001b, Constraining Models
With Vector-Like Fermions from FCNC in K and B Physics,
\textit{Nucl. Phys. B} \textbf{613}, p.285 [hep-ph/01050306].}

\reference{Bartl A, et al., 2001,
\textit{Phys. Rev. D} \textbf{64}, p.076009 [hep-ph/0103324].}

\reference{Bauer CW, Pirjol D and Steward IW, 2001, A Proof of Factorization
for $B\to D\pi$,
\textit{UCSD-01-12} [hep-ph/0107002].}

\reference{Bauer M and Wirbel M, 1989, 
\textit{Z. Phys. C} \textbf{42}, p.671.}

\reference{Bauer CW, Ligeti Z and Luke M, 2001, Precision Determination of $|V_{ub}|$
>From Inclusive Decays,
\textit{UTPT-01-08} [hep-ph/0107074].}

\reference{Beneke M, et al., 2000,
\textit{Nucl Phys. B} \textbf{591}, p.313 [hep-ph/0006124].}

\reference{Beneke M, et al., 2001,
\textit{Nucl Phys. B} \textbf{606}, p.245 [hep-ph/0104110].}

\reference{Beyer M and Melikhov D, 1998,
\textit{Phys. Lett. B} \textbf{436}, p.344 [hep-ph/9807223].}

\reference{Bigi II, et al., 1997,
\textit{Annu. Rev. Nucl. Part. Sci.} \textbf{47}, p.591.}

\reference{Bigi II and Sanda AI, 2000,
\textit{CP Violation}, Cambridge University Press, Cambridge, UK.}

\reference{Bona M, 2001,
\textit{BABAR-CONF-01/71, SLAC-PUB-9045} [hep-ex/0111017].}

\reference{Bortoletto D and Stone S, 1990,
\textit{Phys. Rev. Lett.} \textbf{65}, p.2951.}

\reference{Branco GC, de Gouvea A and Rebelo MN, 2001, Split Fermions in Extra
Dimensions and CP Violation,
\textit{Phys. Lett. B} \textbf{506}, p.115 [hep-ph/0012289].}

\reference{Briere RA, et al., 2001,
\textit{Phys. Rev. Lett.} \textbf{86}, p.3718, [hep-ex/0101032].}

\reference{Browder TE, Honscheid K and Pedrini D, 1996, Nonleptonic Decays and
Lifetimes of b-quark and c-quark Hadrons,
\textit{Ann. Rev. Nucl. Part. Sci.} \textbf{46}, p.395 [hep-ph/9606354].}

\reference{BTeV C, 2000, BTeV An Experiment to Measure Mixing, CP Violation
and Rare Decays of Beauty and Charm at the Fermilab Collider,
[hep-ex/0006037].}

\reference{Buchalla G, Hiller G and Isidori G, 2000,
\textit{Phys. Rev. D} \textbf{63}, p.014015 [hep-ph/0006136].}

\reference{Buras AJ, 1995,
\textit{Nucl. Instrm. and Meth. A}, \textbf{368}, p.1.}

\reference{Buskulic D, et al., 1997,
\textit{Phys. Lett. B} \textbf{395}, p.373.}

\reference{Caprini I, Lellouch L and Neubert M.
\textit{Nucl. Phys. B} \textbf{530}, p.153 [hep-ph/9712417].}

\reference{Chang D, Keung WY and Mohapatra RN, 2001, Models for Geometric CP
Violation With Extra Dimensions,
\textit{Phys. Lett. B} \textbf{515}, p.431 [hep-ph/0105177].}

\reference{Chen S, 2001, Branching Fraction and Photon Energy Spectrum for 
$b\to s\gamma$, [hep-ex/0108032].}

\reference{Ciuchini M, 2001, 
\textit{JHEP} \textbf{0107}, p.23 [hep-ph/0012308].}

\reference{Clegg AB and Donnachie A, 1994
\textit{Z. Phys. C} \textbf{62}, p.455.}

\reference{Debbio LD et al., 1998,
\textit{Phys. Lett. B} \textbf{416}, p.392.}

\reference{Desphande NG, 1994, Theory of Penguins in $B$ Decays, in
\textit{$B$ Decays 2nd Edition} ed. S. Stone, World Scientific, Singapore,
p. 80 and references therein.}

\reference{Dighe A, Dunietz I and Fleischer R, 1998, 
\textit{Phys. Lett. B} \textbf{433}, p.147 [hep-ph/9804254].}

\reference{Du D-S, Dunietz I and Wu D-D, 1986,  
\textit{Phys. Rev. D} \textbf{34}, p.3414.}

\reference{Ellis J, 2000, Highlights of CP 2000,
\textit{Nucl. Phys. Proc. Suppl.} \textbf{99A}, p.331 [hep-ph/0011396].}

\reference{Fleischer R and Mannel T, 1998,  
\textit{Phys. Rev. D} \textbf{57}, p.2752 [hep-ph/9704423].}

\reference{Fleischer R and Buras AJ, 2000, Constraints on $\gamma$ and Strong
Phases from $B\to\pi K$ Decays," presented at ICHEP 2000, Osaka, Japan,
July 2000. To appear in the Proceedings [hep-ph/0008298].} 

\reference{Fulton R, et al., 1990, 
\textit{Phys. Rev. Lett.}, \textbf{64}, p.16.}

\reference{Gaillard M and Lee B, 1974,  
\textit{Phys. Rev. D} \textbf{10}, p.897.}

\reference{Gao Y and W\"urthwein F, 1999,
[hep-ex/9904008].}

\reference{Gavela MB, Hern\'andez P, Orloff J and P\`ene O, 1993, Standard 
Model CP Violation and Baryon Asymmetry,
\textit{Mod. Phys. Lett. A} \textbf{9}, p.795 [hep-ph/9312215].}

\reference{Gilman FJ and Singleton R, 1990,
\textit{Phys. Rev. D} \textbf{41}, p.142.}

\reference{Greub FJ, Ioannissian A and Wyler D, 1995,
\textit{Phys. Lett. B} \textbf{346}, p.149 [hep-ph/9408382].}

\reference{Grinstein B, Isgur N and Wise MB, 1986
\textit{Phys. Rev. Lett.} \textbf{56}, p.258.}

\reference{Gronau M and Wyler D, 1991,
\textit{Phys. Lett. B} \textbf{265}, p.172.}

\reference{Groom DE, et al., 2001, The Particle Data Group,
\textit{The European Physics Journal C} \textbf{15}, p.1.} 

\reference{Grossman Y and Quinn HR, 1997,
\textit{Phys. Rev. D} \textbf{56} p.7259 [hep-ph/9705356].}

\reference{Hashimoto S, et al., 2001,
``Lattice Calculation of the Zero-recoil Form Factor of $\overline{B}
\to D^*\ell^-\bar{\nu}$: Toward a Model Independent Determination of
$|V_{cb}|$," [hep-ph/0110253].}

\reference{Heltsley BK, 2001, CKM Status and Prospects,
\textit{To appear in Proceedings of XXI Physics in Collision, Seoul, Korea,
June, 2001}, [hep-ph/0110260]/}
 
\reference{Hinchliff I and Kersting N, 2001a, Constraining CP Violating
Phases of the MSSM,
\textit{Phys. Rev. D} \textbf{63}, p.015003 [hep-ph/0003090].}

\reference{Hinchliff I and Kersting N, 2001b, CP Violation from Noncommutative
Geometry,
\textit{LBNL-47750} [hep-ph/0104137].}

\reference{Hocker A, Lacker H, Laplace S, Le Diberder F, 2001, A New Approach
to a Global Fit of the CKM Matrik,
\textit{Eur. Phys. J. C} \textbf{21}, p.25 [hep-ph/0104062].}

\reference{Hurth T, 2001, ``Inclusive Rare B Decays,
\textit{CERN-TH/2001-146}, [hep-ph/0106050].}

\reference{Isgur N and Scora D, Grinstein B and Wise MB, 1989, 
\textit{Phys. Rev. D} \textbf{39}, p.799.}

\reference{Isgur N and Wise MB, 1990, 
\textit{Phys. Lett. B} \textbf{237}, p.527.}

\reference{Isgur N and Wise MB, 1994, ``Heavy Quark Symmetry," in 
\textit{$B$ Decays 2nd Edition} ed. S. Stone, World Scientific, Singapore,
p. 231 and references therein.}

\reference{Isgur N and Scora D, 1995, 
\textit{Phys. Rev. D} \textbf{52}, p.2783.}

\reference{Isgur N, 1999, Duality-Violating $1/m_Q$ Effects in Heavy Quark
Decays,
\textit{Phys. Lett. B}, \textbf{448}, p.111 [hep-ph/98113777].}

\reference{Kayser B, 1997, Cascade Mixing and the CP-Violating Angle Beta,
[hep-ph/9709382],}

\reference{Kobayashi M and Maskawa K, 1973,
\textit{Prog. Theor. Phys.} \textbf{49}, p.652.}

\reference{K\"orner JG and Schuler GA, 1988, 
\textit{Z. Phys. C} \textbf{38}, p.511.}

\reference{Leroy O, 2001,
\textit{http://lepbosc.web.cern.ch/LEPBOSC/people.html} and references therein.}

\reference{Ligeti Z and Wise M, 1996,
\textit{Phys. Rev. D} \textbf{53}, p.4937.}

\reference{Lingel K, Skwarnicki T and Smith JG, 1998, Penguin Decays of B
Mesons,
\textit{Ann. Rev. of Nucl. \& Part. Science} \textbf{48}, p.253 [hep-ex/9804015].}

\reference{Luke M, 1990,
\textit{Phys. Lett. B} \textbf{252} p.447.}

\reference{Masiero A and Vives O, 2000, New Physics Behind the Standard Model's
Door?,
\textit{Int. School on Subnuclear Physics, Erice, Italy, 1999} [hep-ph/0003133].}

\reference{Masiero A and Vives O, 2001, New Physics in CP Violation
Experiments,
\textit{Ann. Rev. of Nucl. \& Part. Science} \textbf{51},  [hep-ph/0104027].}

\reference{Mele S, 1999, 
\textit{Phys. Rev. D} \textbf{59}, p.113011.}

\reference{Melikhov D, 1996, 
\textit{Phys. Rev. D} \textbf{53}, p.2460.}

\reference{Muheim F, 2001, Status of the LHCb Experiment,
\textit{Nucl. Instrum and Meth. A} \textbf{462}, p. 233 [hep-ex/0012059].}

\reference{Neubert M, 1991,
\textit{Phys. Lett. B} \textbf{264}, p.455.}

\reference{Neubert M, 1996,
\textit{Int. J. Mod. Phys. A} \textbf{11}, p.4173 [hep-ph/9604412].}

\reference{Neubert M and Stech B,  1997, ``Non-Leptonic Weak Decays of B Mesons, in
\textit{Heavy Flavours, 2nd Edition}, ed. Buras AJ and Lindner M,
World, Scientific, Singapore, p.294 [hep-ph/9705292].}

\reference{Neubert M and Rosner JL, 1998,
\textit{Phys. Rev. Lett.} \textbf{81}, p.5076 [hep-ph/9809311].}

\reference{Neubert M, 2000,
\textit{JHEP} \textbf{7}, p.22 [hep-ph/0006068].}

\reference{Nir Y, 1999, CP Violation In and Beyond the Standard Model,
\textit{IASSNS-HEP-99-96} [hep-ph/9911321].}

\reference{Papavassiliou J and Santamaria A, 2001, Extra Dimensions at the One
Loop Level: $Z\to b\overline{b}$ and $B-\overline{B}$ Mixing,
\textit{Phys. Rev. D} \textbf{63}, p.016002 [hep-ph/016002].}

\reference{Peskin ME, 2000, Theoretical Summary Lecture for EPS HEP99,
to appear in proceedings [hep-ph/0002041].}

\reference{Ramirez C, Donoghue JF, and Burdman G, 1990, 
\textit{Phys. Rev. D} \textbf{41}, p.1496.}

\reference{Richman JD and Burchat PR, 1995,
\textit{Rev. Mod. Phys.} \textbf{67}, p.893 and references therein.}

\reference{Rosner J, 2001, The Standard Model in 2001,
\textit{In these proceedings,} and references therein [hep-ph/0108195].}

\reference{Sachrajda CT, 1999, Lattice B-physics,
\textit{Nucl. Instru. and Meth. A} \textbf{462}, p.23 [hep-lat/9911016].}

\reference{Sharma VA and Weber FV, 1994, Recent Measurements of the Lifetimes
of b Hadrons,
\textit{$B$ Decays 2nd Edition} ed. S. Stone, World Scientific, Singapore,
p. 395.}

\reference{Silva JP and Wolfenstein L, 1997, 
\textit{Phys. Rev. D} \textbf{55}, p.5331 [hep-ph/9610208].}

\reference{Skwarnicki T, 2001, Overview of the BTeV Experiment,
\textit{Nucl. Instrum and Meth. A} \textbf{462}, p. 227.}

\reference{Snyder AE and Quinn HR, 1993, 
\textit{Phys. Rev. D} \textbf{48}, p.2139.}

\reference{Stone S, 2000, Hadronic $B$ Decays to Charm from CLEO, in
\textit{Proc. of XXXth Int. Conf. on High Energy Physics, Aug. 2000, Osaka,
Japan} p.842 [hep-ex/0008070].}

\reference{Tajima H, 2001, ``Belle B Physics Results,"
\textit{To appear in Proceedings of the XX Int. Symp on Lepton and Photon
Interactions at High Energies, July, 2001, Rome, Italy}, [hep-ex/0111037].}

\reference{Wirbel M, et al., 1985,
\textit{Z. Phys. C} \textbf{25}, p.627.}

\reference{Wirbel M, Stech B and Bauer M, 1989, 
\textit{Z. Phys. C} \textbf{29}, p.637.}

\reference{Wise M, 2001, ``Recent Progress in Heavy Quark Physics,"
\textit{To appear in Proceedings of the XX Int. Symp on Lepton and Photon
Interactions at High Energies, July, 2001, Rome, Italy}, [hep-ph/0111167].}

\reference{Wolfenstein L, 1983,
\textit{Phys. Rev. Lett.} \textbf{51}, p.1945.}

\reference{Wolfenstein L and Wu YL, 1994,
\textit{Phys. Rev. Lett.} \textbf{74}, p.2809 [hep-ph/9410253].}
\end{small}

\end{document}